\documentclass[%
 twocolumn,
 amsfonts,amsmath,amssymb,
 aps,
prd,
floatfix,
showpacs,
longbibliography,
nofootinbib,
nobibnotes
]{revtex4-2}

\usepackage[normalem]{ulem}
\usepackage{graphicx}
\usepackage{dcolumn}
\usepackage{bm}
\usepackage{wasysym}
\usepackage{xcolor,soul}
\usepackage{orcidlink}
\definecolor{ReflexBlue}{rgb}{ .0902,.0902,.5882}
\usepackage{hyperref}
\hypersetup{
    colorlinks=true,
    linkcolor=ReflexBlue,
    filecolor=magenta,      
    urlcolor=ReflexBlue,
    citecolor=ReflexBlue,
}
\let\Phi\varPhi

\newcommand{\Rs}{R_{\mathrm{S}}}
\newcommand{\beq}{\begin{equation}}
\newcommand{\eeq}{\end{equation}}

\definecolor{CPcolor}{rgb}{0.2, 0.13, 0.9}
\definecolor{PBcolor}{rgb}{0.9, 0.5, 0.0}
\newcommand{\EE}{{\mathcal{E}}}
\newcommand{\PP}{{\mathcal{P}}}
\newcommand{\TT}{{\mathcal{T}}}
\newcommand{\pt}{p_\perp}

\begin{document}

\title{Slowly rotating anisotropic relativistic stars}
\author{Philip Beltracchi \orcidlink{0000-0002-4226-7796}}
\email{phipbel@aol.com}
\affiliation{}
\author{Camilo Posada \orcidlink{0000-0001-8826-1974}}
\email{camilo.posada@physics.slu.cz}
\affiliation{Research Centre for Theoretical Physics and Astrophysics, Institute of Physics, Silesian University in Opava, Bezru\v{c}ovo n\'{a}m. 13, CZ-746 01 Opava, Czech Republic}
\affiliation{Programa de Matem\'atica, Fundaci\'on Universitaria Konrad Lorenz, 110231 Bogot\'a, Colombia}

\begin{abstract}
The present paper is devoted to a study of the equilibrium configurations of slowly rotating anisotropic stars in the framework of general relativity. For that purpose, we provide the equations of structure where the rotation is treated to second order in the angular velocity. These equations extend those first derived by Hartle for slowly rotating isotropic stars. As an application of the new formalism, we study the rotational properties of Bowers-Liang fluid spheres. A result of particular interest is that the ellipticity and mass quadrupole moment are negative for certain highly anisotropic configurations; thus, such systems are prolate rather than oblate. Furthermore, for configurations with high anisotropy and compactness close to their critical value, quantities like the moment of inertia, change of mass, and mass quadrupole moment approach to the corresponding Kerr black hole values, similar to other ultracompact systems like sub-Buchdahl Schwarzschild stars and analytic rotating gravastars.
\end{abstract}

\maketitle
\section{Introduction}

A fundamental result in general relativity (GR), due to Buchdahl~\cite{Buchdahl:1959zz}, establishes that the maximum size of a spherically symmetric self-gravitating compact object of total mass $M$ and radius $R$, whose mass-energy density is non-negative and decreases monotonically outward, is given by $R=(9/4)M$. The existence of this limit in GR is independent of the equation of state (EOS) describing the star's composition. Thus, in principle, Buchdahl's bound forbids the existence of ultracompact objects with a radius extremely close to that of a nonrotating black hole (BH) given by the Schwarzschild radius.  

One of the assumptions in deriving Buchdahl's limit is that the fluid pressure is isotropic. However, it has been known for some time now that various physical phenomena can lead to local \emph{anisotropies}. For instance, Ruderman~\cite{Ruderman:1972} proposed that nuclear matter may be anisotropic at least at very high densities $\rho > 10^{15}~\mathrm{g}/\mathrm{cm}^3$, typical of the inner core of a realistic neutron star (NS). Anisotropies can also be produced due to the existence of an inner solid core in the star~\cite{Kippenhahn:2012qhp}, pion condensation~\cite{Sawyer:1972cq}, scalar fields minimally coupled to gravity~\cite{Liebling:2012fv}, and strong magnetic fields~\cite{Yazadjiev:2011ks,Folomeev:2015aua}, among others (see~\cite{Herrera:1997plx} for a review). Thus, anisotropic configurations, rather than isotropic, seem to be a more plausible scenario for the description of realistic astrophysical compact objects.

A compact object can evade Buchdahl's bound if it is subjected to anisotropic stresses. Thus, one may expect to have anisotropic configurations with a radius below Buchdahl's radius. The fact that anisotropic spheres can support greater compactness has been known at least since Ref.~\cite{Lemaitre:1933gd}, and the idea was brought to wider attention with the exact solution found by Bowers and Liang~\cite{Bowers:1974tgi} for an anisotropic fluid sphere with uniform density. Further exact anisotropic solutions were found by Refs.~\cite{Bayin:1982vw, Mak:2001eb, Dev:2000gt, Herrera:2004}. The effect of the anisotropy in the dynamical stability of compact objects was studied by Refs.~\cite{Hillebrandt:1976,Dev:2003qd}, and some general properties of anisotropic stars have been studied by a number of authors~\cite{Doneva:2012rd, Yagi:2015hda, Carloni:2017bck, Raposo:2018rjn, Ovalle:2019lbs, Pretel:2020xuo, Pretel:2022plg}.

Despite the progress on the study of static anisotropic stars, little has been explored about their rotational properties. Bayin~\cite{Bayin:1982vw} considered certain models of slowly rotating anisotropic stars, at first order in the angular frequency $\Omega$. More recently, Silva {\it et al.}~\cite{Silva:2014fca} considered slowly rotating anisotropic NSs in GR and scalar-tensor theories, at first order in $\Omega$. There have been also previous approaches aiming to generalize Hartle's second order in $\Omega$ framework~\cite{Hartle:1967he, Hartle:1968si} to anisotropy~\cite{Pattersons:2021lci, Beltracchi:2022vvn}. However, Ref.~\cite{Beltracchi:2022vvn} was predominantly concerned with specific systems and would not be easily applicable to the Bowers-Liang sphere. Although Ref.~\cite{Pattersons:2021lci} considered the Bowers-Liang model in slow rotation, they did not carry out a thorough study of integral and surface properties, as done by~\cite{Chandra:1974, Beltracchi:2023qla} for slowly rotating isotropic constant density stars. Moreover, \cite{Pattersons:2021lci} was not concerned on looking at systems which approach the BH compactness. Furthermore, Ref.~\cite{Pattersons:2021lci} contained some typos and other errors, which marred the results presented there.

Our aim in this work is to improve and extend the results of Ref.~\cite{Pattersons:2021lci} on several accounts. First of all, following Hartle's methods, we derive the equations of structure for slowly rotating anisotropic relativistic masses, at second order in the angular velocity $\Omega$. As an application of the new formalism, we study in detail the equilibrium configurations of slowly rotating Bowers-Liang spheres, by solving the extended structure equations for such configurations. Thus, our analysis can be considered an extension of those presented by Refs.~\cite{Chandra:1974,Beltracchi:2023qla} for isotropic homogeneous masses. Furthermore, our treatment completes the results of~\cite{Pattersons:2021lci} in various directions; for instance, we determine all the metric and energy-momentum tensor perturbation functions, including quantities like the moment of inertia, ellipticity, and mass quadrupole moment which were not addressed by Ref.~\cite{Pattersons:2021lci}. We also paid attention to highly anisotropic configurations, which can remain nonsingular close to the Schwarzschild limit.

The structure of the paper is as follows. In Sec.~\ref{sect:hartle}, we derive the equations of structure of slowly rotating anisotropic spheres, at the second order in $\Omega$. The formalism developed in this section is model independent apart from very basic assumptions such as regularity. In Sec.~\ref{sect:bl}, we briefly discuss the Bowers-Liang solution for anisotropic stars with uniform density. The rotational perturbations, as well as the surface and integral properties of slowly rotating Bowers-Liang spheres, obtained from the numerical integration of the extended structure equations, are presented in Sec.~\ref{sect:results}. In Sec.~\ref{sect:concl}, we summarize our main conclusions and open questions for future work. In Appendix~\ref{app:dM}, we give a detailed derivation of a generalized formula to compute the change of mass. In Appendix~\ref{app:grav}, we depict a brief analysis of the Bowers-Liang spheres in the gravastar limit. In Appendix~\ref{app:table} we provide a table with some of the numerical results of surface and integral properties of slowly rotating Bowers-Liang spheres in GR. Throughout the paper we employ geometrical units $(c=G=1)$, unless stated otherwise, and signature $(-,+,+,+)$.
\section{Hartle formalism for relativistic masses with anisotropic pressure}
\label{sect:hartle}
\subsection{A nonrotating anisotropic stellar model is computed}
The field equations, at the zeroth order in $\Omega$, provide the relation between the mass of the star and its central energy density. The starting point is a nonrotating and spherically symmetric configuration in the standard Schwarzschild form
\beq\label{metric_stat}
ds^2 = -e^{2\nu(r)}dt^2 + e^{2\lambda(r)}dr^2 + r^2 (d\theta^2 + \sin^2 d\phi^2),
\eeq
\noindent where $\nu$ and $\lambda$ are functions of the radial coordinate $r$, only. Let us consider the most general static spherically symmetric energy-momentum tensor (EMT), which may have anisotropic stresses. Specifically, it has a radial pressure $p_r$, as well as a “transverse pressure" $p_\perp$ corresponding to the stress along the local plane perpendicular to $p_r$. Thus, a locally anisotropic EMT takes the general form
\beq\label{emt}
T^{\mu}_{\,\,\nu}=\mathrm{diag}\left(-\rho, ~p_{r},~p_{\perp},~p_{\perp}\right),
\eeq
\noindent where $\rho$ is the energy density and $p_{\perp}=p_{\theta}=p_{\phi}$. Note that the form \eqref{emt} is completely general, regardless of the EOS. This configuration is Segre type [(11)1,1], where the EMT has two degenerate eigenvalues with spacelike eigenvectors, a distinct eigenvalue with a spacelike eigenvector, and a distinct eigenvalue with a timelike eigenvector~\cite{Beltracchi:2022vvn}. The Einstein equations $G^{\mu}_{\,\,\nu}=8\pi T^{\mu}_{\,\,\nu}$ for this spacetime geometry and matter distribution give 
\begin{subequations}\label{EE_stat}
\beq\label{EE_stat_t}
e^{-2\lambda}\left(\frac{2\lambda'}{r}-\frac{1}{r^2}\right)+\frac{1}{r^2}=8\pi\rho,
\eeq
\beq\label{EE_stat_r}
e^{-2\lambda}\left(\frac{2\nu'}{r}+\frac{1}{r^2}\right)-\frac{1}{r^2}=8\pi p_{r},
\eeq
\beq\label{EE_stat_th}
e^{-2\lambda}\left[\nu'' + (\nu')^2 - \nu'\lambda' +   
\frac{(\nu'-\lambda')}{r}\right]=8\pi p_{\perp},
\eeq
\end{subequations}
\noindent where a prime $\equiv d/dr$. The system \eqref{EE_stat} consists of three equations with five unknowns. Therefore, we are required to provide two EOSs which connect the behavior of $\rho$, $p_r$, and $p_\perp$. Equation~\eqref{EE_stat_t} can be integrated to give the standard relation
\beq
e^{-2\lambda}=1-\frac{2m(r)}{r},
\eeq
\noindent where
\beq\label{mass}
\frac{dm}{dr}=4\pi\rho(r)r^2
\eeq
gives $m(r)$, the total mass enclosed in the radius $r$. The total mass of the configuration is $M=m(R)$, where $R$ denotes the stellar radius. From the conservation of the EMT, $\nabla_{\mu}T^{\mu}_{\,\,\,\nu}=0$, we obtain the equation for hydrostatic equilibrium as   
\beq\label{tov}
\frac{dp_r}{dr}=-\left(\rho + p_{r}\right)\nu' + \frac{2}{r}(p_{\perp}-p_r),
\eeq
which generalizes the Tolman-Oppenheimer-Volkoff (TOV) equation to anisotropic configurations. In the isotropic case, i.e. $\pt=p_r$, Eq.~\eqref{tov} reduces to the standard TOV equation.

In the exterior vacuum spacetime, $\rho=p=0$; thus, the spacetime geometry is described by Schwarzschild's exterior solution
\beq
e^{2\nu(r)}=e^{-2\lambda(r)}=1-\frac{2M}{r},\quad r>R.
\eeq
Unless there is a junction layer at the surface, the interior and exterior geometries are matched at the boundary $\Sigma=R$, such that
\beq
[\nu]=0,\quad [\nu']=0,\quad [\lambda]=0,
\eeq
where $[f]$ indicates the difference between the value of $f$ in the vacuum exterior and its value in the interior, evaluated at $\Sigma$, i.e., $[f]=f^{+}\vert_{\Sigma}-f^{-}\vert_{\Sigma}$.
\subsection{The rotational perturbations of the metric and the EMT are specified.}
The standard Hartle perturbative metric for slowly rotating systems reads~\cite{Hartle:1967he,Hartle:1968si}
\begin{align}
&ds^2 =  - e^{2\nu_0(r)} \Big[ 1+2 h_0(r) + 2 h_2(r)\, P_2(\cos\theta)  \Big]  dt^2 \nonumber\\ & 
 +e^{2\lambda_0(r)}\left\{1+\frac{2 e^{2\lambda_0(r)}}{r}\Big[ m_0(r) + m_2(r)\, P_2(\cos\theta) \Big] \right\}  dr^2 \nonumber\\& 
+ r^2 \Big[ 1+2 k_2(r)\, P_2(\cos\theta)  \Big] \Big[ d\theta^2 + \sin^2\!\theta\,\big(d\phi - \omega(r) dt \big)^2 \Big].
\label{kingmet}
\end{align}
Here, $P_l(\cos\theta)$ is the Legendre polynomial of the order $l$, $\nu_0(r)$ and $\lambda_0(r)$ are the metric functions of the nonrotating solution, and $h_l(r)$, $k_l(r)$, and $m_l(r)$ are the monopole ($l=0$) and quadrupole ($l=2$) perturbations of the order $\Omega^2$ in rotation, respectively. The function $\omega(r)$ is the contribution at the first order in $\Omega$, which gives rise to the inertial frame dragging. The condition $k_0(r)=0$ corresponds to Hartle's choice of gauge~\cite{Hartle:1967he}.

Originally, Hartle's metric was paired with a perfect fluid EMT in the form
\begin{align}
T^{\mu\nu} = (\EE + \PP) u^\mu u^\nu + \PP g^{\mu\nu},
\label{Tfluid}
\end{align}
where $\EE$ and $\PP$ are the energy density and radial pressure in the comoving frame of the rotating fluid (which due to the perfect fluid nature is degenerate with the transverse pressure), respectively, and $u^\mu$ is its four-velocity with components
\begin{align}
u^t &= \left(-g_{tt}-2\Omega g_{t\phi}-\Omega^2 g_{\phi \phi}\right)^{-1/2}, \nonumber\\
u^\phi &= \Omega u^t,\quad u^r = u^\theta=0.
\label{4vhartle}
\end{align}
We can expand $\EE$ and $\PP$, at the order of $\Omega^2$, in the following form:
\begin{align}
\EE &= \rho(r) + \EE_0(r) + \EE_2(r) P_2(\cos\theta),\\
\PP &= p(r) + \PP_0(r) + \PP_2(r) P_2(\cos\theta) ,
\end{align}
where  $\EE_0(r)$, $\EE_2(r)$, $\PP_0(r)$, and $\PP_2(r)$ are monopole and quadrupole perturbation functions of order $\Omega^2$, where for normal systems the rotation parameter $\Omega$ is the fluid angular velocity\footnote{The angular velocity $\Omega$ loses such a simple interpretation for vacuum energy-type solutions, such as the pure vacuum Hartle-Thorne solution~\cite{Hartle:1967he, Hartle:1968si} and de Sitter-like solutions~\cite{Uchikata:2014kwm,Uchikata:2015yma,Pani:2015tga,Uchikata:2016qku} because the four-velocity drops out the EMT \eqref{Tfluid}.}. 
In many cases, it is conventional to use fractional changes~\cite{Hartle:1968si}
\begin{align}
& \PP = p(r) + (\rho+p)\left[p_{0}^{*}(r) + p_{2}^{*}(r) P_2(\cos\theta)\right],\label{fracP0}\\
& \EE = \rho(r) + \frac{d\rho}{dp}(\rho+p)\left[p_{0}^{*}(r) + p_{2}^{*}(r) P_2(\cos\theta)\right].\label{fracE0}
\end{align}
One convenient way to write the EMT for a rotating anisotropic configuration is
\beq
T^\mu_{~\nu}=(\mathcal{E}+\mathcal{T})u^\mu u_\nu+\mathcal{T} \delta^\mu_{~\nu}-(\mathcal{T}-\mathcal{P}) k^\mu k_\nu, \label{anisotropicEMTcov}
\eeq
where $\TT$ is the transverse pressure in the comoving frame of the rotating fluid, $u$ is a normalized four-velocity, and $k$ is a normalized vector along the distinct axis (spacelike eigenvector with a distinct eigenvalue). The eigenvalues of the EMT are $\mathcal{E}$, $\mathcal{P}$, and $\mathcal{T}$ twice, so this is still Segre type [(11)1,1]. The reason for the choice is that our EOSs, between the eigenvalues, are supposed to be the same whether it is rotating or not, so the eigenvalue structure should be the same (it would be strictly speaking possible to have four distinct eigenvalues for a rotating axisymmetric system). We can expand the eigenvalues in a similar fashion to Hartle~\cite{Hartle:1967he}, 
\begin{align}
    \EE &= \rho(r) + \EE_0(r) + \EE_2(r) P_2(\cos\theta),\\
 \PP &= p_r(r) + \PP_0(r) + \PP_2(r) P_2(\cos\theta) ,\\
 \TT &= p_\perp(r)+\TT_0(r)+\TT_2(r)P_2(\cos\theta),
\end{align}
where we have zeroth-order terms and second-order monopole and quadrupole terms in rotation. Systems like this can have two EOSs, which may be written in the general form 
\beq
F(\EE,\PP,\TT)=0,\qquad G(\EE,\PP,\TT)=0.
\label{EOSgen}
\eeq
Geometrically speaking, these two EOSs can be interpreted as 2D surfaces in 3D space. Thus, allowed configurations, satisfying both, will typically be space curves at the intersection of the 2D surfaces. Suppose one of the variables is one-to-one with a space curve parameter $z$. Call that variable $A$, and call the others $B$ and $C$. Taking derivatives with respect to $z$, we obtain
\begin{align}
    \frac{\partial F}{\partial A}+\frac{\partial F}{\partial B}\frac{\partial B/\partial z}{\partial A/\partial z}+\frac{\partial F}{\partial C}\frac{\partial C/\partial z}{\partial A/\partial z}=0,\\
     \frac{\partial G}{\partial A}+\frac{\partial G}{\partial B}\frac{\partial B/\partial z}{\partial A/\partial z}+\frac{\partial G}{\partial C}\frac{\partial C/\partial z}{\partial A/\partial z}=0.
     \label{Geospdz}
\end{align}
Notice that $\frac{\partial B/\partial z}{\partial A/\partial z}=\frac{\partial B}{\partial A}$ and $\frac{\partial C/\partial z}{\partial A/\partial z}=\frac{\partial C}{\partial A}$. We can effectively replace $z$ with $A$ because they are one to one. We can then solve to obtain 
\begin{align}
    \frac{\partial C}{\partial A}=\frac{(\frac{\partial F}{\partial B}\frac{\partial G}{\partial A}-\frac{\partial F}{\partial A}\frac{\partial G}{\partial B})}{(\frac{\partial F}{\partial C}\frac{\partial G}{\partial B}-\frac{\partial F}{\partial B}\frac{\partial G}{\partial C})},
    \label{yuckyEOS1}\\
    \frac{\partial B}{\partial A}=\frac{(\frac{\partial F}{\partial C}\frac{\partial G}{\partial A}-\frac{\partial F}{\partial A}\frac{\partial G}{\partial C})}{(\frac{\partial F}{\partial B}\frac{\partial G}{\partial C}-\frac{\partial F}{\partial C}\frac{\partial G}{\partial B})}.
    \label{yuckyEOS2}
\end{align}
Considering that $\EE_x,\PP_x$, and $\TT_x$ are small perturbations, we can write
\begin{align}
    B_x=\frac{\partial B}{\partial A} A_x,\qquad   C_x=\frac{\partial C}{\partial A} A_x. 
    \label{yuckyEOS3}
\end{align}
For the four-vectors in Eq.~\eqref{anisotropicEMTcov}, we use the same four-velocity as the standard Hartle formalism \eqref{4vhartle}, and the explicit form we choose for $k$ is
\beq
k^\mu=\left(0,\frac{1-W\Omega-X\Omega^2}{\sqrt{g_{rr}}},Z\Omega+Y\Omega^2,0\right),
\eeq
in $(t,r,\theta,\phi)$ coordinates. The components of $k$ are chosen to satisfy the following properties:
\begin{enumerate}
\item The vector $k$ is normalized:  
\begin{align}
k^\mu k_\mu &= 1-2W \Omega+(W^2-2X+r^2 Z^2)\Omega^2+O(\Omega^3),\nonumber\\
& = 1.
\end{align}
Canceling the first-order term requires $W=0$, while $X=r^2Z^2/2$ cancels the second-order term.
\item In the nonrotating limit, $k$ points purely along $r$. In anisotropic axisymmetric systems, it is possible for there to be $T^r_{~\theta}$-type terms, but those must disappear when the system degenerates to spherical symmetry with this type of metric.
\item The vector $k$ has no $t$ or $\phi$ components as those lead to $T^r_{~t}$ and $T^r_{~\phi}$-type components which are incompatible with this axisymmetric metric format.
\end{enumerate}
\subsubsection{Moment of inertia and dragging of the inertial frames}
The $t\phi$, or $\phi t$, components of the Einstein equations are first order in $\Omega$. They can be cleaned up with the shorthand functions 
\beq
\varpi\equiv \Omega-\omega    
\eeq
and 
\beq
j\equiv e^{-(\nu_0+\lambda_0)}=e^{-\nu_0}\sqrt{1-\frac{2m}{r}},    
\eeq
to derive the frame-dragging equation
\begin{align}
\frac{1}{r^4}\frac{d}{dr}\left(r^4 j \frac{d\varpi}{dr}\right)+\frac{4j'\varpi}{r}\frac{\rho+\pt}{\rho+p_r}=0. \label{newomega}
\end{align}
In the isotropic case $p_\perp=p_r$, we recover the original Hartle equation [see Eq.~(46) in \cite{Hartle:1967he}]. We note that Eq.~\eqref{newomega} is equivalent to Eq.~(72) from Ref.~\cite{Pattersons:2021lci}.

In the region $r>R$, $\epsilon=p=0$ and $j(r)=1$. Thus, Eq.~\eqref{newomega} can be easily integrated to give
\begin{equation}\label{omegaout}
\varpi(r)^{+}=\Omega - \frac{2J}{r^3},
\end{equation}
\noindent where $J$ is an integration constant associated with the angular momentum of the star \cite{Hartle:1967he}. For the interior solution, Eq.~\eqref{newomega} is integrated outward from the origin, with the boundary conditions
\begin{subequations}
\beq
\varpi(0)=\varpi_{\mathrm{c}}=\mathrm{const}.,
\eeq

\beq
\left(\frac{d\varpi}{dr}\right)_{r=0}=0.
\label{bc_omega}
\eeq
\end{subequations}
Regularity of the solution at the surface demands that $[\varpi]=[\varpi']=0$. Thus, one integrates numerically Eq.~\eqref{newomega} and obtains the surface value $\varpi(R)$. Then, and only then, one can determine the angular momentum $J$ and the angular velocity $\Omega$ as
\begin{equation}\label{wsurf}
J=\frac{1}{6}R^4\left(\frac{d\varpi}{dr}\right)_{r=R},\quad \Omega = \varpi(R) + \frac{2J}{R^3}.
\end{equation}
Once the angular momentum and angular velocity are determined, the relativistic moment of inertia can be obtained from the relation $I = J/\Omega$.

We also find first-order terms in $\Omega$ by looking at the $r\theta$ component of the Einstein equation (or, equivalently,  $\theta r$), which leads to the condition
\begin{widetext}
\begin{align}
&\frac{8\pi r^{7/2}(Z \Omega+Y\Omega^2)(p_r-p_\perp)}{(r-2m)^{1/2}} = 3\Omega ^2 \sin\theta\cos\theta\left[r 
   \frac{d}{dr}\left(h_2+k_2\right)+h_2 \left(r
 \frac{d\nu_0}{dr}-1\right)-\frac{m_2}{r-2m} \left(r\frac{d\nu_0}{dr}+1\right)\right].\label{rthetaEEQ}
\end{align}
\end{widetext}
This requires $Z=0$ in order to balance the first-order terms in $\Omega$. Notice that, in the isotropic case, the left side vanishes.\footnote{Also, in the case of isotropy the same vectors which diagonalize the metric in the $r\theta$ block will diagonalize the energy-momentum tensor, so there is no point in introducing $Y$.}Alternatively, if we have $Y=0$, the left side must vanish as well. In either case this gives a relation between the quadrupole functions and leads to the natural introduction of the auxiliary function $v_2=h_2+k_2$. 

In \cite{Pattersons:2021lci} it is assumed that $T^{r}_{~\theta}=T^{\theta}_{~r}=0$ . However, while this must be true for spherically symmetric systems, or for perfect fluid stationary axisymmetric systems in this type of coordinate system, we stress that this is not strictly necessary for anisotropic axisymmetric systems.
\subsubsection{Monopole perturbations: The $l=0$ sector}
The monopole sector of the second-order $tt$ perturbation in the Einstein equations results in the following differential equation
\beq
\frac{dm_0}{dr} = 4\pi r^2 \EE_0 +\frac{j^2 r^4}{12}\left(\frac{d\varpi}{dr}\right)^2 - \frac{r^3\varpi^2}{3}\frac{dj^2}{dr}\frac{\rho+\pt}{\rho+p_r}. 
\label{newm0}
\eeq
In the isotropic case, $\pt=p_r$, Eq.~\eqref{newm0} reduces to Eq.~(97) in \cite{Hartle:1967he}, Eq.~(15a) in \cite{Hartle:1968si}, and Eq.~(18) in \cite{Chandra:1974}. The $\varpi'$ term already matches the isotropic case. and the $\varpi$ term will match when isotropy is applied because the fraction will go to 1. The other difference is that we write $\EE_0$ explicitly rather than the fractional changes [see  Eq.~\eqref{fracE0} for the relation in the isotropic case]. 
We note that Eq.~\eqref{newm0} is equivalent to Eq.~(61) from Ref.~\cite{Pattersons:2021lci} using the definitions from their Eqs.~(48) and (55) to express the $\EE_0$ term.

The $rr$ perturbation equation gives a differential equation for $h_0$ in the form
\beq
\frac{dh_0}{dr} = 4\pi r e^{2\lambda}\PP_0+\frac{(1+8\pi r^2 p_r)}{(r-2m)^2}m_0-\frac{e^{2\lambda}r^3 j^2 \varpi'^2}{12}.\label{newh0}
\eeq
Equation \eqref{newh0} agrees with Eq.~(19) from Ref.~\cite{Chandra:1974} and Eq.~(98) from Ref.~\cite{Hartle:1967he}, when the isotropic pressure condition is applied. There does not seem to be a corresponding differential equation for $h_0$ in Ref.~\cite{Pattersons:2021lci}.

Finally, we use $(8\pi T^\theta_{~\theta}-G^\theta_{~\theta})+(8\pi T^\phi_{~\phi}-G^\phi_{~\phi})=0$, or the monopole perturbation to the $r$ component of $\nabla_\mu T^\mu_{~~\nu}=0$, to derive
\begin{widetext}
\beq
\PP_0'=-(\EE_0+\PP_0)\nu_0'+\frac{2(\TT_0-\PP_0)}{r}+\frac{\rho+\pt}{3}\frac{d}{dr}\left(\frac{r^3\varpi^2j^2}{r-2m}\right)
+\frac{(\rho+p_r)}{r-2m}\left[\frac{1}{12}r^4j^2\varpi'^2 - 4\pi r^2 \PP_0 - \frac{\left(1+8\pi r^2 p_r\right)}{(r-2m)}m_0\right].
\label{newp0}
\eeq
\end{widetext}
 In the isotropic case, we can write $p_\perp=p_r=p$, and $\PP_0=(\rho+p)p_{0}^{*}$; thus, $\PP_0'=(\rho+p)p_0^*{}'+p_0^*(\rho+p)'$. The term with $\TT_0$ goes to 0 when isotropy is enforced.  It is possible to cancel the $-(\EE_0+\PP_0)\nu_0'$ term in Eq.~\eqref{newp0} with the $p_0^*(\rho+p)'$ term using the zeroth-order isotropic Einstein equations\footnote{Specifically, the isotropic TOV equation can be rewritten as
 \beq 
 (\rho+p)\nu_0'=-\frac{dp}{dr}\nonumber.
 \eeq
 Using this expression and the relationship between $\EE_0,\PP_0$, and $p_0^*$, which applies in the isotropic case, we obtain
 \begin{equation*}
  -(\EE_0+\PP_0)\nu_0'= \frac{dp}{dr}\left(\frac{d\rho}{dp}+1\right)p_0^*=(\rho'+p')p_0^*,
 \end{equation*}
 where we applied the chain rule in the last equality.}, then divide the entire equation by $\rho+p$ to recover Eq.~(100) in \cite{Hartle:1967he}, or Eq.~(15b) in \cite{Hartle:1968si}. There seem to be typos in the corresponding Eq.~(62) of \cite{Pattersons:2021lci}, for instance: $1-2m$ appears instead of $r-2m$ in the denominator of the $m_0$ term.
 
 In principle, one can integrate Eq.~\eqref{newomega} to find the frame dragging, at least numerically. One can then numerically integrate Eqs.~\eqref{newm0}, \eqref{newh0}, and \eqref{newp0}, but whether it is more convenient to solve them simultaneously will depend on the specific EOS~\eqref{EOSgen}. In general, Eqs.~\eqref{newm0} and \eqref{newp0} are integrated outwards with the conditions $m_0(0)=\PP_0(0)=0$, at the origin, respectively. We find it convenient to define the auxiliary function
\beq
H_0\equiv h_0-h_c,
\label{hc_eqn}
\eeq
where $h_c$ is an undetermined constant. We recognize that $H_0$ also follows the differential equation \eqref{newh0}, thus it can be integrated with the boundary condition $H_0(0)=0$. We then find the corresponding value of $h_c$ by continuity, and the values of $H_0(r\rightarrow R^-)$ and the exterior $h_0(r\rightarrow R^+)$, since the background $g_{tt}$ is nonzero in this case.

In the vacuum exterior where $\rho=p=0$, Eqs.~\eqref{newm0} and \eqref{newh0} can be integrated explicitly giving
\beq
m_0^{+}=\delta M-\frac{J^2}{r^3},
\label{m0out}
\eeq
\beq
h_0^{+}=-\frac{m_0}{r-2M},
\label{h0out}
\eeq
where $\delta M$ is a constant of integration associated with the change of mass induced by the rotation. In his approach, Hartle~\cite{Hartle:1967he} assumed the continuity of $m_0$ and $h_0$, at the surface, i.e., $[h_0]=[m_0]=0$. This condition was employed to determine the constant $\delta M$.

In \cite{Reina:2014fga}, it was argued that for systems where the density at the surface is nonzero, a discontinuity appears in $m_0$ and a modified formula
\begin{multline}
\delta M_{\mathrm{mod}}=m_0(R)+\frac{J^2}{R^3}\\
+8\pi R^3 \left(\frac{R}{2M}-1\right)\rho(R)p_0^*(R),
\label{dmMOD}  
\end{multline}
is presented. In the anisotropic case, it can be shown that the modified change of mass takes the general form
\beq
\delta M_{\mathrm{mod}}=m_0(R)+\frac{J^2}{R^3}+4\pi\rho(R)R^2 \xi_{0},
\label{dmMODfin}
\eeq
where $\xi_0$ is the spherical deformation parameter
\begin{align}
    \xi_0=-\frac{\PP_0}{p_r'(r\rightarrow R^-)}.
    \label{xi0def}
\end{align}
A derivation of Eq.~\eqref{dmMODfin} is presented in Appendix~\ref{app:dM}. 
\subsubsection{The quadrupole perturbations: The $l=2$ sector}
\label{Sec:quad}
The quadrupole perturbations can be treated as follows. 
Recall from Eq.~\eqref{rthetaEEQ} that $Z$ must vanish. Thus, it is convenient to introduce an auxiliary function $\Upsilon(r)$ related to $Y$ as 
\beq
(p_r-\pt)Y=\Upsilon(r)\sin\theta\cos\theta, 
\label{ups_first}
\eeq
such that
\begin{align}
    &8\pi r^3 \Upsilon(r)=3 e^{-\lambda} \Bigg[r 
   \frac{d}{dr}\left(h_2+k_2\right)+h_2 \left(r
 \frac{d\nu_0}{dr}  -1\right)\nonumber\\&-\frac{m_2}{r-2m} \left(r \frac{d\nu_0}{dr}+1\right)\Bigg].
 \label{upsdef}
\end{align} 
The left-hand side goes to zero for isotropic systems, minor algebra then produces Eq.~(122) of Ref.~\cite{Hartle:1967he}, or Eq.~(20) of Ref.~\cite{Chandra:1974}. The analogous Eq.~(21a) from Ref.~\cite{Hartle:1968si} requires more algebra such as substituting $m_2$.

In terms of the auxiliary function $\Upsilon$, the $\theta$ component of the EMT conservation equation becomes
\begin{widetext}
\beq
3 \TT_2+(\rho+p_\perp)\left(3 h_2+ e^{-2\nu_0}\varpi^2 r^2\right) = 
(p_r-p_\perp)\left(\frac{3e^{2\lambda}}{r}\right)m_2 + e^{-\lambda}\left[r\Upsilon(4+r\nu_0') + r^2 \Upsilon'\right].
\label{constheta}
\eeq    
\end{widetext}
In the isotropic case, everything on the right-hand side goes to 0, thus, Eq.~\eqref{constheta} reduces to Eq.~(91) of Ref.~\cite{Hartle:1967he}, or Eq.~(15) of Ref.~\cite{Chandra:1974}. It is noteworthy that Ref.~\cite{Pattersons:2021lci} uses the isotropic formula, but even if we assume $T^r_{~\theta}=0$, or, equivalently $\Upsilon=0$, then there is an extra term with $m_2$ compared to the isotropic case.

Using the difference between the $\theta\theta$ and $\phi\phi$ components $(8\pi T^\theta_{~\theta}-G^\theta_{~\theta})-(8\pi T^\phi_{~\phi}-G^\phi_{~\phi})=0$, we obtain a fairly simple equation which could eliminate $m_2$ or $h_2$:
\begin{equation}
    m_2=(r-2m)\left[\frac{j^2 r^4 \varpi'^2}{6}-\frac{r^3\varpi^2}{3}\left(\frac{dj^2}{dr}\right)\frac{\rho+\pt}{\rho+pr}-h_2\right].\label{m2item}
\end{equation}
In the isotropic case, Eq.~\eqref{m2item} reduces to Eq.~(120) from Ref.~\cite{Hartle:1967he} or Eq.~(23a) from Ref.~\cite{Hartle:1968si}. 

Using Eq.~\eqref{m2item} to replace $m_2$, the $rr$ component of the Einstein field equations can be written as
\begin{widetext}
\begin{multline}
\frac{j^2 r^2}{3}\left[16\pi(\rho+\pt)(1+2r\nu_0')\varpi^2+\left(\frac{3}{2}+8\pi r^2p_r\right)\left(\frac{d\varpi}{dr}\right)^2\right]=
-8\pi \PP_2+\frac{2(r-2m)}{r^2}\left[\frac{dh_2}{dr}+(1+r\nu_0')\frac{dk_2}{dr}\right] \\
+\frac{4}{r^2}\left[h_2(4\pi r^2 p_r-1)-k_2\right].
\label{quadA}
\end{multline}
\end{widetext}
Equation~\eqref{quadA} reduces to Eq.~(22) from Ref.~\cite{Chandra:1974} and Eq.~(124) from Ref.~\cite{Hartle:1967he} when Eq.~(\ref{m2item}) is used and isotropy is enforced.

The quadrupole perturbation to the $r$ component of $\nabla_\mu T^\mu_{~~\nu}=0$ gives the following equation:
\begin{widetext}
\begin{multline}
\frac{d\PP_2}{dr}=\frac{2(\TT_2-\PP_2)}{r}-(\EE_2+\PP_2)\nu_0'+\frac{2}{3}r\varpi e^{-2\nu_0}(\rho+\pt)\left[2\varpi(1+2r\nu_0')-r\left(\frac{d\varpi}{dr}\right)\right]+\frac{j^2r(1+r \nu_0')}{8\pi}\left(\frac{d\varpi}{dr}\right)^2\\
-\left(\frac{3}{4\pi r^2}\right)\frac{d}{dr}\left(h_2+k_2\right)-(\rho+p_r)\frac{dh_2}{dr}+2(\pt-p_r)\frac{dk_2}{dr}-\left(\frac{3\nu_0'}{2\pi r^2}\right)h_2.
    \label{quadC}
\end{multline}
\end{widetext}
Analogous equations to this one are not usually written in the literature, because they are unnecessary for isotropic systems.

When formulated in this manner, the quadrupole sector is dependent on the three-metric perturbation functions $h_2,k_2$, and $m_2$, three EMT eigenvalue perturbation functions $\EE_2,\PP_2$, and $\TT_2$, and an EMT eigenvector perturbation function $\Upsilon$ for a total of seven functions (in contrast to five in the isotropic case). Thus, Eqs.~\eqref{upsdef}--\eqref{quadC} plus two EOSs make this, in principle, a well-defined system.

In the vacuum exterior, the functions $h_2$ and $k_2$ are specified by~\cite{Hartle:1967he,Hartle:1968si}
\beq\label{h2out}
h_{2}^{+}=\frac{J^2}{Mr^3}\left(1 + \frac{M}{r}\right) + KQ_{2}^{\;2}(\zeta),
\eeq
\beq\label{v2out}
v_{2}^{+}=h_{2}^{+}+k_{2}^{+}=-\frac{J^2}{r^4} + K\frac{2M}{\left[r(r-2M)\right]^{1/2}}Q_{2}^{\;1}(\zeta),
\eeq
\noindent respectively, where $K$ is a constant of integration and $Q_{n}^{\;m}$ are the associated Legendre functions of the second kind with argument $\zeta\equiv (r/M) -1$. 

To match with the exterior solution, one treats the interior equations by solving Eqs.~\eqref{upsdef}-\eqref{quadC} for both, the particular case with nonzero $\varpi$ and the homogeneous case with $\varpi=0$ and then choosing the true functions
\beq
\text{true} = \text{particular} + C*\text{homogeneous},
\label{split}
\eeq
where $C$ is the same constant for all the functions. The integration constants $C$ and $K$ can be determined by matching $h_2$ and $k_2$ to the exterior solutions given by Eqs.~\eqref{h2out} and \eqref{v2out}. Once the constant $K$ has been determined, one can compute the mass quadrupole moment of the star, as measured at infinity, from the relation~\cite{Hartle:1968si}
\beq
Q=\frac{J^2}{M}+\frac{8}{5}K M^3.
\eeq
Following the standard Hartle conventions~\cite{Hartle:1967he}, the EMT eigenvalue perturbations $\EE_2,\PP_2$, and $\TT_2$ go to zero at the origin. The function $\Upsilon$, which describes the eigenvector perturbation, must go to zero at the origin for well-behaved systems, because it is proportional to a factor of $p_r-p_\perp$ which is zero there.

In addition to the behavior in the exterior, some constraints can be derived for the behavior near $r=0$. For well-behaved systems, the initial behavior of the relevant zeroth- and first-order terms in $\Omega$ are as follows: 
\begin{align}
m &= \frac{4\pi \rho_c}{3}r^3,\\
\nu_0 &= \nu_c +\frac{2\pi}{3}(\rho_c + 3p_c)r^2,\\
p_r &= p_\perp=p_c,\\
\rho &= \rho_c,\\
\varpi &= \varpi_c,
\end{align}
where the subscript “$c$" denotes the value of the quantity at the center of the configuration. We can then find series solutions near the origin for the particular equations~\eqref{upsdef}, \eqref{constheta}, \eqref{quadA}, and \eqref{quadC}, as follows:
\beq
k_{2}^{\text{(P)}} = k_2^a r^2+k_2^b r^4,
\label{k2p_expa}
\eeq
\beq 
h_{2}^{\text{(P)}} = -k_2^a r^2+h_2^b r^4,
\label{h2p_expa}
\eeq
\begin{multline}    
\Upsilon^{\text{(P)}} = \Bigg[\frac{3(h_2^b+k_2^b)}{2\pi}-e^{-2\nu_c}\varpi_c^2(p_c+\rho_c)-\\
k_2^a(3p_c+\rho_c)\Bigg]r,
\label{yp_expa}
\end{multline}
\begin{multline}
\PP_{2}^{\text{(P)}} = \frac{1}{6}\Bigg[\frac{3(h_2^b+k_2^b)}{\pi}+4k_2^a \rho_c\\
-4e^{-2\nu_c}\varpi_c^2(p_c+\rho_c)\Bigg]r^2,
\label{p2p_expa}
\end{multline}
\begin{multline}
\TT_{2}^{\text{(P)}} = \Bigg[\frac{5(h_2^b+k_2^b)}{2\pi}-2e^{-2\nu_c}\varpi_c^2(p_c+\rho_c)\\
-\frac{2}{3}k_2^a(6p_c+\rho_c)\Bigg]r^2,    
\label{T2p_expa}
\end{multline}
where $h_{2}^{b}$, $k_{2}^{a}$, and $k_{2}^{b}$ are arbitrary constants. The homogeneous equations initial behavior constraints can be found by setting $\varpi_c=0$ in the previous equations. In the isotropic case, we require $\PP_2=\TT_2$, which implies that $\Upsilon=0$ and we recover the established relations between the coefficients for that case [see Eq.~(56) in~\cite{Miller:1977}].

Besides the quadrupole moment, there are two other shape parameters associated with the quadrupole sector, namely, the perturbation function $\xi_2$ defined as
\beq
\xi_2=-\frac{\PP_2}{p_r'(r\rightarrow R^-)}
\label{xi2def}
\eeq    
and the ellipticity of the isobaric surfaces~\cite{Chandra:1974}
\beq
\epsilon(r)=-\frac{3}{2r}\left[\xi_2(r) + rk_2(r)\right].
\label{elipticitydef}
\eeq
\section{Bowers-Liang anisotropic fluid sphere}
\label{sect:bl}
\begin{figure}
\includegraphics[width=\linewidth]{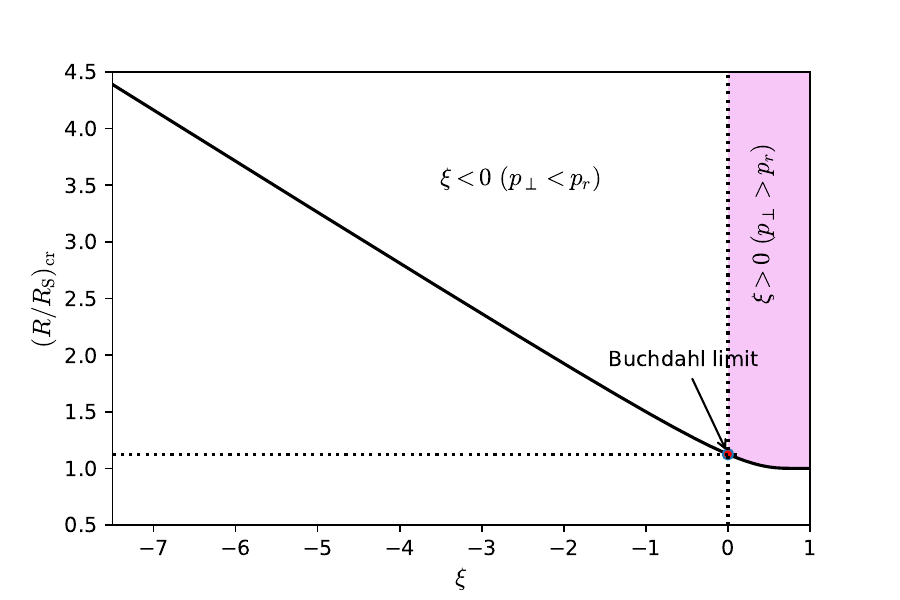}
\caption{The critical ratio of the stellar radius $R$ to the Schwarzschild radius $\Rs\equiv 2M$, as a function of the parameter $\xi$, for an anisotropic uniform density configuration. In the isotropic case, when $\xi=0$, we recover the standard Buchdahl limit $R=(9/8)\Rs$.}
\label{fig:beta_cr}
\end{figure}

The Bowers-Liang (BL) solution~\cite{Bowers:1974tgi}, for an anisotropic mass with constant density $\rho_{0}$, is described by the following EOS
\beq\label{pt}
p_{\perp} = p_{r} + \sigma f(p_{r},r)(\rho_0 + p_{r})r^2,
\eeq
where $\sigma$ is a constant that measures the “strength" of the anisotropy and
\beq\label{fpr}
f(p_{r},r) = \frac{\rho_0 + 3p_{r}}{1-2m(r)/r}.
\eeq
\noindent By solving Eq.~\eqref{tov}, the radial pressure for the BL model gives
\beq\label{pr}
p_{r}=\rho_{0}\left[\frac{\left(1-2Mr^2/R^3\right)^q - (1-2M/R)^q}{3\left(1-2M/R\right)^q - \left(1-2Mr^2/R^3\right)^q}\right], 
\eeq
where\footnote{We note that Ref.~\cite{Bowers:1974tgi} missed the $\pi$ factor in $\xi$. It is awkward that we employ $\xi$ here to indicate the anisotropy parameter, while the same letter is used to denote the perturbation functions $\xi_0$ in Eq.~\eqref{xi0def} and $\xi_2$ in Eq.~\eqref{xi2def}. However, we are using the previously established notation and the subindexes differentiate the various expressions, so we hope the awkwardness is not a serious one.} 
\beq
q = \frac{1}{2} - \frac{3\sigma}{4\pi}=\frac{1}{2}(1-\xi),\quad\quad \xi\equiv \frac{3\sigma}{2\pi}. 
\eeq
There is a divergence in the pressure at the origin
\begin{align}
p_r(0)=p_\perp(0)=\rho_0\left[\frac{1-(1-2M/R)^q}{3(1-2M/R)^q -1}\right],\label{prorigin}
\end{align}
 when we reach the critical radius
\beq\label{radcrit}
  R_{\text{cr}} =\frac{\Rs}{1-(1/3)^{1/q}},
\eeq
where $\Rs\equiv 2M$ is the Schwarzschild radius. This critical radius can be thought of as the smallest size for BL spheres, of a given anisotropy, to remain nonsingular. In the limit of weak anisotropy, $\vert\sigma\vert\ll 1$, Eq.~\eqref{radcrit} gives approximately
\beq\label{rcrit_expa}
\left(\frac{R}{\Rs}\right)_{\text{cr}}\simeq\frac{9}{8}\left[1 - \frac{\log(3)}{4}\xi + \mathcal{O}(\xi^2)\right].
\eeq
On the other hand, when $\xi=1$, corresponding to the limit of very strong anisotropy, we reach the Schwarzschild limit $R_{\text{cr}} =\Rs$. In Fig.~\ref{fig:beta_cr}, we display the ratio $(R/\Rs)_{\text{cr}}$ as a function of the anisotropy parameter $\xi$. Note that, for $\xi<0$, $p_{\perp}\le p_r$, while, for $\xi>0$, $p_{\perp}\ge p_r$. In the strict isotropic case $\xi=0$, Eq.~\eqref{radcrit} or~\eqref{rcrit_expa} reduces to $R_{\text{cr}}=(9/8)\Rs$ (red dot in Fig.~\ref{fig:beta_cr}), which corresponds to the well-known Buchdahl limit~\cite{Buchdahl:1959zz}.

\begin{figure}[ht!]
\centering
\includegraphics[width=\linewidth]{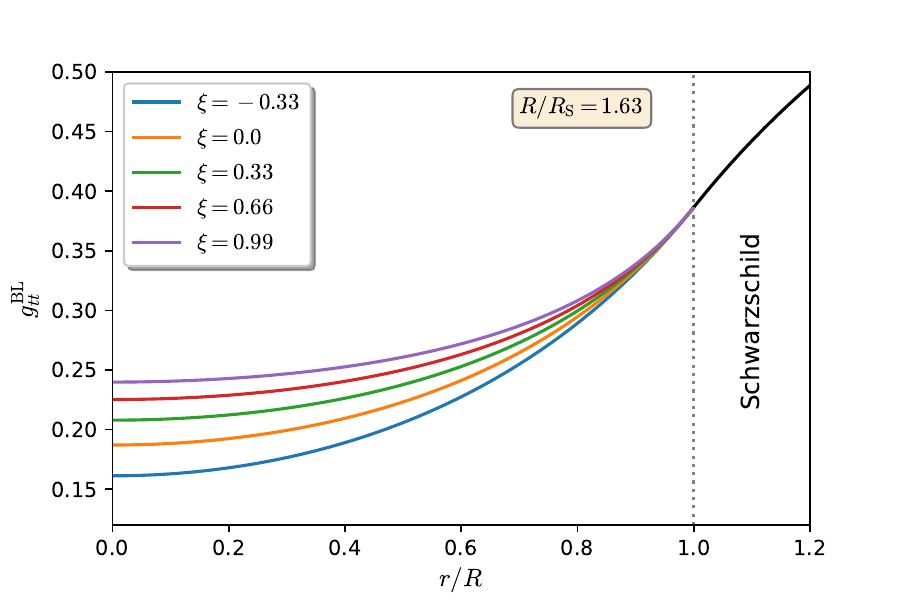}
\caption{Radial profiles of the metric element $g_{tt}$~\eqref{gttBLx}, for the Bowers-Liang model, for different values of the anisotropy parameter $\xi$. We consider a configuration with $R/\Rs=1.63$. Note the continuous matching at the surface $r=R$ with the exterior Schwarzschild solution.}
\label{fig:gttBL}
\end{figure}

The $g_{tt}$ metric component can be obtained from the integration of Eq.~\eqref{EE_stat_r}, which gives\footnote{We note that Bowers and Liang~\cite{Bowers:1974tgi} did not write explicitly the metric element $g_{tt}$.}
\beq\label{gtt_BL}
e^{2\nu}=\frac{1}{2^{1/q}}\left[3\left(1-\frac{2M}{R}\right)^q - \left(1-\frac{2Mr^2}{R^3}\right)^q\right]^{1/q}.
\eeq
Following Refs.~\cite{Chandra:1974,Beltracchi:2023qla}, we introduce the variables
\beq
y(r) = 1-\frac{2m(r)}{r},\quad y_{1}=1-\frac{2M}{R},
\eeq
together with the normalized radial coordinate $x\equiv r/R$ and the parameter $\beta\equiv R/\Rs$. In terms of $x, y$, and $\beta$, the BL solution, Eqs.~\eqref{pt}, \eqref{pr}, and \eqref{gtt_BL}, takes the form
\beq\label{prx}
p_{r}=\rho_{0}\left[\frac{y^q - y_{1}^q}{3y_{1}^q - y^q} \right],
\eeq
\beq\label{ptx}
p_{\perp} = \rho_{0}\left[\tilde{p}_{r} + \frac{3\sigma(1+3\tilde{p}_{r})(1+\tilde{p}_{r})x^2}{8\pi\beta(1-x^2/\beta)}\right],
\eeq 
\beq\label{gttBLx}
e^{2\nu} = \frac{1}{2^{1/q}}\left(3y_{1}^q - y^q\right)^{1/q},
\eeq
where, $y=1-x^2/\beta$, and $\tilde{p}_{i}=p_{i}/\rho_{0}$. In the isotropic case $\xi=0$, we have $q=1/2$, and Eqs.~\eqref{prx} and \eqref{gttBLx} reduce to
\beq
p_{r}=\rho_{0}\left(\frac{\sqrt{y} - \sqrt{y_{1}}}{3\sqrt{y_{1}} - \sqrt{y}}\right),
\eeq
\beq
e^{2\nu} = \frac{1}{4}\left(3\sqrt{y_{1}} - \sqrt{y}\right)^2,
\eeq
respectively, which corresponds to Schwarzschild's interior solution with uniform density~\cite{Schwarzschild:1916inc}. In Fig.~\ref{fig:gttBL}, we show the radial profiles of the $g_{tt}$ metric component~\eqref{gttBLx}, for the BL model, for various values of the anisotropy parameter $\xi$. We consider a configuration with $\beta=1.63$. For a given radius $r<R$, an increase in the anisotropy increases the function $g_{tt}$. Observe how the various curves match smoothly, at the boundary $r=R$, with the exterior Schwarzschild solution.

\begin{figure}[ht!]
\centering
\includegraphics[width=\linewidth]{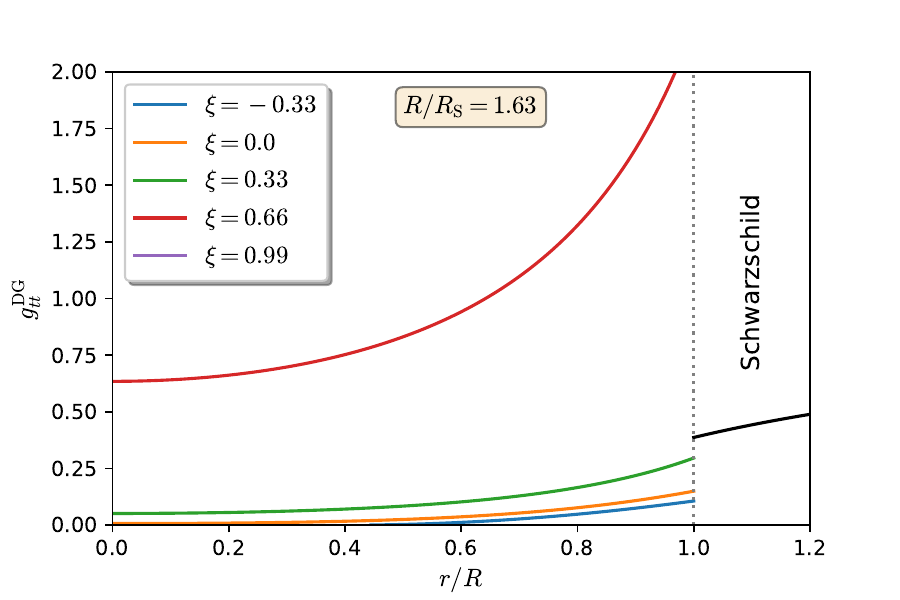}
\caption{Profiles of the metric element $g_{tt}$ for the Bowers-Liang model, as suggested by Dev and Gleiser~\cite{Dev:2003qd}, for different values of $\xi$. We consider a configuration with $R/\Rs=1.63$. Note that the DG metric does not give the correct matching, at the boundary, with the exterior Schwarzschild solution.}
\label{fig:gttDG}
\end{figure}

Dev and Gleiser~\cite{Dev:2003qd} suggested the metric element $g_{tt}$, for the BL model, in the following form [see Eq.~(101) in~\cite{Dev:2003qd}]
\beq
g_{tt}^{\text{DG}} = \frac{1}{4}\left(3y_{1}^{2q} - y^{2q}\right)^{1/q},
\label{gtt_DG}
\eeq
which has the wrong powers of $q$ and the incorrect factor $(1/4)$. Let us recall that this factor is crucial to make the correct matching, at the boundary, with the exterior Schwarzschild solution. Note that the problem is not only a choice of gauge. As we show in Fig.~\ref{fig:gttDG}, the Dev-Gleiser metric~\eqref{gtt_DG} does not match with the exterior Schwarzschild solution.

\begin{figure*}[ht!]
\centering
\includegraphics[width=0.495\linewidth]{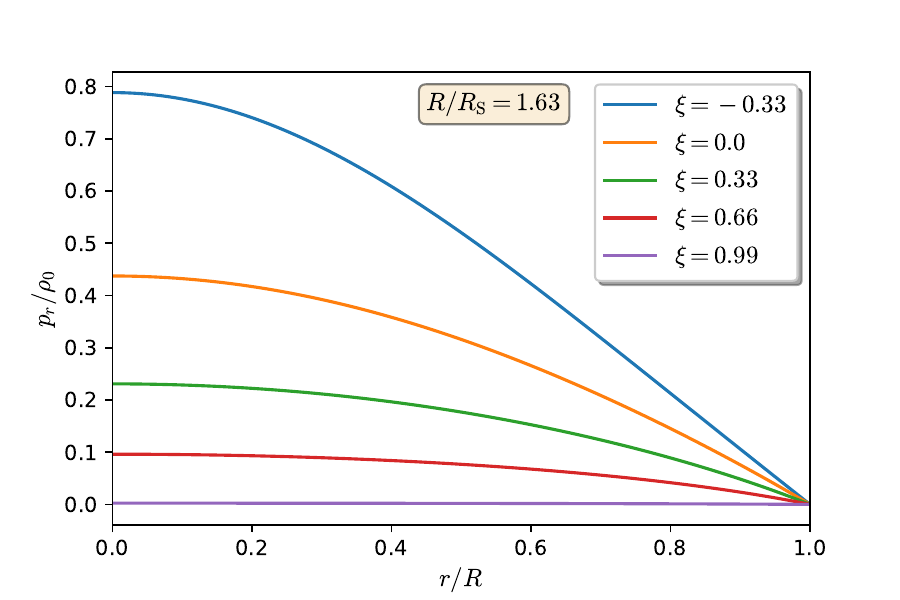}
\includegraphics[width=0.495\linewidth]{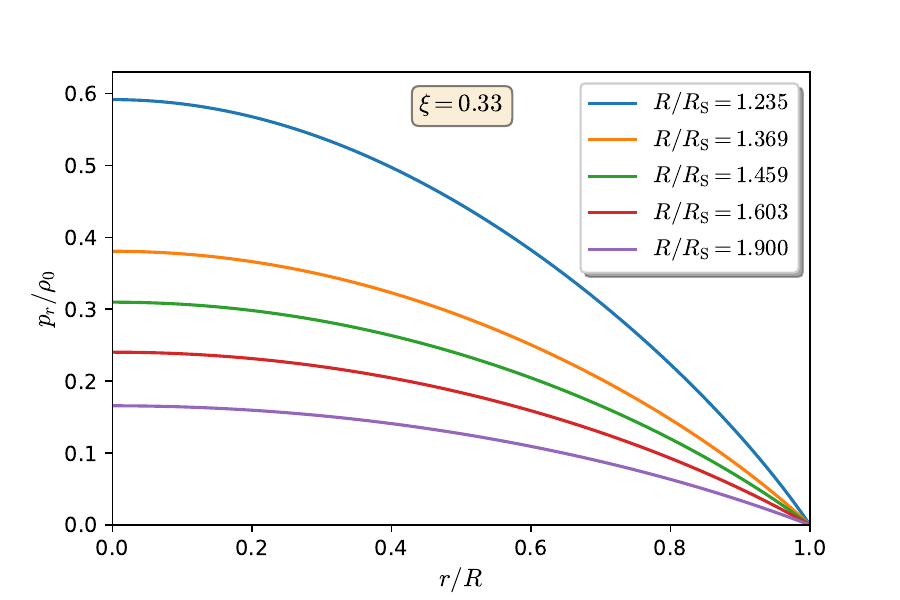}
\caption{{\bf Left panel}: profiles of the radial pressure for the Bowers-Liang model, for a configuration with $R/\Rs=1.63$, for different values of the anisotropy parameter $\xi$. {\bf Right panel}: profiles of the radial pressure, for $\xi=0.33$, for various values of the parameter $R/\Rs$. The pressure is measured in units of the energy density $\rho_0$.}
\label{fig:pr}
\end{figure*}

In the left panel in Fig.~\ref{fig:pr} we plot the radial pressure profile \eqref{prx}, for a configuration with $\beta=1.63$, for various values of the anisotropy parameter $\xi$. Note that the radial pressure decreases monotonically as we move outward from the center up to the surface. We observe that as $\xi$ increases the pressure is suppressed. In the limit as $\xi\to 1$, the radial pressure approaches zero. In the right panel of the same figure, we plot the radial pressure, but now for a fixed value of the anisotropy parameter $\xi$, for varying $\beta$. Observe how as $\beta$ decreases, or the compactness increases, the interior pressure is enhanced.

\begin{figure*}[ht!]
\centering
\includegraphics[width=0.495\linewidth]{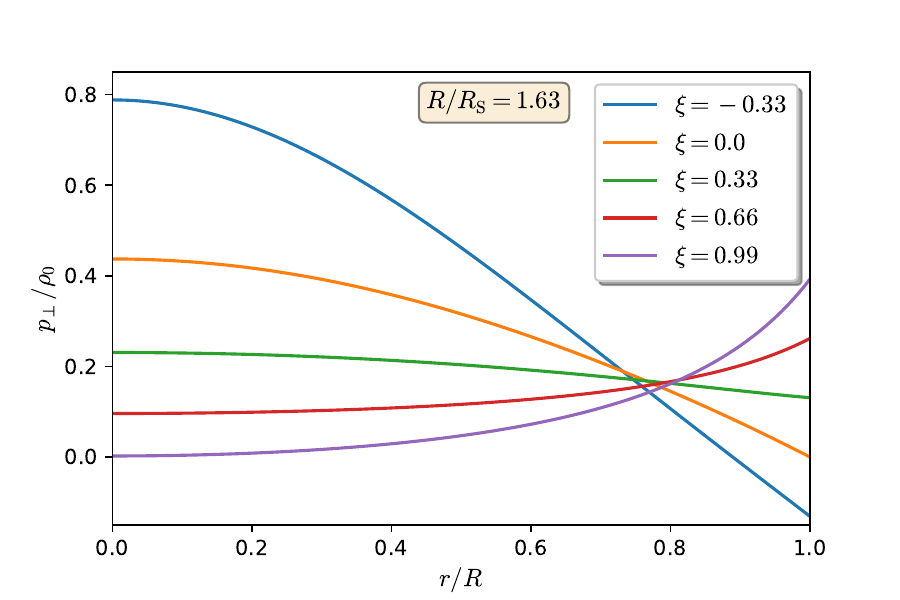}
\includegraphics[width=0.495\linewidth]{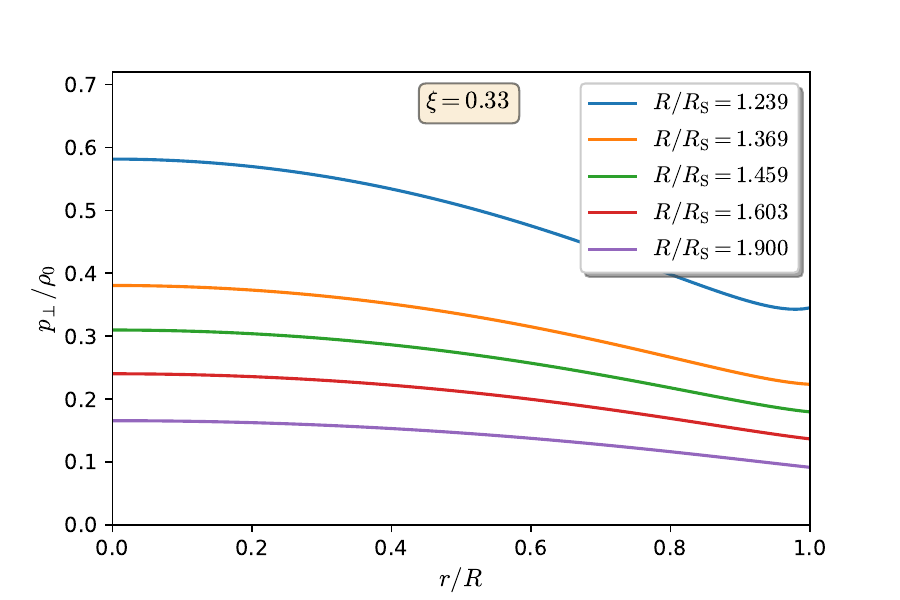}
\caption{{\bf Left panel:} radial profiles of the transverse pressure $\pt$ for the Bowers-Liang model, for a configuration with $R/\Rs=1.63$, for different values of the anisotropy parameter $\xi$. {\bf Right panel:} radial profiles of $\pt$, for $\xi=0.33$, for various values of the parameter $R/\Rs$.}
\label{fig:pt}
\end{figure*}

In the left panel in Fig.~\ref{fig:pt}, we display the profiles of the transverse pressure \eqref{ptx}, for a configuration with $\beta=1.63$, for various values of the anisotropy parameter $\xi$. Note that for the nonpositive value $\xi=-0.33$, the transverse pressure becomes \emph{negative} in a certain interior region. We also observe that $\pt$ does not vanish at the surface, except for the isotropic case. In the right panel of the same figure, we plot the transverse pressure $\pt$, for a fixed value of $\xi$, for varying $\beta$. We observe that, as $\beta$ decreases, the interior transverse pressure increases, and again, it does not vanish at the surface.

\subsection{Dominant energy condition Bowers-Liang sphere}
\label{sec:DEC}
One interesting case study of the Bowers-Liang model is the smallest sphere for which the dominant energy condition (DEC) is satisfied. The radial pressure, as given by Eq.~\eqref{prorigin}, is always greatest at the origin (see Fig.~\ref{fig:pr}). The transverse pressure at the surface of the star is
\beq
p_\perp(R)=\frac{3\sigma \rho_0}{8\pi(\beta-1)},
\eeq
which is the other possibility for the global maximum of one of the pressures for configurations near the DEC bound. Ensuring that $p_\perp(R)\le \rho$ and $p(0)\le \rho$ gives the following conditions for extremizing the compactness and the anisotropy:
\beq
\beta\ge\left[1-\left(\frac{1}{2}\right)^{1/q}\right]^{-1},\qquad \beta\ge\frac{1}{2}\left(\frac{5}{2}-q\right).
\eeq
\begin{figure}[ht!]
\includegraphics[width=\linewidth]{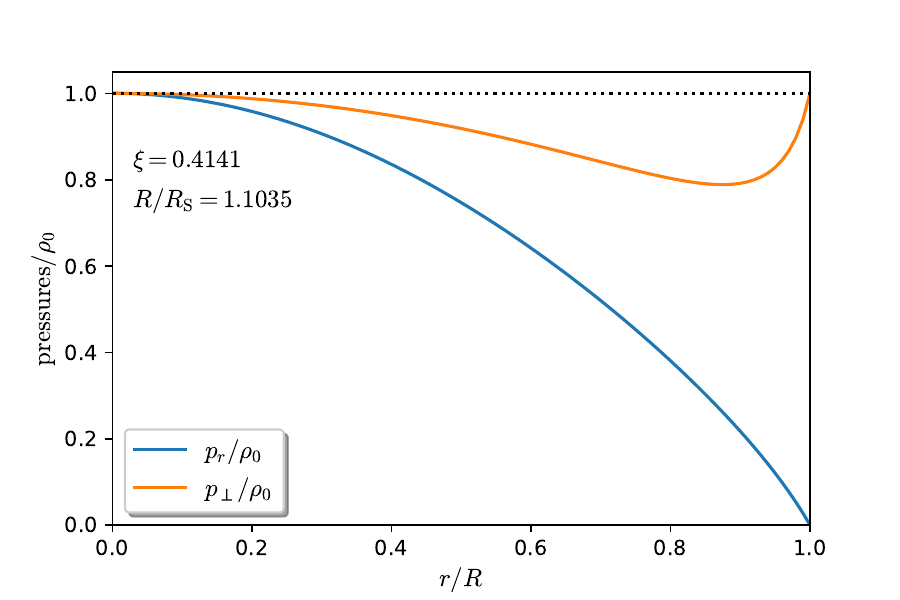}
\caption{Pressure profiles for the smallest anisotropic Bowers-Liang sphere for which the DEC is satisfied.}
\label{fig:DEC}
\end{figure}
The smallest $\beta$ compatible with both of these conditions is $\beta_{\text{DEC}}=1.1035$ for $q_{\text{DEC}}=0.2929$, or, equivalently, $\sigma_{\text{DEC}}=0.8674,~\xi_{\text{DEC}}=0.4141$. 
In Fig.~\ref{fig:DEC}, we display the profiles of radial pressure $p_r$ and transverse pressure $p_{\perp}$ for the smallest Bowers-Liang sphere that satisfies the DEC. Note that $\beta_{\text{DEC}}<\beta_{\text{Buchdahl}}=9/8$, so this particular test case is below the Buchdahl limit. However, it is rather mundane in the sense that it satisfies the DEC, the strong energy condition, and is nonsingular; while many ultracompact systems, like the Schwarzschild star~\cite{Mazur:2015kia}, may violate one or more of these conditions.

\section{Numerical results}
\label{sect:results}
In this section, we present our results for surface and integral properties of slowly rotating anisotropic Bowers-Liang spheroids, at second order in the angular velocity $\Omega$. We numerically integrated the extended Hartle structure equations, derived in Sec.~\ref{sect:hartle}, for various values of the parameters $\xi$ and $R/\Rs$. By doing so, we model the whole sequence of slowly rotating BL spheroids in adiabatic and quasistationary contraction. As a consistency check, we include results for the isotropic case $\xi=0$, which are in very good agreement with the ones reported by Refs.~\cite{Chandra:1974, Beltracchi:2023qla} for slowly rotating isotropic homogeneous masses. We employ dimensionless variables where the units in which the dimensions are expressed are given in the corresponding descriptions.

It will be convenient for the following to set the EOS for the Bowers-Liang sphere, Eq.~\eqref{EOSgen}, in the following form
\begin{align}
F(\EE,\PP,\TT)=\EE-\rho_0=0,\\
G(\EE,\PP,\TT)=\TT-g(\PP)=0,
\end{align}
where the first equation is the constant density condition and the second is analogous to Eq.~\eqref{pt}. Because $p_r$ is monotonic in $r$, we can use $r$ as our space curve parameter and $p_r$ as the thermodynamic variable $A$ in Eqs.~\eqref{EOSgen}-\eqref{yuckyEOS3}. Thus, using Eq.~\eqref{Geospdz}, we obtain
\beq
\frac{dg}{d\PP}=\frac{d p_\perp/d r}{d p_r/ d r}=\frac{d p_\perp}{d p_r}.
\label{eosPDs}
\eeq
Using the fact that $\PP_x$ and $\TT_x$ are small perturbations, together with Eqs.~\eqref{yuckyEOS3} and \eqref{eosPDs}, we can use the substitution 
\beq
\TT_x=\frac{\partial p_\perp}{\partial p_r}\PP_x,\quad\EE_x=0,
\eeq
where the second expression can be derived either by employing Eq.~\eqref{yuckyEOS2} with $p_r=A, p_\perp=C$, and $\rho=B$ or directly from the uniform density condition.

\subsection{Frame dragging for the Bowers-Liang sphere}

To determine the dragging of the inertial frames, we numerically integrated Eq.~\eqref{newomega} with the boundary condition~\eqref{bc_omega}, starting from $x_0$ (or rather from $x_0 + \epsilon$, with a cutoff value $\epsilon\sim 10^{-7}$), up to the stellar surface $x=1$, for various values of the parameter $R/\Rs$.

As a consistency check, we also set up and solved the equations using units of $r/\Rs$, rather than $x=r/R$, and tested several values of the cutoff. We follow the same conventions as Refs.~\cite{Chandra:1974,Beltracchi:2023qla}, where $\varpi$ is measured in units of $J/\Rs^3$; thus, it will be convenient henceforth to introduce the following quantities:
\beq
\widetilde{\varpi}\equiv \frac{\varpi}{(J/\Rs^3)},\quad \widetilde{\Omega}\equiv \frac{\Omega}{(J/\Rs^3)}.
\eeq
\begin{figure*}[ht!]
\centering
\includegraphics[width=.495\linewidth]{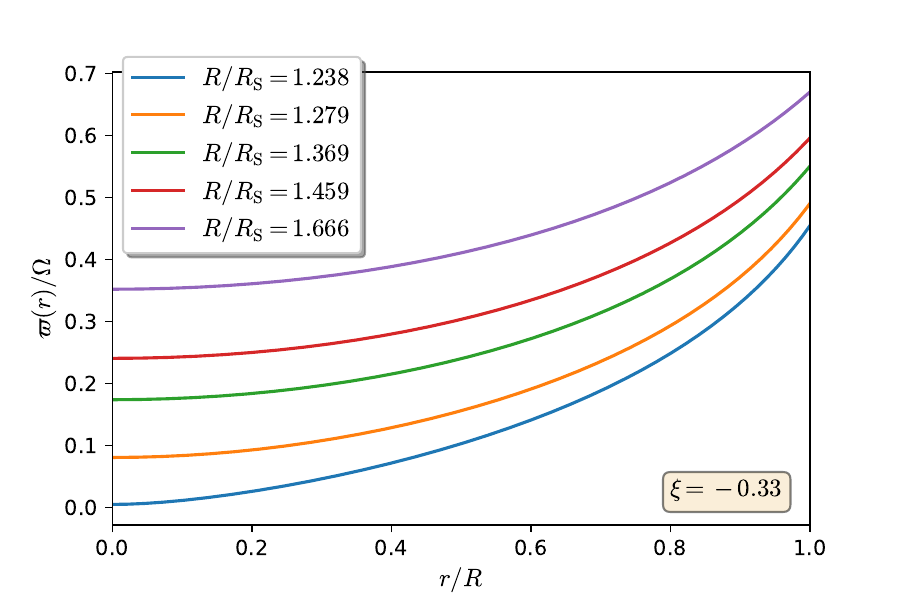}
\includegraphics[width=.495\linewidth]{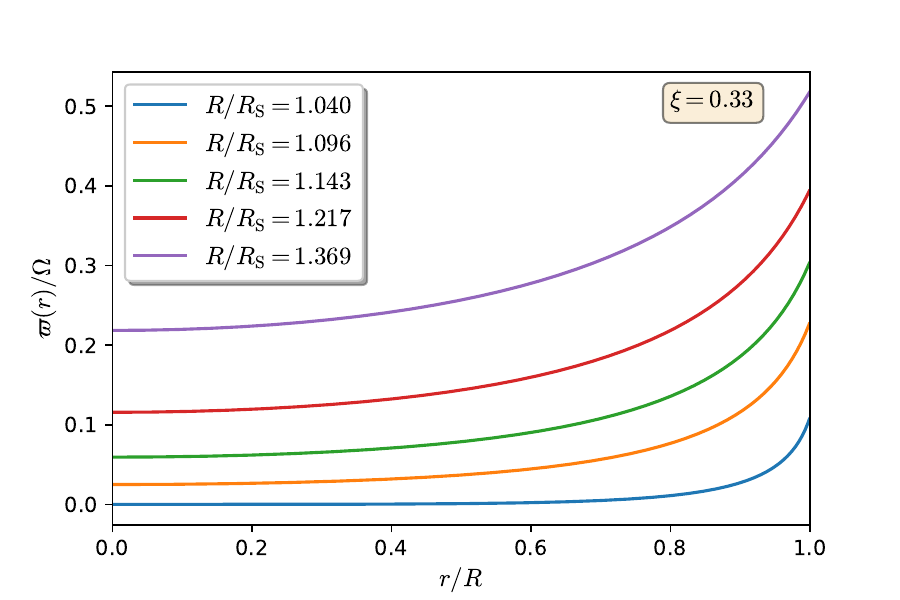}
\includegraphics[width=.495\linewidth]{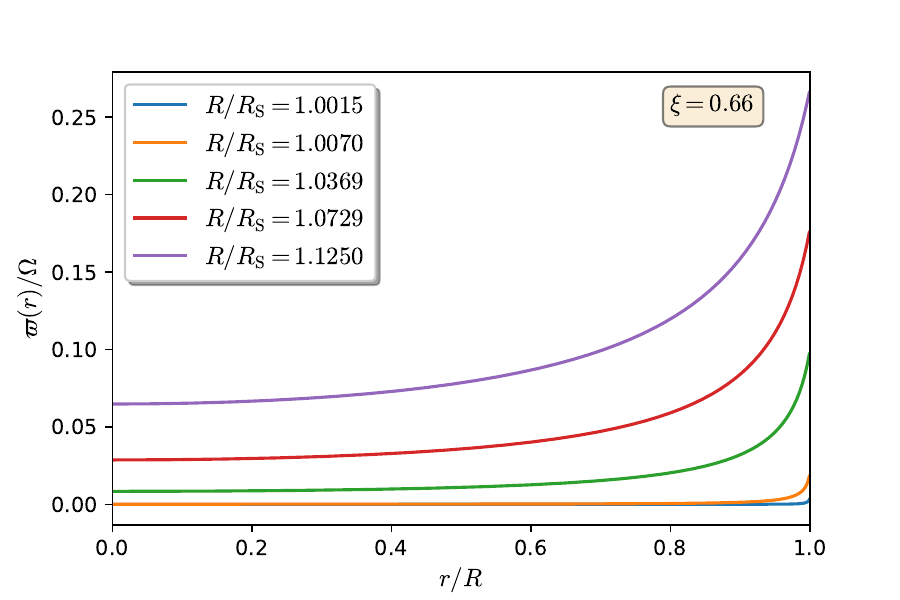}
\includegraphics[width=.495\linewidth]{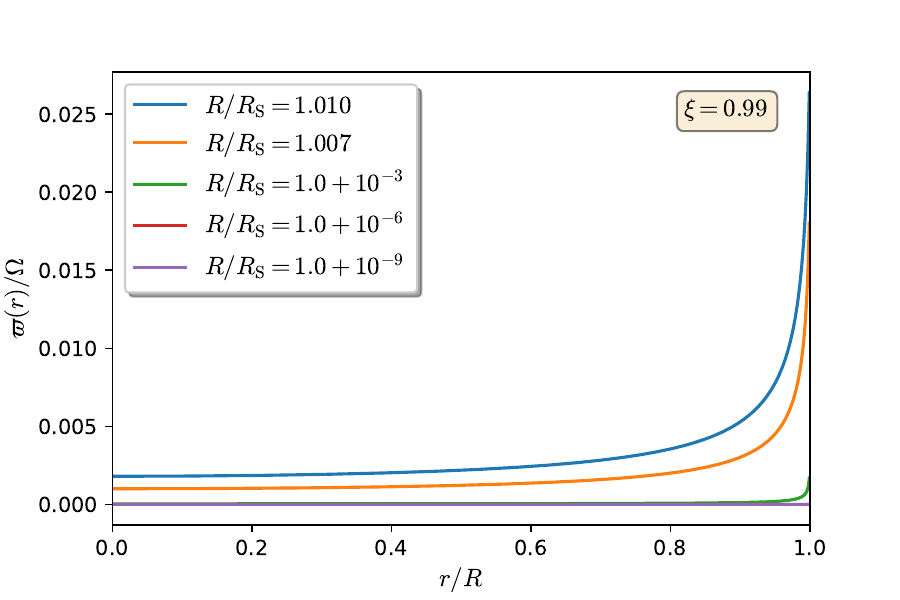}
\caption{Radial profiles of the $\varpi$ function, in the unit $\Omega$, for various values of the parameter $R/\Rs$. We display the profiles for $\xi=\{-0.33, 0.33, 0.66, 0.99\}$. Observe that, for the critical compactness, $\varpi\to 0$ at the origin; thus, $\omega=\Omega$. For highly anisotropic spheres near to their critical compactness (which approaches the Schwarzschild limit), $\varpi\approx0$ and $\Omega\approx\omega$ hold throughout the interior.}
\label{fig:omega_xis}
\end{figure*}
In Fig.~\ref{fig:omega_xis}, we display the radial profiles of the function $\varpi/\Omega$, for various values of the parameter $\beta=R/\Rs$. We consider $\xi=\{-0.33, 0.33, 0.66, 0.99\}$. We observe that $\varpi\to 0$, at the origin, when $\beta$ approaches the critical value $\beta_{\mathrm{cr}}$; thus, in this limit, the frame-dragging function $\omega$ approaches the value of the angular velocity $\Omega$. We see that $\varpi$ monotonically increases from the origin to the surface, although as $\beta\rightarrow1$ it is rather flat.

\begin{figure*}[ht]
\centering
\includegraphics[width=0.495\linewidth]{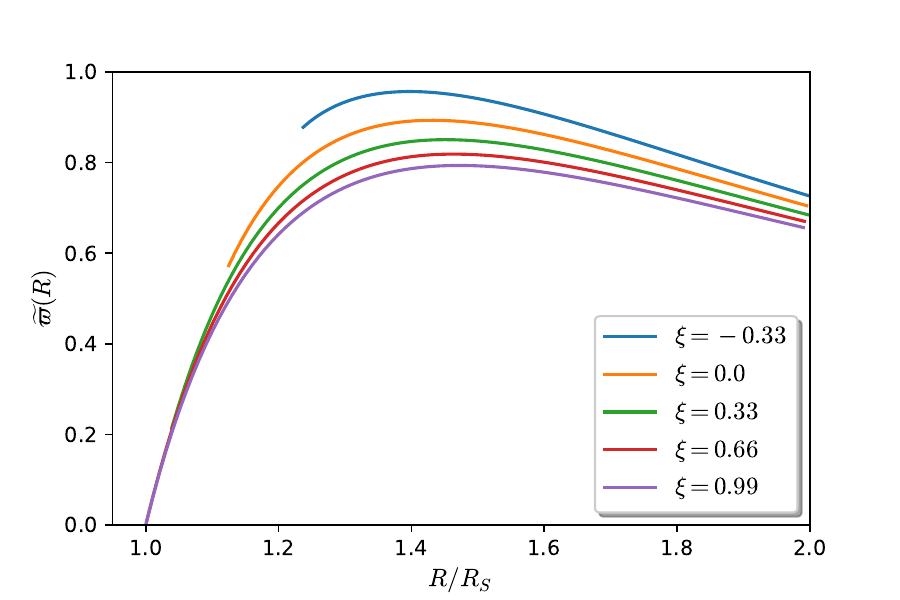}
\includegraphics[width=0.495\linewidth]{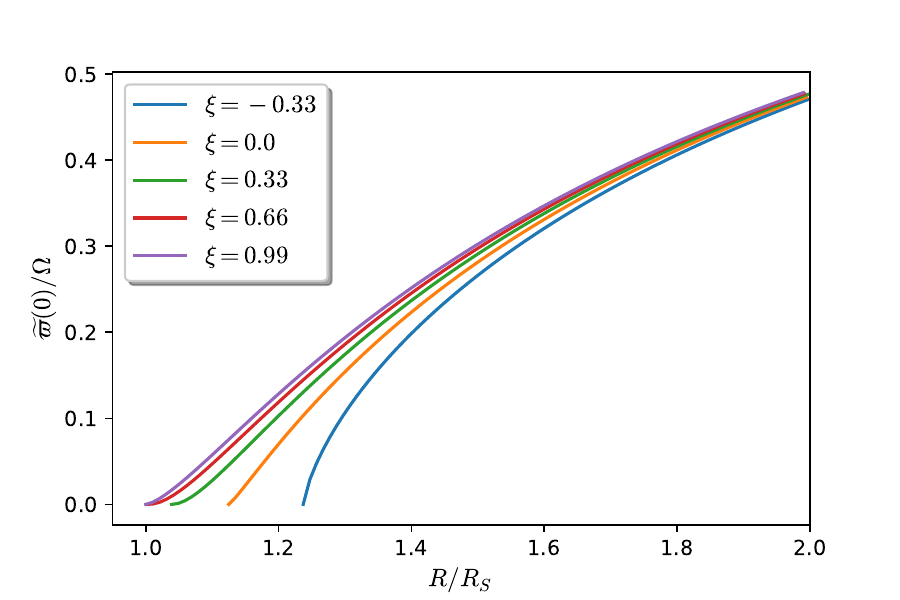}
\caption{{\bf Left panel:} surface value of the $\widetilde{\varpi}$ function, against the parameter $\beta$, for various values of the anisotropy parameter $\xi$. For configurations with high anisotropy, in the limit as $R\to \Rs$, $\widetilde{\varpi}(R)\to 0$. {\bf Right panel:} central value of $\varpi$, in the unit $\Omega$, as a function of $\beta$. Observe the approach of $\varpi(0)$ to zero, as the compactness approaches the critical value where the central pressure diverges.}
\label{fig:w1x}
\end{figure*}

In the left panel in Fig.~\ref{fig:w1x}, we present the surface value $\widetilde{\varpi}_{1}=\widetilde{\varpi}(x_1)$, as a function of $R/\Rs$, for various values of the anisotropy $\xi$. We measure $\widetilde{\varpi}$ in units of its central value $\widetilde{\varpi}_{c}$. In the isotropic case $\xi=0$, we recover with very good agreement the results reported by Refs.~\cite{Chandra:1974, Beltracchi:2023qla} for slowly rotating Schwarzschild stars. We observe that, for highly anisotropic configurations $\xi\to 1$, $\widetilde{\varpi}_{1}\to 0$, similar to other ultracompact objects like Kerr BHs and sub-Buchdahl Schwarzschild stars~\cite{Beltracchi:2023qla}. In the right panel of the same figure, we display the central value $\widetilde{\varpi}(0)$, in units of the angular frequency $\Omega$, as a function of the parameter $R/\Rs$, for the same values of $\xi$ as in the left panel. Observe that $\varpi(0)$ approaches zero whenever the compactness approaches the critical value (where the central pressure diverges). Thus, $\omega(0)$ approaches to the angular velocity of the fluid $\Omega$. This behavior was also found for Schwarzschild stars at the Buchdahl limit~\cite{Chandra:1974}. In~\cite{Beltracchi:2023qla}, we concluded that the only boundary condition that allowed for regular monopole perturbations on the infinite pressure surface for sub-Buchdahl Schwarzschild stars was $\varpi=0$ there. Hartle~\cite{Hartle:1967he} argued that $\varpi>0$, however, this explicitly involves the assumption $e^{\nu}>0$ which is violated in all these particular situations. We also observe that the increase in anisotropy enhances $\varpi(0)/\Omega$ at a given radius $R$.

\begin{figure}[ht]
\centering
\includegraphics[width=\linewidth]{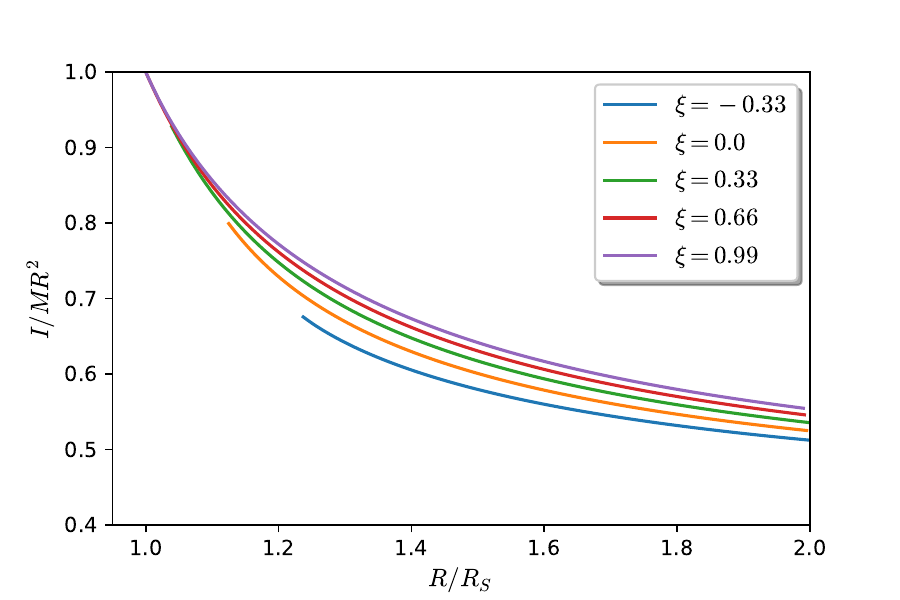}
\caption{Normalized moment of inertia $I/MR^2$, as a function of $R/\Rs$, for various values of the anisotropy parameter $\xi$. At any given radius, increasing $\xi$ increases the moment of inertia. For stars with high anisotropy, we see that as $R/R_s\rightarrow 1$ the normalized moment of inertia also goes to 1. The $\xi=0$ curve corresponds to the Schwarzschild star (above the Buchdahl limit where all pressures are finite).}
\label{fig:moI}
\end{figure}

In Fig.~\ref{fig:moI}, we show the normalized moment of inertia $I/MR^2$, as a function of the parameter $R/\Rs$, for varying $\xi$. In the case $\xi=0$, we recover with very good agreement the results reported by Refs.~\cite{Chandra:1974, Beltracchi:2023qla} for slowly rotating isotropic homogeneous stars. We observe that, for a given radius $R$, as the anisotropy parameter $\xi$ increases, the moment of inertia also increases. In the limit of maximum anisotropy $\xi\to 1$, it is possible for higher compactness to be maintained with finite central pressure. Moreover, for such stars, as $R\to R_{\text{cr}}$, $I\to MR^2$ similar to other ultracompact objects like Kerr BHs, rotating sub-Buchdahl Schwarzschild stars~\cite{Beltracchi:2023qla}, or rotating gravastars~\cite{Beltracchi:2021lez}. Although for highly anisotropic stars the parameter $R/\Rs$ can get very close to 1, we point out that the strict BH limit cannot be taken from these configurations, for any value of the central energy density, because their critical radii are always greater than the Schwarzschild radius [see Eq.~\eqref{radcrit} and Table~\ref{tab:crit_radius}].
\begin{table}[h]
    \centering
    \begin{tabular}{c|c}
    \hline\hline
       $\xi$ & $\vert 1-(R/\Rs)_{\text{cr}}\vert$ \\
       \hline
      $-0.33$ & $2.371\times 10^{-1}$ \\
       $0.0$ & $1.25\times 10^{-1}$ \\
       $0.33$ & $3.912\times 10^{-2}$ \\
       $0.66$ & $1.56\times 10^{-3}$\\
       $0.99$ & $3.76\times 10^{-96}$\\
       \hline\hline
    \end{tabular}
    \caption{List of the critical ratio $(R/\Rs)_{\mathrm{cr}}$, where the central pressure diverges, for some selective values of the anisotropy parameter $\xi$.}
    \label{tab:crit_radius}
\end{table}
\subsection{Monopole perturbations for the Bowers-Liang sphere}
In order to study the spherical deformations of the stars ($l=0$ sector), we numerically integrated Eqs.~\eqref{newm0}--\eqref{newp0}, with the conditions $m(0)=\PP_0(0)=0$, at the origin (or rather from $x_0 + \epsilon$, with a cutoff value $\epsilon\sim 10^{-6}$). Following the conventions of Refs.~\cite{Chandra:1974, Beltracchi:2023qla}, it will be convenient to introduce the quantities 
\begin{subequations}
\beq
\widetilde{h_0}\equiv \frac{h_0}{(J^2/\Rs^4)},\quad \widetilde{m_0}\equiv \frac{m_0}{(J^2/\Rs^3)},
\eeq
\beq
\widetilde{\TT_0}\equiv\frac{\TT_0}{(J^2/\Rs^6)},\quad\widetilde{\PP_0}\equiv \frac{\PP_0}{(J^2/\Rs^6)},\quad \widetilde{p_0^{*}}\equiv \frac{p_0^{*}}{(J^2/\Rs^4)}.
\eeq
\end{subequations}

\begin{figure}[ht]
\centering
\includegraphics[width=\linewidth]{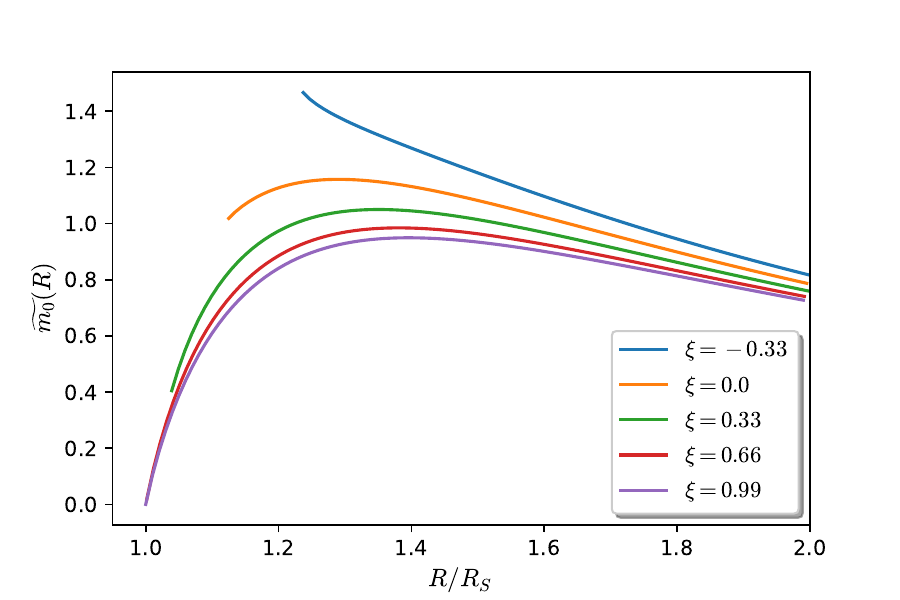}
\caption{Surface value $\widetilde{m_{0}}(R)$, against $R/\Rs$, for various values of the anisotropy parameter $\xi$. Many of the curves have a local maximum (except for the one with negative $\xi$). Observe the approach of $\widetilde{m_{0}}(R)$ to 0, as we approach the Schwarzschild limit. Increasing the anisotropy, at any plotted radius, decreases $\widetilde{m_{0}}(R)$.}
\label{fig:m0s}
\end{figure}

In Fig.~\ref{fig:m0s}, we display the surface value $\widetilde{m_{0}}(R)$, as a function of $R/\Rs$, for various values of the anisotropy parameter $\xi$. The $\xi=0$ curve corresponds to the isotropic case, which is in very good agreement with the one reported by us in~\cite{Beltracchi:2023qla} for Schwarzschild stars. Observe that, for a given radius $R$, as the anisotropy parameter $\xi$ increases, the function $\widetilde{m_0}(R)$ decreases. Furthermore, in the limit of high anisotropy $\xi\to 1$, $\widetilde{m_0}(R)\to 0$, similar to what we found for slowly rotating sub-Buchdahl Schwarzschild stars in the gravastar limit~\cite{Beltracchi:2023qla}.

\begin{figure*}[ht]
\centering
\includegraphics[width=0.495\linewidth]{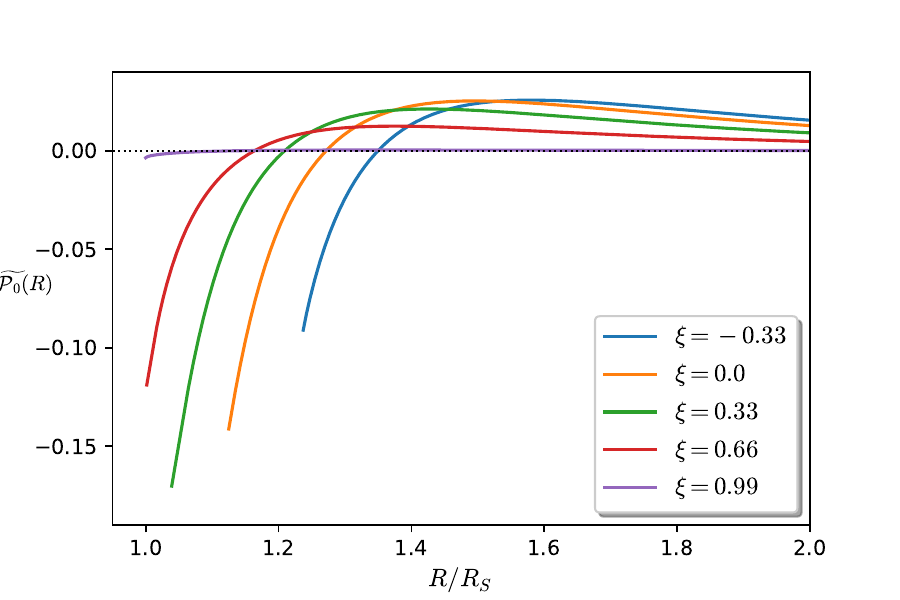}
\includegraphics[width=0.495\linewidth]{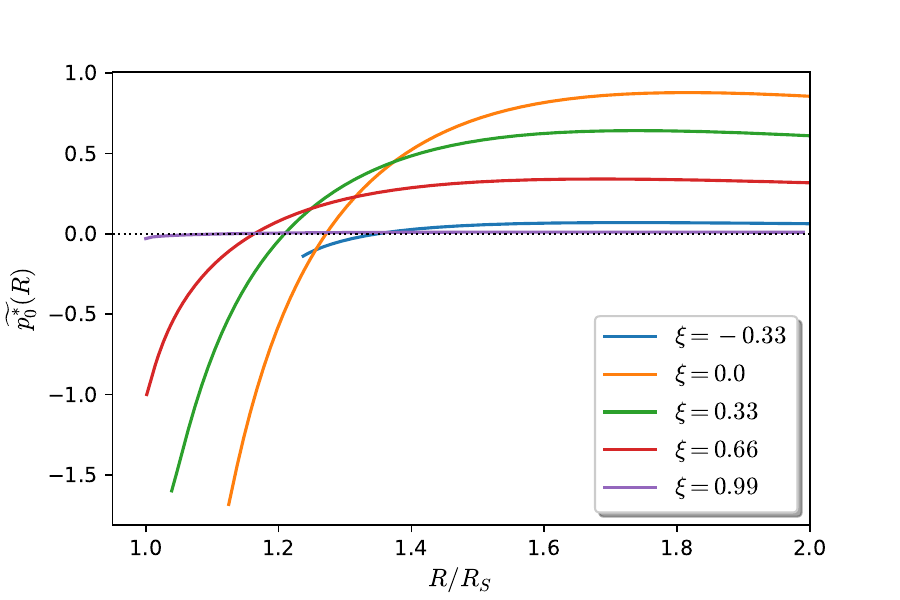}
\caption{{\bf Left panel:} surface value of the perturbation function $\widetilde{\PP_{0}}(R)$, against $R/\Rs$, for different values of the anisotropy parameter $\xi$. Observe that in the limit of maximum anisotropy $\xi\to 1$, the function $\widetilde{\PP_{0}}(R)$ gets flattened. {\bf Right panel}: the same data as the left panel, but now in terms of the associated perturbation function $\widetilde{p_{0}^{*}}(R)$.}
\label{fig:p0s}
\end{figure*}

The left panel in Fig.~\ref{fig:p0s} shows the surface value $\widetilde{\PP_{0}}(R)$, as a function of the parameter $R/\Rs$, for various values of the anisotropy $\xi$. Observe that the various curves show a local maximum and then decrease as $R\to R_{\mathrm{cr}}$. A different behavior is shown by the higher anisotropy curve $\xi=0.99$, which gets flattened to nearly zero for every radius $R$. In the right panel of the same figure, we display the same data as the left panel, but now in terms of the associated perturbation function $p_0^{*}=\PP_0/(\rho+p_r)$. In the isotropic case, $\xi=0$, we recover with very good agreement the results reported by Refs.~\cite{Reina:2015jia,Beltracchi:2023qla} for Schwarzschild stars above the Buchdahl limit, where the radial pressure is finite.  

\begin{figure}[ht]
\centering
\includegraphics[width=\linewidth]{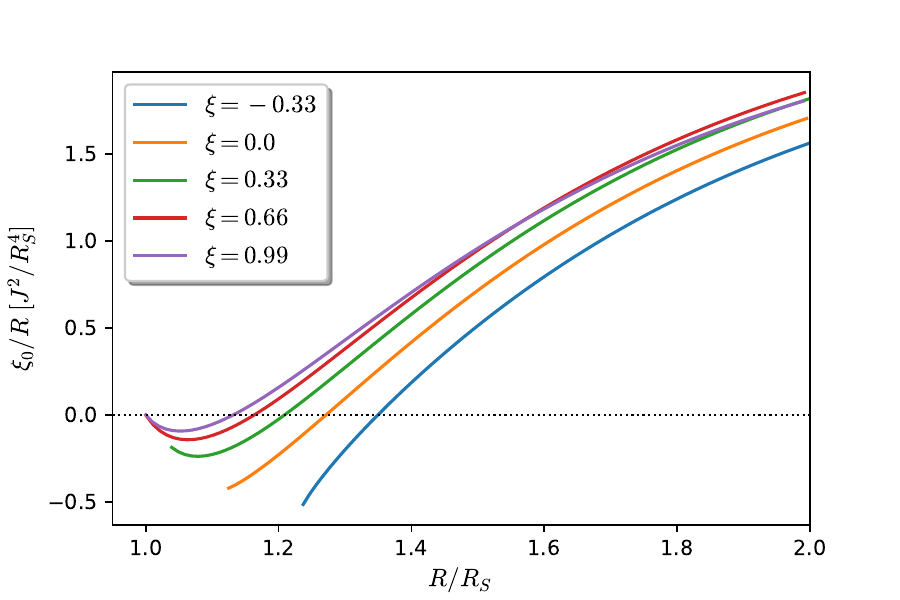}
\caption{Surface value of the deformation function $\xi_0/R$ (in units of $J^2/\Rs^4$), as a function of $R/\Rs$, for various values of the anisotropy parameter $\xi$. Note that $\xi_0\to 0$ for high compactness. The various curves switch sign between $R/\Rs=1$ and $R/\Rs=1.4$. The ones with positive anisotropy have a local minimum somewhere below the Buchdahl limit.}
\label{fig:xi0s}
\end{figure}

In Fig.~\ref{fig:xi0s} we plot the surface value of the deformation function $\xi_0/R$, as a function of the parameter $R/\Rs$, for different values of the anisotropy $\xi$. The function $\xi_{0}/R$ is measured in units of $J^2/\Rs^4$. At large radii, $\xi_0$ tends to be positive but it typically becomes negative at some smaller radius. For the positive anisotropy cases, we observe that $\xi_0/R$ shows a local minimum near the critical compactness. Moreover, in the limit as $R\to R_{\text{cr}}$, the function $\xi_0/R$ approaches zero. The $\xi=0$ curve corresponds to the isotropic case, which is in very good agreement with the results we reported for Schwarzschild stars above the Buchdahl limit \cite{Beltracchi:2023qla}. It is interesting, that for this particular case, a local minimum appears beyond the Buchdahl limit (see Fig. 11 in~\cite{Beltracchi:2023qla}). It is possible that there would also be a local minimum in $\xi_0/R$ for $\xi=-0.33$ below its critical radius.

\begin{figure}[h!]
    \includegraphics[width=\linewidth]{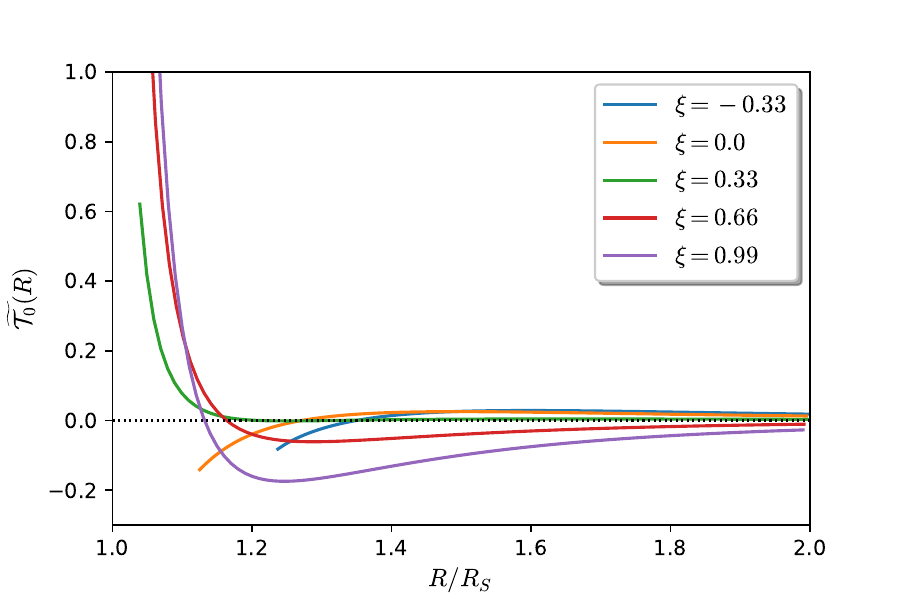}
    \caption{Surface values of the transverse pressure perturbations $\widetilde{\TT_0}$, as a function of $R/\Rs$, for different anisotropic parameters. For the isotropic case $\xi=0$, we have $\widetilde{\TT_0}=\widetilde{\PP_0}$. For some of the cases with positive anisotropy, there is a local minimum. For the highly anisotropic compact cases, $\widetilde{\TT_0}$ does not fit on the scale of this plot. For instance, when $\xi=0.99$ and $R/\Rs=1.001$, we have $\widetilde{\TT_0}=315.42$.}
    \label{fig:T0s}
\end{figure}

In Fig.~\ref{fig:T0s}, the surface value of the transverse pressure perturbations $\widetilde{\TT_0}$ are plotted, against $R/\Rs$, for various values of the anisotropy parameter $\xi$. Let us recall that, in the isotropic case $\xi=0$, $\TT_0=\PP_0$. Observe that for some of the curves with positive anisotropy, there is a local minimum. For the highly anisotropic configurations, the function $\widetilde{\TT_0}(R)$ grows rapidly as $\beta\to\beta_{\text{cr}}$.

\begin{figure}[ht!]
    \centering
    \includegraphics[width=\linewidth]{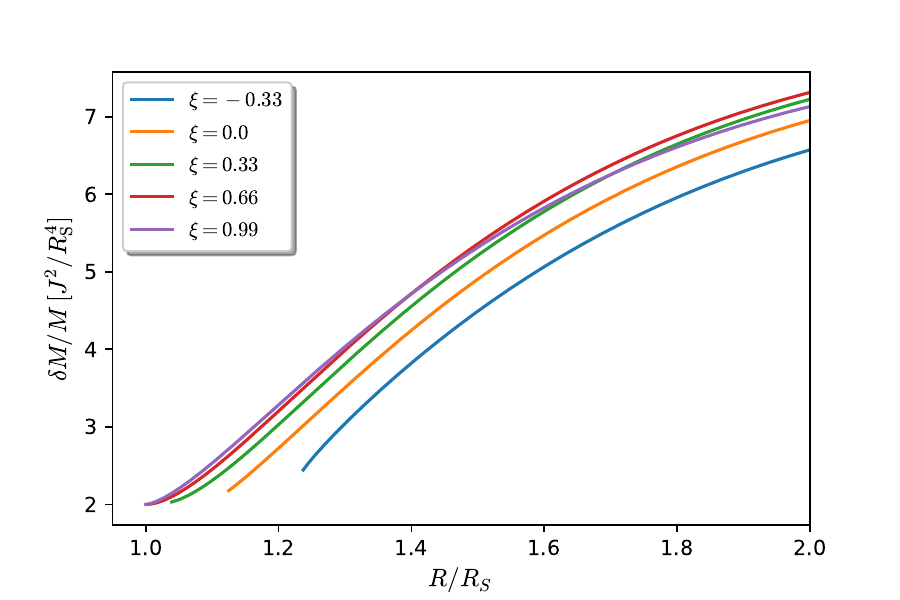}
    \caption{Change in mass $\delta M/M$ (in units of $J^2/\Rs^4$), as a function of the parameter $R/\Rs$, for various values of the anisotropy parameter $\xi$. For highly anisotropic configurations, we observe that $\delta M/M\to 2$ as $R\to\Rs$.}
    \label{fig:dm}
\end{figure}

We can use Eq.~\eqref{dmMODfin} to compute the change in mass $\delta M$. We plot the results in Fig.~\ref{fig:dm}, where we measure $\delta M/M$ in units of $J^2/\Rs^4$. In the isotropic case $\xi=0$, we recover with very good agreement the results reported by Refs.~\cite{Reina:2015jia, Beltracchi:2023qla} for uniform density stars. Note that $\delta M/M$ seems to increase as $R/\Rs$ increases. For highly anisotropic configurations, in the limit as $R\to R_{\text{cr}}$, $\delta M/M\to 2J^2/\Rs^4$, the same value as a Kerr BH and sub-Buchdahl Schwarzschild stars in the gravastar limit~\cite{Beltracchi:2023qla}.~\footnote{We stress again that the strict BH limit cannot be taken from any of these configurations; all of them are limited by their critical radii.} This allows regularity of $h_0$ in the exterior because the numerator of Eq.~\eqref{h0out} will go to 0 with the denominator.

\begin{figure*}[ht]
\centering
\includegraphics[width=0.495\linewidth]{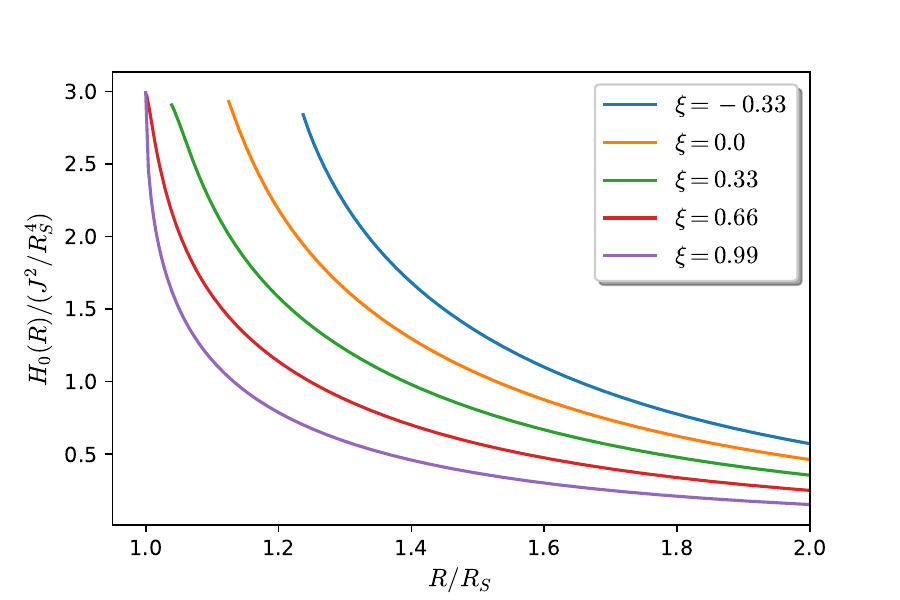}
\includegraphics[width=0.495\linewidth]{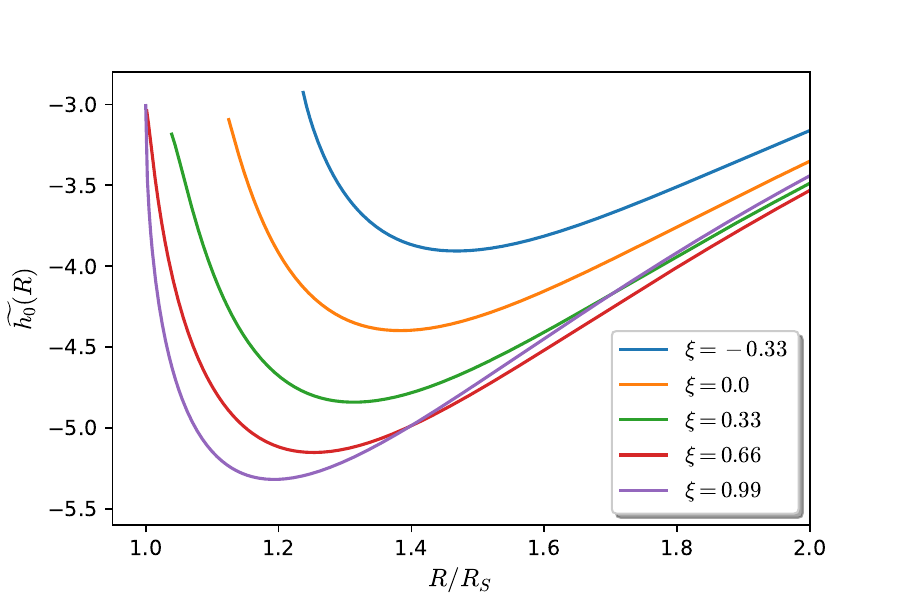}
\caption{{\bf Left panel:} surface value $H_0(R)$ (in units of $J^2/\Rs^4$), as a function of $R/\Rs$, for various values of the anisotropic parameter $\xi$. As $R$ approaches the critical value $R_{\text{cr}}$, $H_0(R)$ increases. At a given radius, increasing anisotropy tends to decrease $H_0(R)$. Observe that in the $R\rightarrow \Rs$ limit, $H_0(R)\to 3J^2/\Rs^4$. {\bf Right panel:} surface value $\widetilde{h_0}(R)$, as a function of $R/\Rs$, for the same values of $\xi$ as in the left panel. Observe that all of the curves have a local minimum. For the highly anisotropic configurations, as $R\to\Rs$, $\widetilde{h_{0}}(R)\to-3$, while the others approach approximately $-3$ as we approach their critical radii.}
\label{fig:H0}
\end{figure*}

Recall that finding the true $h_0$ function requires a few steps. First, find $H_0$ as a solution to Eq.~\eqref{newh0} with an assumed integration constant. Then, compute the necessary integration constant to make it match with the exterior solution
\beq
h_0 = \frac{1}{r-2M}\left(\frac{J^2}{r^3} - \delta M\right),\quad r>R,
\eeq
so the value from the exterior is given. We will not have a mismatch with the $h$ functions here because the background $g_{tt}$ is nonzero and finite. The values of $H_0(R)$, in units of $J^2/\Rs^4$, are plotted in the left panel in Fig.~\ref{fig:H0}, as a function of $R/\Rs$. We consider various values of the anisotropy parameter $\xi$. Observe that, for the various curves, as the compactness increases the surface value $H_0(R)$ grows monotonically, approaching the value of about $3J^2/\Rs^4$, with minor variations for different anisotropies. In the right panel of the same figure, we plot the surface values $\widetilde{h_0}(R)$, as a function of the parameter $R/\Rs$, for the same values of anisotropy as in the left panel. We observe that the various curves have a local minimum; moreover, as they approach their critical radii $R_{\text{cr}}$, $\widetilde{h_0}\to -3$ with minor variations for different anisotropies again. Note that, for highly anisotropic configurations, their critical radii are very close to the Schwarzschild radius, and $H_0(R)$ and $\widetilde{h_0}(R)$ take values of almost exactly $\pm 3$. It is noteworthy that we found the same behavior for sub-Buchdahl Schwarzschild stars when their compactness approaches the BH limit~\cite{Beltracchi:2023qla}.

\begin{figure}[ht]
    \centering
    \includegraphics[width=\linewidth]{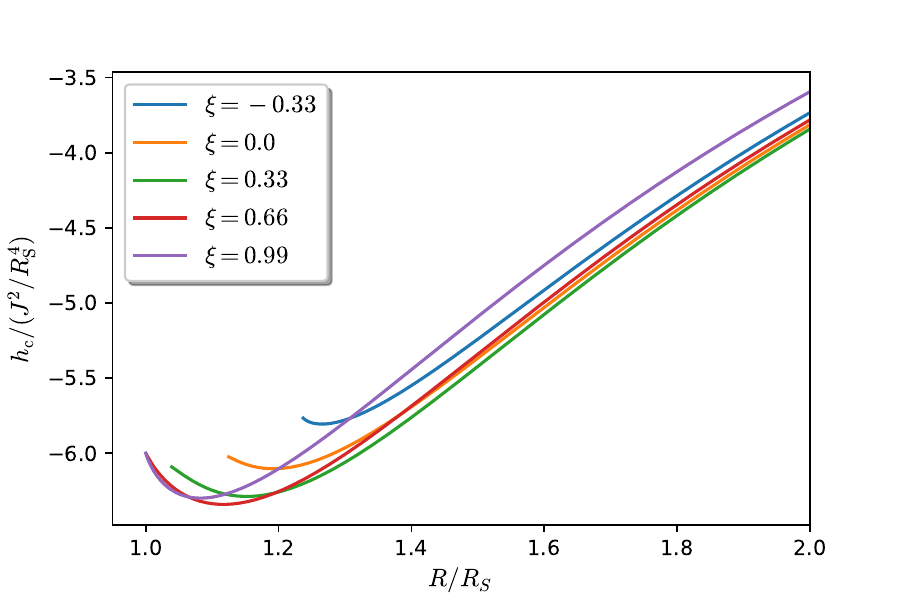}
    \caption{Profiles of $h_{\mathrm{c}}$, as a function of $R/\Rs$, for various values of the anisotropic parameter $\xi$. The constant $h_c$ is measured in units of $J^2/\Rs^4$. Note that the various curves have a local minimum. For configurations with high anisotropy, as $R\rightarrow\Rs$, $h_c\rightarrow-6J^2/\Rs^4$.}
    \label{fig:hc}
\end{figure}

Finally, from Eq.~\eqref{hc_eqn}, we can determine the necessary constant $h_c$ to match the functions $h_0$ and $H_0$. In Fig.~\ref{fig:hc}, we display the value of the constant $h_c$, in units of $J^2/\Rs^4$, as a function of $R/\Rs$, for various values of $\xi$. One interesting observation is that as we approach the critical radius $R\to R_{\text{cr}}$, we have 
\beq
H_0(R)=\frac{3J^2}{\Rs^4},\quad h_0(R)=-\frac{3J^2}{\Rs^4},\quad h_c=-\frac{6J^2}{\Rs^4},
\eeq
which matches the behavior of the sub-Buchdahl Schwarzschild stars in the gravastar limit~\cite{Beltracchi:2023qla}.

\subsection{Quadrupole perturbations for the Bowers-Liang sphere}
The quadrupole perturbations of the slowly rotating BL spheres are determined by Eqs.~\eqref{upsdef}, \eqref{constheta}, \eqref{quadA},  and \eqref{quadC}. As we discussed in Sec.~\ref{Sec:quad}, the standard approach to integrate this system of equations is to separate the general solution into a homogeneous part ($\varpi=0$), and a particular solution ($\varpi\neq 0$) as given by Eq.~\eqref{split}. Thus, approximating the quadrupole functions as regular series solutions, for the homogeneous part we obtain the following expressions for the series coefficients near the origin:
\begin{multline}
h_2^{\text{(H)}}=- r^2 \\
+ \frac{4}{7}\frac{(y_1-1)\left[3(q+2)y_{1}^{q} - 2\left(1+6q\right)\right]}{\left(1+6q\right)\left(3y_1^{q}-1\right)}\frac{r^4}{R^2},
\end{multline}
\begin{multline}
k_2^{\text{(H)}} = r^2\\
-\frac{3}{14}\frac{(y_1-1)\left[(9+22q)y_{1}^{q} - 3(1+6q)\right]}{(1+6q)\left(3y_{1}^{q}-1\right)}\frac{r^4}{R^2},
\end{multline}
\beq
\Upsilon^{\text{(H)}}=-\frac{9}{4\pi}\frac{(y_1-1)(2q-1)y_{1}^{q}}{(1+6q)\left(3y_{1}^{q}-1\right)}\frac{r}{R^2},
\eeq
\beq
\PP_2^{\text{(H)}}=\frac{6}{\pi}\frac{q(y_1-1)y_{1}^{q}}{(1+6q)\left(3y_{1}^{q}-1\right)}\frac{r^2}{R^2}+\mathcal{O}(r^4),
\eeq
\beq
\TT_2^{(\text{H})}=\frac{3(1-4q)(y_1-1)y_1^q}{\pi(1+6q)(3y_1^q-1)}\frac{r^2}{R^2},
\eeq
where we have set $k_2^a=1$. These solutions are compatible with the general expressions~\eqref{k2p_expa}--\eqref{T2p_expa}. For the particular solutions, we set $w_c=1$ and $k_2^a=0$ to obtain the following behaviors near the origin:
\beq
k_2^\text{(P)}=-\frac{2^{1/q}(46q+5)(y_1-1) y_1^q}{7(6q+1)\left(3
   y_1^q-1\right)^{\tilde{q}}}\frac{r^4}{R^2},
\label{k2p_exp}
\eeq
\beq
h_{2}^\text{(P)}=\frac{2^{\tilde{q}} (16q-1) (y_1-1) y_1^q}{7(6q+1)\left(3
   y_1^q-1\right)^{\tilde{q}}}\frac{r^4}{R^2},
\label{h2p_exp}
\eeq
\beq
\Upsilon^\text{(P)}=\frac{3}{\pi}\frac{2^{\frac{1}{q}-2}(2q-1)(y_1-1)y_1^q}{\left(6q+1\right)\left(3y_1^{q}-1\right)^{\tilde{q}}}\frac{r}{R^2},
\label{ups_exp}
\eeq
\beq
\PP_2^\text{(P)}=\frac{2^{\tilde{q}}(y_1-1)q~y_1^q}{\pi(6q+1)\left(3y_1^q-1\right)^{\tilde{q}}}\frac{r^2}{R^2},
\label{pp2_exp}
\eeq
\beq
\TT_2^\text{(P)}=\frac{2^{1/q} (4q-1)(y_1-1)y_1^q}{\pi\left(6q+1\right)\left(3y_1^q-1\right)^{\tilde{q}}}\frac{r^2}{R^2},
\label{tt2_exp}
\eeq
where $\tilde{q}\equiv 1+1/q$. Equations~\eqref{k2p_exp}--\eqref{tt2_exp} are also compatible with the general expressions~\eqref{k2p_expa}--\eqref{T2p_expa}.

We numerically integrated the quadrupole perturbation equations from the origin (or rather some cutoff value $x_0=10^{-7}$). It is worthwhile to mention that, when we approached too close to the critical radius $R_{\mathrm{cr}}$, we found numerical instabilities in the respective codes; thus, we stopped the integration at the value $R_{\mathrm{stop}}=R_{\mathrm{cr}}+\delta$, with $\delta=10^{-4}$.

Following the conventions used in~\cite{Beltracchi:2023qla}, we introduce the quantities
\begin{subequations}
\beq
\widetilde{h_{2}}\equiv \frac{h_2}{(J^2/\Rs^4)},\quad
\widetilde{k_{2}}\equiv \frac{k_2}{(J^2/\Rs^4)},
\eeq
\beq
\widetilde{m_{2}}\equiv \frac{m_2}{(J^2/\Rs^3)},\quad
\quad \widetilde{p_{2}^{*}}\equiv \frac{p_{2}^{*}}{(J^2/\Rs^4)},
\eeq
\beq
\widetilde{\TT_2}\equiv\frac{\TT_2}{(J^2/\Rs^6)},\quad\widetilde{\PP_2}\equiv \frac{\PP_2}{(J^2/\Rs^6)}.
\eeq
\end{subequations}

\begin{figure}[ht!]
\includegraphics[width=\linewidth]{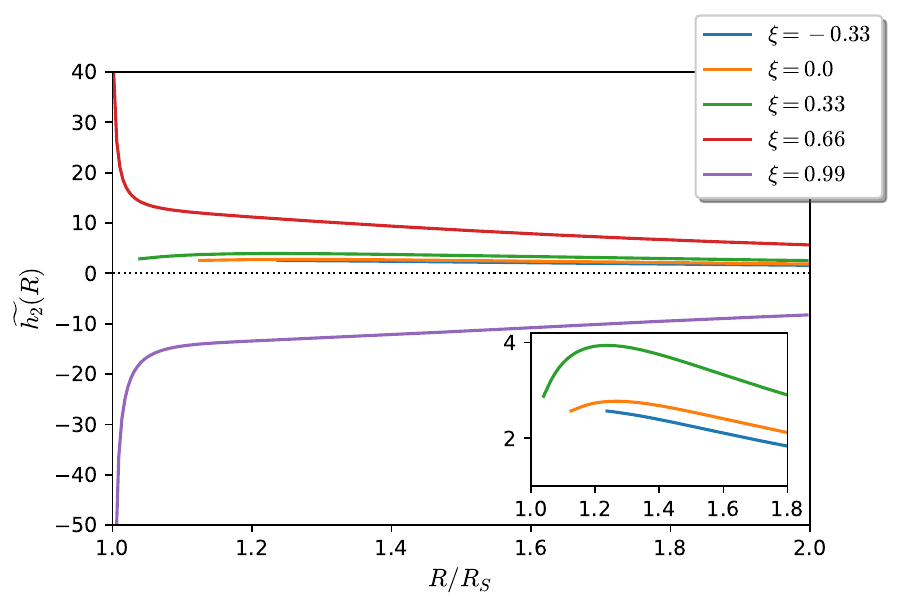}
\caption{Surface value $\widetilde{h_2}(R)$, as a function of $R/\Rs$, for various values of the anisotropy parameter $\xi$. The inset enlarges the curves $\xi=\{-0.33,0,0.33\}$. Observe the rapid increase (decrease) in the curves for $\xi= 0.66~(0.99)$, which do not fit on the scale of this plot.}
\label{fig:h2s}
\end{figure}

In Fig.~\ref{fig:h2s}, we plot the surface value $\widetilde{h_2}(R)$, as a function of $R/\Rs$, for different values of the anisotropy parameter $\xi$. In the isotropic case $\xi=0$, we recover with very good agreement the results presented by us for Schwarzschild stars above the Buchdahl limit~\cite{Beltracchi:2023qla}. Observe how the curve $\xi=0.66$ grows rapidly as we approach the critical radius, while the extremely anisotropic curve $\xi=0.99$ takes negative values and decreases rapidly as $R\to R_{\text{cr}}$.

\begin{figure}[ht!]
\includegraphics[width=\linewidth]{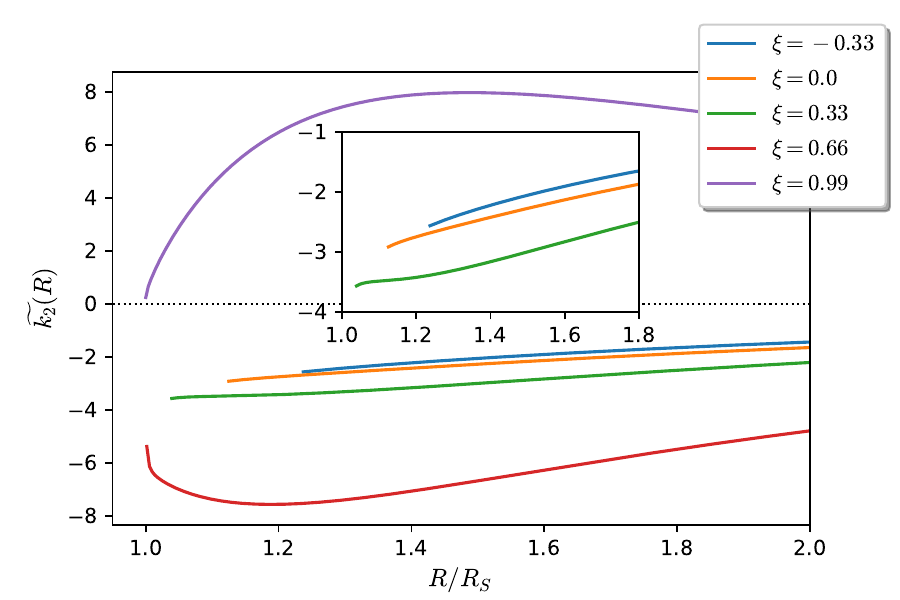}
\caption{Surface value $\widetilde{k_2}(R)$, as a function of $R/\Rs$, for various values of the anisotropy parameter $\xi$. The inset enlarges the curves $\xi=\{-0.33, 0, 0.33\}$.}
\label{fig:k2s}
\end{figure}

In Fig.~\ref{fig:k2s}, we display the surface value $\widetilde{k_2}(R)$, as a function of $R/\Rs$, for various values of the anisotropy $\xi$. In the isotropic case $\xi=0$, we recover with very good agreement the results presented by us for Schwarzschild stars above the Buchdahl limit (see Fig. 14 in~\cite{Beltracchi:2023qla}). Note that the function $\widetilde{k_2}(R)$ takes negative values, except for the highly anisotropic case $\xi=0.99$, which is positive and seems to approach zero as $R\to R_{\text{cr}}$.

\begin{figure}[ht!]
\includegraphics[width=\linewidth]{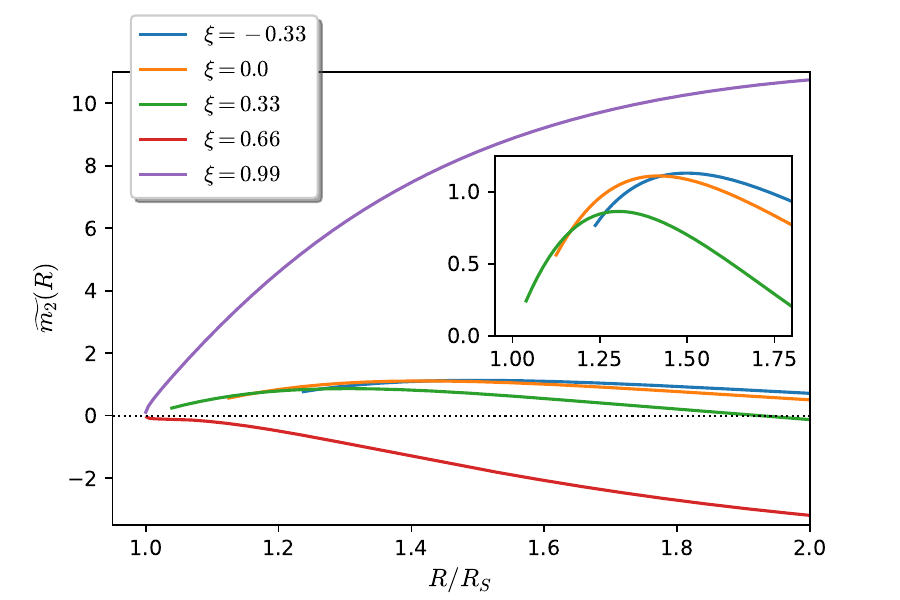}
\caption{Surface value $\widetilde{m_2}(R)$, as a function of $R/\Rs$, for various values of the anisotropy parameter $\xi$. The inset enlarges the curves $\xi=\{-0.33, 0, 0.33\}$. For highly anisotropic configurations, as $R\to R_{\text{cr}}$, $m_2(R)\to 0$.}
\label{fig:m2s}
\end{figure}

In Fig.~\ref{fig:m2s}, we display the value at the boundary of the perturbation function $\widetilde{m_2}(R)$, as a function of $R/\Rs$, for various values of the parameter $\xi$. First of all, in the isotropic case $\xi=0$, we recover with very good agreement the results presented by us for Schwarzschild stars above the Buchdahl bound (see Fig.~13 in~\cite{Beltracchi:2023qla}). Note that, for the case $\xi=0.66$, $\widetilde{m_2}(R)$ takes negative values. For highly anisotropic configurations, $\widetilde{m_2}(R)\to 0$, as $R\to R_{\text{cr}}$. It is interesting that we found the same behavior for Schwarzschild stars in the gravastar limit~\cite{Beltracchi:2023qla}. 
\begin{figure*}[ht!]
\includegraphics[width=0.495\linewidth]{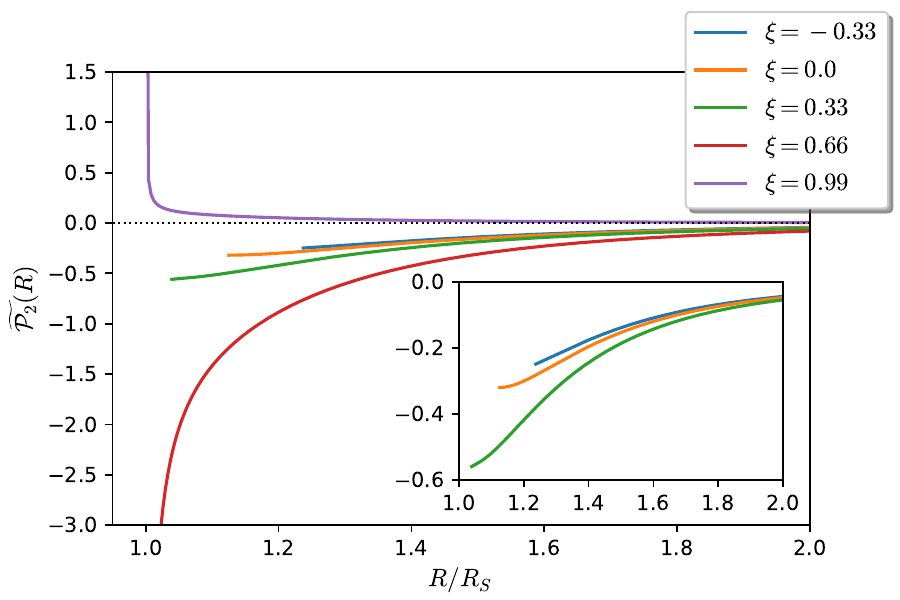}
\includegraphics[width=0.495\linewidth]{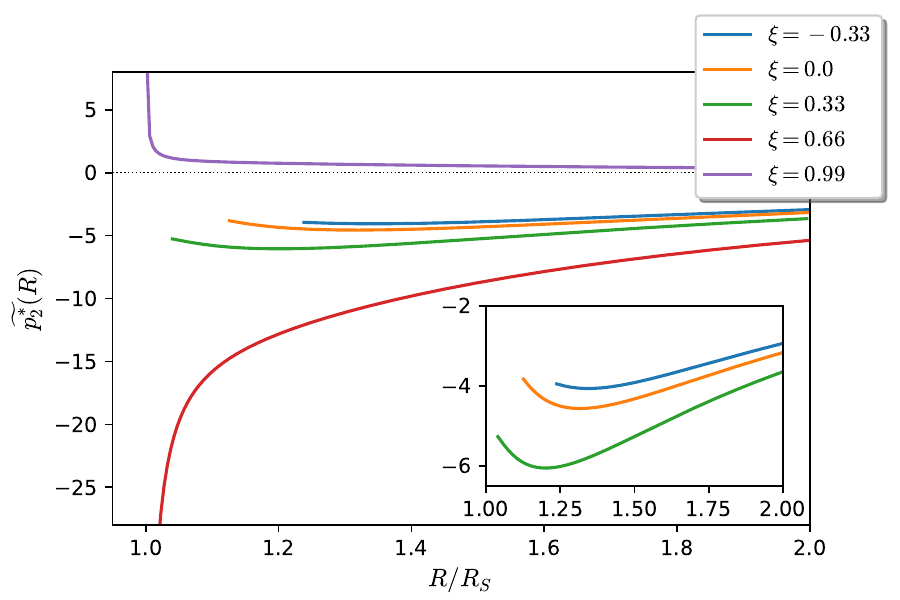}
\caption{{\bf Left panel:} surface value of the perturbation function $\widetilde{\PP_2}(R)$, as a function of $R/\Rs$, for various values of the anisotropy parameter $\xi$. {\bf Right panel}: the same data as in the left panel, but now in terms of the associated perturbation function $p_{2}^{*}=\PP_2/(\rho+p_{r})$, as a function of $R/\Rs$. The insets enlarges the curves $\xi=\{-0.33, 0, 0.33\}$.}
\label{fig:PP2s}
\end{figure*}

In the left panel in Fig.~\ref{fig:PP2s} we show the profiles of the perturbation function $\widetilde{\PP_2}(R)$, evaluated at the boundary, as a function of the parameter $R/\Rs$. The various curves correspond to different values of the anisotropy parameter $\xi$. The inset enhances the cases $\xi=\{-0,33, 0, 0.33\}$. Observe that the various profiles take negative values, except for the ultra-anisotropic case $\xi=0.99$, which is positive. In the right panel of the same figure, we plot the same data as in the left panel, but now in terms of the auxiliary perturbation function $\widetilde{p_{2}^{*}}$. Note that, for the highly anisotropic configurations $\xi=0.66 (0.99)$, the function $\widetilde{p_{2}^{*}}$ decreases (grows) rapidly as the corresponding stellar radii approaches their critical value.

\begin{figure}[ht!]
\includegraphics[width=\linewidth]{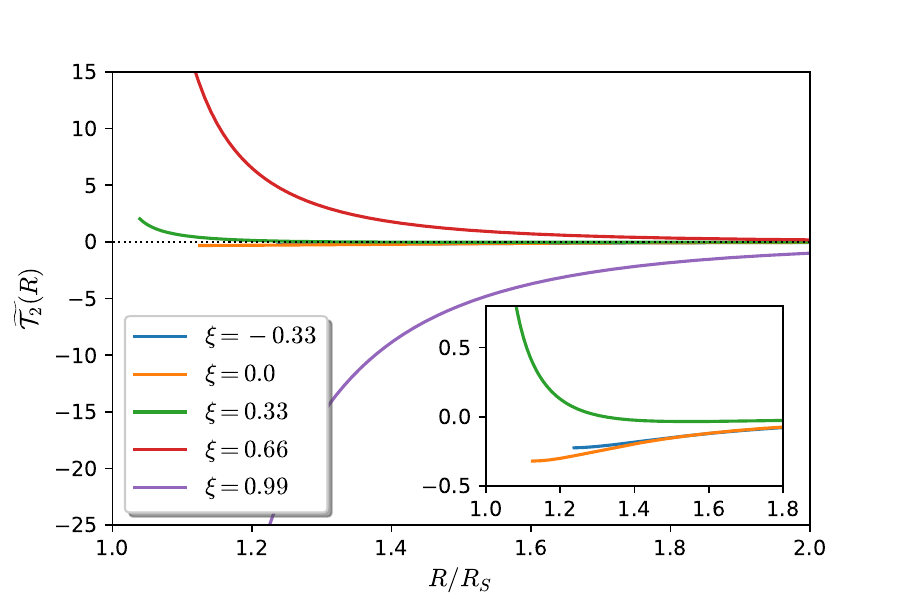}
\caption{Surface value $\widetilde{\TT_2}(R)$, as a function of $R/\Rs$, for various values of the anisotropy parameter $\xi$. The inset enlarges the curves $\xi=\{-0.33, 0, 0.33\}$. As we approach the critical radius for $\xi=0.66,0.99$, the values for $\TT_2$ get extremely large; for instance, $\widetilde{\TT_2}(\xi=0.99,R/R_S=1.001)>100000$.}
\label{fig:T2s}
\end{figure}

In Fig.~\ref{fig:T2s}, we display the surface perturbation function $\widetilde{\TT_2}(R)$, as a function of $R/\Rs$, for different anisotropies $\xi$. In the isotropic case, $\widetilde{\TT_2}(R)=\widetilde{\PP_2}(R)$; thus, the $\xi=0$ curve agrees with the one in Fig.~\ref{fig:PP2s}. Observe that, for the highly anisotropic configurations, the function $\widetilde{\TT_2}(R)$ grows rapidly (out of the plot $y$ scale) as $R\to R_{\mathrm{cr}}$.

\begin{figure}[ht!]
\includegraphics[width=\linewidth]{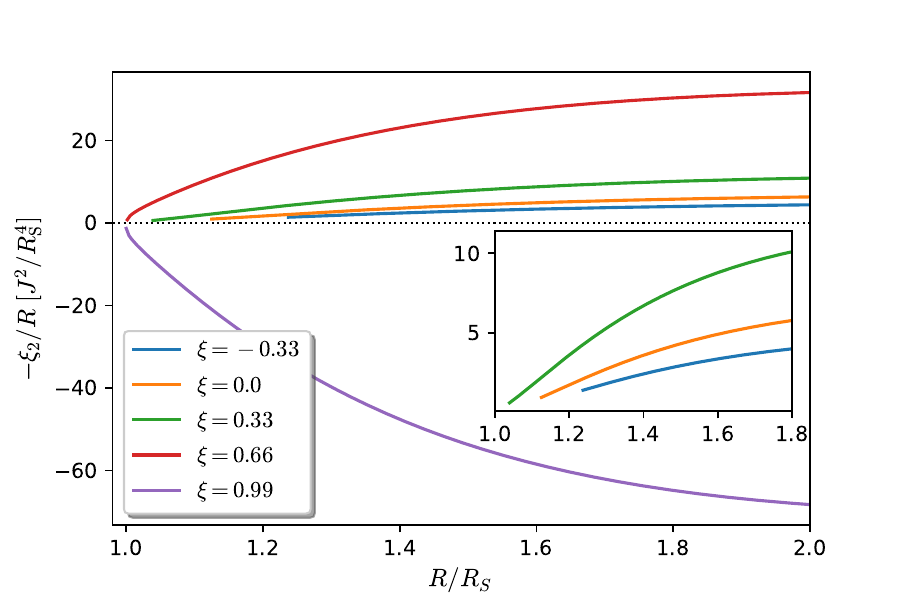}
\caption{The quadrupole deformation function $-\xi_{2}(R)/R$, as a function of $R/\Rs$, for various values of the anisotropy parameter $\xi$. We measure $-\xi_{2}(R)/R$ in units of $J^2/\Rs^4$. The inset enlarges the curves $\xi=\{-0.33, 0, 0.33\}$.}
\label{fig:xi2Rs}
\end{figure}

Figure~\ref{fig:xi2Rs} shows the function $-\xi_{2}(R)/R$ (measured in units of $J^2/\Rs^4$), which determines the $l=2$ deformation of the boundary, as a function of the parameter $R/\Rs$. The various curves correspond to different values of the anisotropy parameter $\xi$. In the isotropic case $\xi=0$, we recover with excellent agreement the results reported by Refs.~\cite{Chandra:1974, Beltracchi:2023qla}, for isotropic uniform density configurations. Observe that, for the ultra-anisotropic case $\xi=0.99$, the function $-\xi_{2}(R)/R$ takes negative values.

\begin{figure}[ht!]
\includegraphics[width=\linewidth]{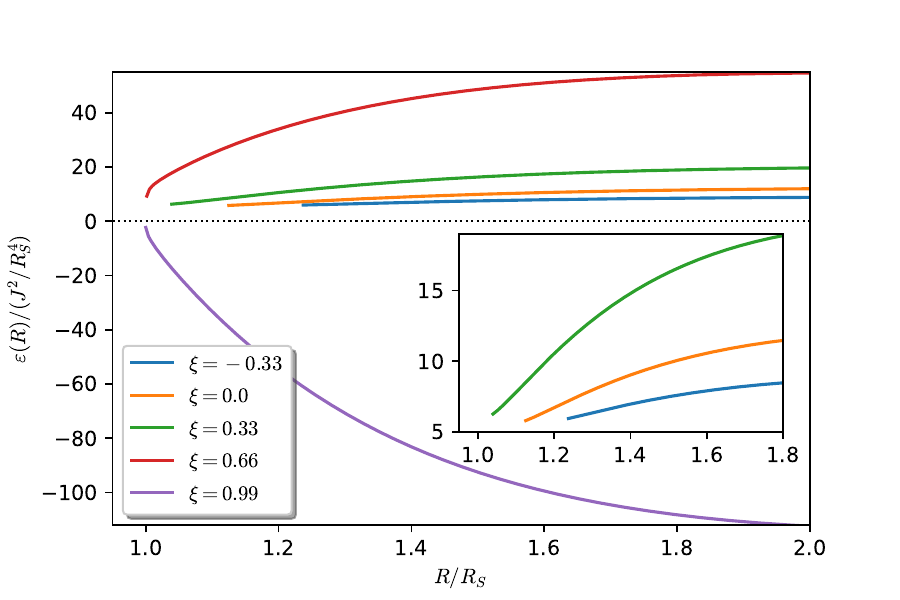}
\caption{Surface value of the ellipticity $\epsilon(R)$, as a function of $R/\Rs$, for various values of the anisotropy parameter $\xi$. The ellipticity is measured in units of $J^2/\Rs^4$. The inset enlarges the curves $\xi=\{-0.33, 0, 0.33\}$.}
\label{fig:elips}
\end{figure}

The ellipticity of the bounding surface $\epsilon$, as defined in~\eqref{elipticitydef}, is plotted in Fig.~\ref{fig:elips} against $R/\Rs$, for various values of the parameter $\xi$. We measure $\varepsilon$ in units of $J^2/\Rs^4$. In the isotropic case $\xi=0$, we corroborate the results reported by Refs.~\cite{Chandra:1974,Beltracchi:2023qla} for Schwarzschild stars above the Buchdahl limit. It is interesting that the highly anisotropic configuration $\xi=0.99$ shows a \emph{negative} ellipticity. Thus, such a configuration, when it is set into slow rotation, becomes \emph{prolate} rather than oblate. For static spherically symmetric systems, Eq.~\eqref{tov} can be interpreted as having terms corresponding to a pressure gradient force $dp_{r}/dr$, gravitational force $(\rho+p_r)\nu'$, and a contribution due to the anisotropy $2(p_\perp-p_r)/r$. Physically, we also expect centrifugal-type forces to be present in the rotating configuration, as well as corrections to the pressure gradient, gravitational, and anisotropy forces. Some of the terms in Eqs.~\eqref{newp0} and \eqref{quadC} have the corresponding forms to be corrections to one of those forces; others might correspond to the centrifugal forces, and for some, it is not immediately clear which force they should be interpreted as. In any case, it seems likely that the prolate deformation stems mostly from a contribution of the anisotropy forces, although as we show in the next subsubsection we see that the interplay between the forces is rather complicated and there are certain configurations in which perturbation theory seems to fail.

\begin{figure}[ht!]
\includegraphics[width=\linewidth]{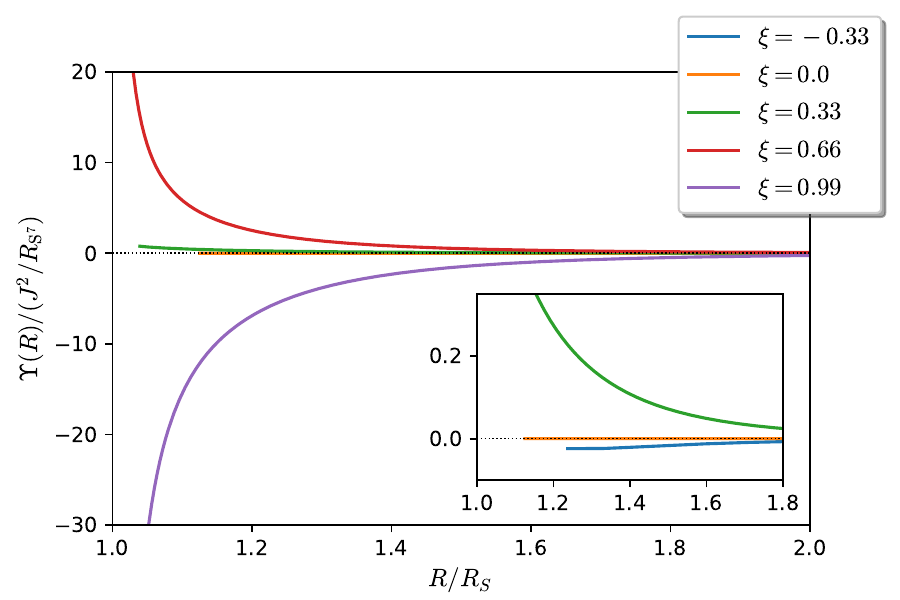}
\caption{Profiles of the auxiliary function $\Upsilon$, evaluated at the surface, as a function of $R/\Rs$, for various values of the anisotropy parameter $\xi$. We measure $\Upsilon(R)$ in units of $J^2/\Rs^7$. The inset enlarges the curves $\xi=\{-0.33, 0, 0.33\}$. Notice that in the isotropic $\xi=0$ case $\Upsilon$ is identically zero, but for nonzero anisotropies it takes nonzero values.}
\label{fig:Ys}
\end{figure}

In Fig.~\ref{fig:Ys} we show the surface value of the auxiliary function $\Upsilon(R)$ (see Eq.~\eqref{upsdef}), as a function of $R/\Rs$, for different values of the anisotropy parameter $\xi$. We measure $\Upsilon(R)$ in units of $J^2/\Rs^7$. Observe that, in the isotropic case $\xi=0$, $\Upsilon$ vanishes identically; however, for nonzero values of the anisotropy, $\Upsilon$ is different from zero.

\begin{figure}[ht!]
\includegraphics[width=\linewidth]{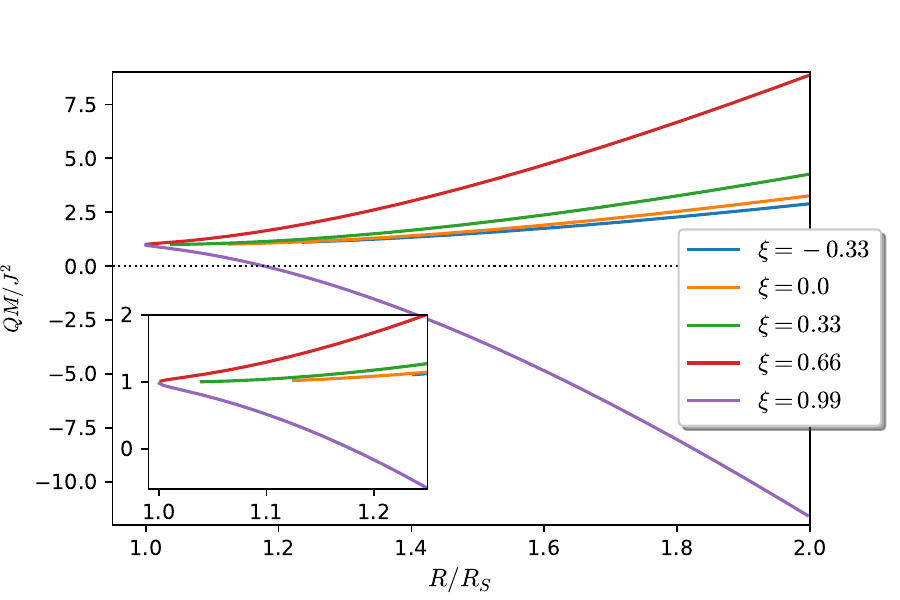}
\caption{Kerr factor $\widetilde{Q}\equiv QM/J^2$, as a function of $R/\Rs$, for various values of the anisotropy parameter $\xi$. The inset enlarges the region $R/\Rs=(1, 1.2]$. Observe how, in the limit as $R\to R_{\text{cr}}$, the Kerr factor for the various curves approaches the BH value $\widetilde{Q}\to 1$, with only minor variation between anisotropies.}
\label{fig:Q}
\end{figure}

The “Kerr factor" $\widetilde{Q}\equiv QM/J^2$ is plotted in Fig.~\ref{fig:Q}, as a function of $R/\Rs$, for various values of the anisotropy parameter $\xi$. This Kerr factor is relevant because it tells how much it deviates the exterior Hartle-Thorne metric away from the Kerr metric. In the isotropic case $\xi=0$, we have a very good agreement with the results reported by Refs.~\cite{Chandra:1974, Beltracchi:2023qla} for Schwarzschild stars above the Buchdahl bound. Observe that, as the various configurations approach their critical radii, $\widetilde{Q}\to 1$, with slight variations depending on the anisotropy. However, we stress that the strict BH limit cannot be taken from these configurations, because their radii are limited by their corresponding critical radii, which are always less than the Schwarzschild radius. It is noteworthy that for $\xi=0.99$ we are approaching the Kerr value from below, and for radii  $R/R_S>1.18$ we have a \emph{negative} quadrupole moment, again indicating that we have highly anisotropic configurations which are prolate rather than oblate.

It should be noted that the highly anisotropic cases $\xi=0.66$, and $\xi=0.99$ have large, and in some cases possibly diverging surface values for the $h_2$, $\PP_2$, $\TT_2$, and $\Upsilon$ perturbation functions, when we approach the corresponding critical radius. This indicates that perturbation theory may be unreliable for high anisotropy spheres near to their critical radius, or that the allowable amount of rotation $\Omega$, before the Hartle method no longer gives satisfactory results, is extremely small. We observe that the $m_2$ and $k_2$ functions remain finite; for $m_2$, this likely is due to the $r-2m$ factor, which is very small near the critical radius. For $k_2$, it seems that the integration constant $K$ is approaching zero at the same rate as the Legendre function [see Eqs.~\eqref{h2out} and \eqref{v2out}] is diverging to end up with a finite residue (if $K=0$, then we take the limit $\tilde{k_2}\to -4$, but we are approaching a different constant). However, the $\xi=0.33$ case remains well behaved with finite surface values, for all the perturbation functions, as we are approaching the critical radius, which is below the Buchdahl radius. 

Finally, in Table~\ref{tab:dec}, we present the main surface and integral properties for the BL sphere that satisfies the DEC, which was discussed in Sec~\ref{sec:DEC}. Note that, while the spherically symmetric version of this BL sphere satisfies the DEC, the rotating version fails to do so, because both $\TT_0$ and $\TT_2$ are positive at the surface of the configuration; thus, $\TT=p_\perp+\TT_0+\TT_2P_2(\cos\theta)$ will be larger than $\EE=\rho$, since $\rho=p_\perp$ on the surface for the DEC sphere. 

\begin{table}[h]
\centering
\begin{tabular}{|c c c c c c c|}
\hline
\hline
\multicolumn{7}{|c|}{Surface properties of the limiting DEC BL sphere}\\
\hline
$R/\Rs$ & $\xi$ & $I/MR^2$ & $\delta M/M$ & $\widetilde{Q}$ & $\varpi(R)$ & $\varepsilon$ \\
\hline
$1.1035$ & $0.4141$ & $0.8388$ & $2.3753$ &  1.0666 & 0.4696 & 9.3969 \\
\hline\hline
\end{tabular}
\caption{Surface and integral properties for the limiting BL sphere that satisfies the DEC. The various quantities are given in the same units as Table~\ref{tab:surf}.}
\label{tab:dec}
\end{table}

\subsubsection{The onset of negative quadrupole moment and ellipticity}

From Figs.~\ref{fig:xi2Rs},~\ref{fig:elips}, and \ref{fig:Q}, we found that for $\xi=0.99$ we have the opposite behavior to what we would expect, corresponding to prolate rather than oblate deformation. We consider that this point deserves further analysis. 

One possible explanation behind this behavior would be the breakdown of either the perturbative formalism or the numerics. However, there are several prolate configurations, for instance, $\xi=0.99,R=1.5$, where the interior perturbation functions are small (so perturbation theory works well) and monotonic (so numeric problems with oscillatory or rapidly varying functions should not be an issue). Even in cases quite close to divergences, there is little sensitivity to the $\epsilon$ integration parameters over a wide range, and consistency check codes written with a variety of schemes [four first-order ordinary differential equations (ODEs), two second-order ODEs, using units of $r/\Rs$ and $r/R$, etc.] all give very similar results (within 1\% error). This stability and good behavior strongly disfavor numerical instabilities or formalism breakdown being the cause of the prolate behavior.

In any event, information about exactly when and how prolate configurations appear in the results is worth investigating. For that purpose, we carried out a thorough study of the onset of negative mass quadrupole moment and ellipticity. Let us first discuss the former.

\begin{figure*}[ht!]
\includegraphics[width=.495\linewidth]{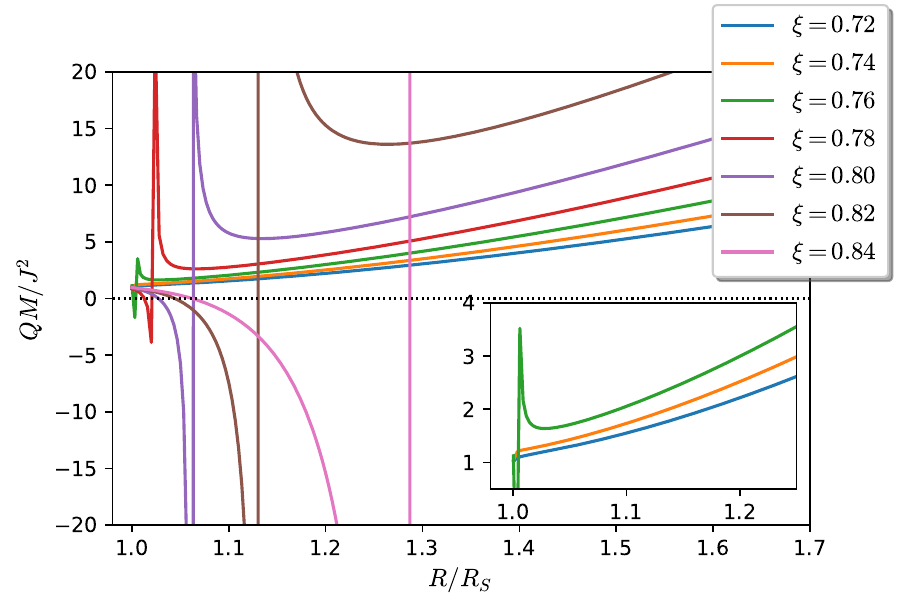}
\includegraphics[width=.495\linewidth]{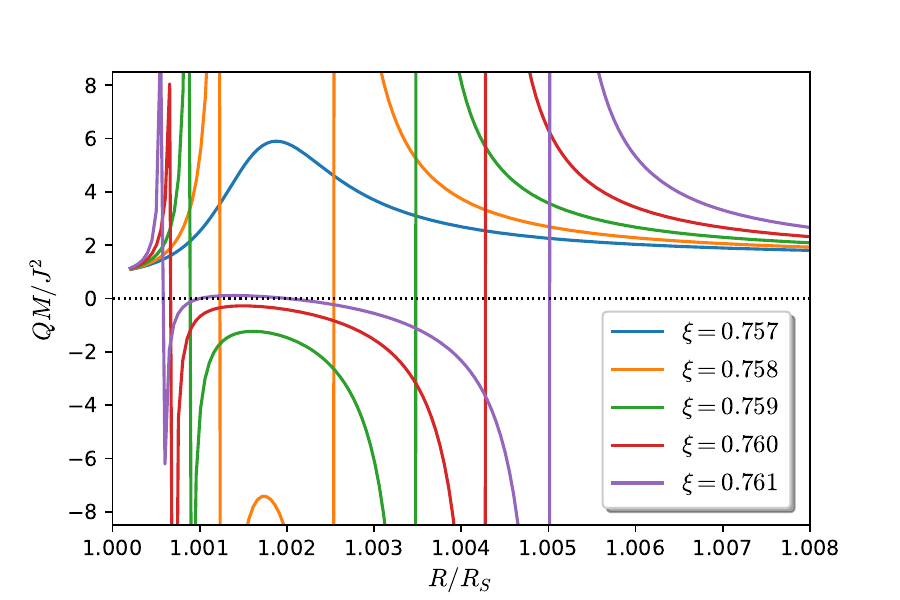}
\caption{{\bf Left panel:} Kerr factor $\widetilde{Q}\equiv QM/J^2$, as a function of $R/\Rs$, for various values of the anisotropic parameter $\xi$ in the range $\xi=[0.76, 0.84]$. The inset enhances the curves $\xi=\{0.72, 0.74, 0.76\}$. {\bf Right panel:} ‘Kerr' factor $\widetilde{Q}$, as a function of $R/\Rs$, for $\xi$ in the range $\xi\in [0.757, 0.761]$. Observe the emergence of regions with a negative mass quadrupole moment for $\xi>0.757$. In the limit as $R\to R_{\mathrm{cr}}$, $\widetilde{Q}\to 1$ corresponding to the Kerr BH value.}
\label{fig:Q_poles}
\end{figure*}

In the left panel in Fig.~\ref{fig:Q_poles} we plot the Kerr factor $\widetilde{Q}\equiv QM/J^2$, as a function of $R/\Rs$, for various values of the anisotropy parameter $\xi$ in the range $[0.72,0.84]$. We observe that the curves for $\xi=0.72, 0.74$ are continuous, and well-behaved, and they seem to approach to 1 as the compactness approaches its corresponding critical value. For $\xi=0.76$, we observe that the curve is also continuous, but it shows a kink when we approach the critical radius. For configurations with $\xi>0.76$, we observe that $\widetilde{Q}$ becomes large as we approach the critical compactness. Furthermore, we also observe the appearance of regions of negative quadrupole moment together with divergences at certain locations; the latter may indicate the failure of the extended Hartle's slow rotation framework in such a regime.

In the right panel in Fig.~\ref{fig:Q_poles} we plot the Kerr factor again, but now enhancing in the region where $\xi\in[0.757, 0.761]$. We observe that, while the quadrupole moment remains positive for $\xi=0.757$, there is a “peak" close to $\beta\simeq~1.0018$, where $\widetilde{Q}\simeq 6$. At $\xi=0.758$, there is a region of negative quadrupole moment centered around $\beta=1.002$, which is bounded by two divergences at $\beta=1.0012$ and $\beta=1.0025$, which we call $D_A$ and $D_B$, respectively. These divergences move away from $\beta=1.002$ as $\xi$ increases. At $\xi=0.761$, the peak between the two divergences, located at $\beta=1.0006$ and $\beta=1.0051$, respectively, breaks into the positive region. Thus, we have an upward and downward $x$-axis crossing, between the divergences, at $\beta\simeq~1.0009$ and $\beta\simeq~1.002$, which we call $Z_A$ and $Z_B$, respectively (see the right panel in Fig.~\ref{fig:Q_poles} for an illustration of this process). As we continue to increase $\xi$, the divergences and zero crossings move farther apart. 

Before we reach $\xi=0.77$, we lose track of $D_A$ and $Z_A$ because they get too close to the critical radius and the numerical solutions are unreliable. This occurs at $\xi=0.768$, when $D_A$ is at $\beta=1.0002$ and $Z_A$ is at $\beta=1.00025$. Beyond this point, it is possible that $D_A$ and $Z_A$ go below the critical radius, collide with each other and annihilate, or stay above the critical radius in the region we cannot probe. In Table~\ref{tab:zeros} we list the results of the various divergences and $x$-axis crossings of the quadrupole moment, for different values of the anisotropy parameter $\xi$ in the range $\xi\in[0.758, 0.99]$.
 
The next interesting feature is the “acceleration" in outward movement of $D_B$ as $\xi$ increases and the subsequent disappearance of $D_B$ close to $\xi=0.89$. Beyond this point, we find no instances of a positive quadrupole moment at higher radii and no divergence, because it appears to have moved off to $\beta=\infty$. The $Z_B$ zero crossing stays in the range of $\beta$ where it is easy to track. It can still be seen clearly in Fig.~\ref{fig:Q} for $\xi=0.99$. 

It is noteworthy that, despite this peculiar behavior of the mass quadrupole moment showing divergences, zero crossings, and negative values, as we approach the critical radius for the different values of $\xi$, we observe that $\widetilde{Q}\to 1$, or very nearly, thus approaching the Kerr BH value. One possible caveat to this is the influence of the inner divergence, or upward zero crossing, after they cross into the region which we cannot probe.
\begin{table}[h]
\centering
\begin{tabular}{c|c|c|c|c}
\hline\hline
$\xi$ & $D_A$ & $Z_A$ & $Z_B$ & $D_B$ \\
\hline
0.758 & 1.0012 & X & X & 1.0025 \\
0.759 & 1.00086 & X & X & 1.0036\\
0.760 & 1.0006 &X &X &1.0043 \\
0.761 & 1.00054 &1.00095 &1.002 & 1.0050\\
0.763 & 1.00043 & 1.00054 & 1.00332 & 1.00657 \\
0.767 & 1.00023&1.00028 & 1.0054 & 1.0097\\
0.768 & 1.0002 & 1.00025 & 1.006 & 1.0106\\
0.770 &X &X &1.00705 & 1.0123\\
0.780 &X & X& 1.0129& 1.0239\\
0.790 & X& X&1.0197 & 1.0386\\
0.800&X & X&1.0274 & 1.0602\\
0.810&X & X&1.0357 & 1.0903\\
0.820&X & X& 1.0443 & 1.133\\
0.830&X & X& 1.0532& 1.1958\\
0.840&X & X& 1.0618& 1.289\\
0.850&X & X& 1.0711 & 1.4412\\
0.860&X & X& 1.080 & 1.708\\
0.870& X& X& 1.0885& 2.276\\
0.880&X & X& 1.097 & 4.170\\
0.885&X & X& 1.099 & 8.760\\
0.886&X & X& 1.102 & 11.66\\
0.887& X& X& 1.103 & 17.55\\
0.888& X& X& 1.104 & 36.64\\
0.890&X & X& 1.1050 & X \\
0.990&X &X & 1.174 & X \\
\hline\hline
\end{tabular}
\caption{Table of the various locations of features in $\widetilde{Q}$. X indicates the points that are not observed. We point out that some of our alternative codes which we constructed in different ways to our main code, for consistency checks, give slightly different answers for some of these, at and above the last decimal place shown. Note that the ellipticity also shows divergences at $D_A$ and $D_B$ but nothing special happens to ellipticity at $Z_A$ and $Z_B$.}
\label{tab:zeros}
\end{table}

\begin{figure*}[ht!]
\includegraphics[width=0.495\linewidth]{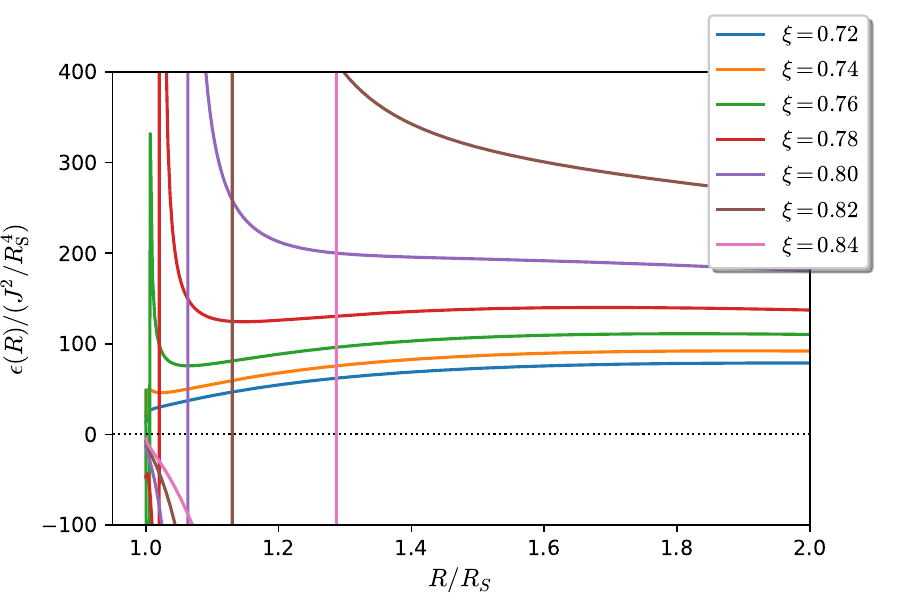}
\includegraphics[width=0.495\linewidth]{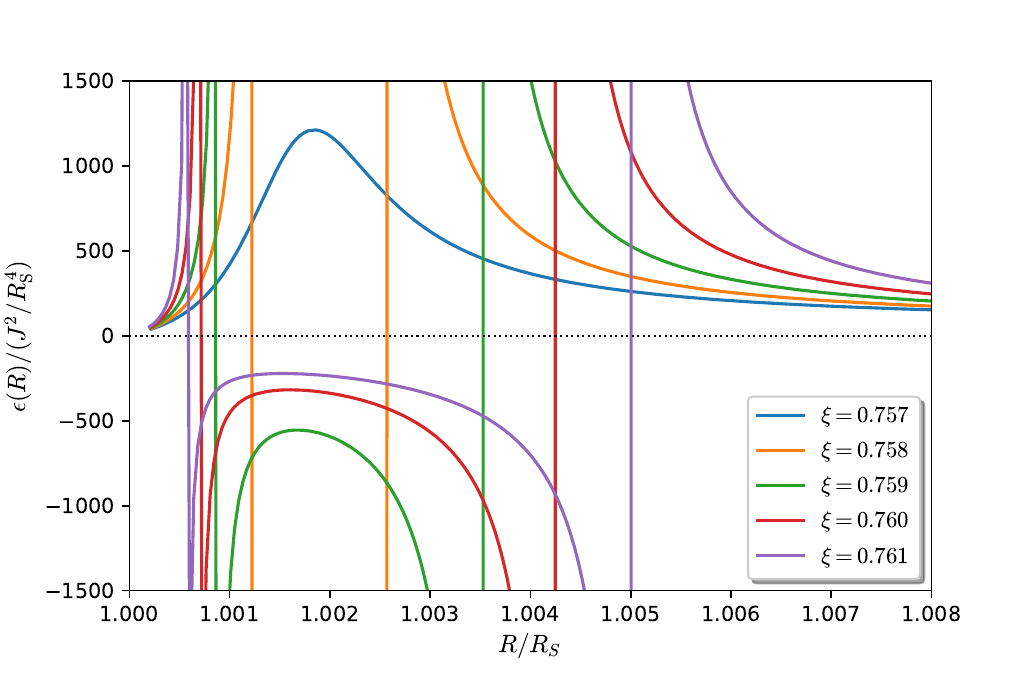}
\caption{{\bf Left panel:} ellipticity (in units of $J^2\Rs^4$), as a function of $R/\Rs$, for various values of the anisotropy parameter $\xi$ in the range $\xi\in [0.72,0.84]$. For highly anisotropic configurations, we observe that the ellipticity becomes large, for compactness near the critical value. Furthermore, for configurations with $\xi>0.76$ we observe the emergence of a region with negative ellipticity. {\bf Right panel:} ellipticity (in units of $J^2\Rs^4$), as a function of $R/\Rs$, enlarged for values of $\xi\in[0.757, 0.761]$. Note that the divergences occur at the same values of compactness as the divergences of $\widetilde{Q}$ (see Fig.~\ref{fig:Q_poles}).}
\label{fig:elips_poles}
\end{figure*}

We discuss now the case for the ellipticity. In the left panel in Fig.~\ref{fig:elips_poles}, we plot the ellipticity, in units of $J^2/\Rs^4$, as a function of $R/\Rs$, for various anisotropy parameters in the range $\xi\in[0.72,0.84]$. Observe that the curve with $\xi=0.72$ is well behaved, even up to its critical radius. For configurations with $\xi=0.74$, we observe that the curve is continuous and well behaved, but it shows a kink near to the critical radius. For greater values of the anisotropy parameter $\xi$, we observe that the ellipticity becomes large near the critical radius, showing a divergence there.

In the right panel in Fig.~\ref{fig:elips_poles}, we plot the ellipticity again, but now enhancing in the range $\xi\in$[0.757, 0.761] . Similarly to the mass quadrupole moment $\widetilde{Q}$, we observe that the ellipticity spikes near $\beta=1.0018$, at $\xi=0.758$, and divergences bounding a negative ellipticity region appear. Furthermore, the region with negative $\widetilde{Q}$ expands as $\xi$ increases. It is interesting that the divergences in the ellipticity occur at the same locations as the divergences in $\widetilde{Q}$, namely,  $D_A$ and $D_B$. This can be explained by the fact that divergences in the quadrupole moment are caused by divergences of the integration constant $K$, which, in turn, lead to divergences of the $k_2$ term in the expression for the ellipticity \eqref{elipticitydef}. However, unlike $\widetilde{Q}$, the ellipticity does not have any intersections with 0, so there are no $Z$ points. This means that there are configurations where the ellipticity of the boundary is negative but the mass quadrupole moment is positive. For instance, for $\xi>0.768$, ellipticity approaches a negative value as we approach our cutoff above the critical radius, but the quadrupole moment approaches $\approx1$. 

Despite the divergences and the regions with negative values, some of the curves in Fig.~\ref{fig:elips_poles} show recognizable features in ellipticity behavior. For instance, in isotropic Schwarzschild stars, there is a local maximum in ellipticity at $\beta\approx2.32$~\cite{Chandra:1974, Beltracchi:2023qla}; this local maximum moves inward as the anisotropy increases and can still be seen in the cases $\xi=\{0.72, 074, 0.76, 0.78\}$.

\section{Conclusions}
\label{sect:concl}

In this work, we have extended the Hartle formalism~\cite{Hartle:1967he}, for slowly rotating relativistic stars, to anisotropic systems. Our objective was twofold: First, we wanted to complete, and improve, previous attempts to generalize Hartle's formalism which contained typos and other errors \cite{Pattersons:2021lci}, casting doubts on the results presented there, or were limited to specific examples and not widely applicable \cite{Beltracchi:2022vvn}. Thus, here we provide the extended Hartle equations of structure and present their derivation. As a second objective, using this new formalism, we studied the rotational properties of anisotropic configurations with constant density, as described by the Bowers-Liang solution. 

While some of the structure equations we derive, such as Eqs.~\eqref{newomega} and \eqref{newm0}, agree with those presented in \cite{Pattersons:2021lci}, a critical difference relies in our treatment of the off-diagonal terms in the $r\theta$ block. While the $T^r_{~\theta}$ and $T^{\theta}_{~r}$ terms must be identically 0 for stationary axisymmetric perfect fluid systems, or for static spherically symmetric anisotropic systems, this is not {\it a priori} necessary for axisymmetric anisotropic systems. In order to treat this, we introduced the function $\Upsilon$ which can be thought of, mathematically, as a second order in $\Omega$ deviation of the vector which diagonalizes the $r\theta$ block of the EMT vs the $r\theta$ block of the metric. The addition of $\Upsilon$ leads to additional modifications of the quadrupole sector. We also found that $\Upsilon$ vanishes in the isotropic case, which is expected, but for general anisotropic cases, it is nonzero.

We then applied the new extended Hartle formalism to compute the rotational perturbations, as well as the integral and surface properties of slowly rotating Bowers-Liang fluid spheres. The Bowers-Liang solution provides a simple closed-form anisotropic analytic model which can be more compact than the Buchdahl limit, thus allowing us to explore the behavior of systems below this bound, especially those that approach the Schwarzschild limit. For this particular BL model, we found that quantities like $\widetilde{Q}$ and $\delta M$ both approach the values associated with the sub-Buchdahl Schwarzschild star~\cite{Beltracchi:2023qla}, the analytic rotating gravastar~\cite{Beltracchi:2022vvn}, and the Hartle-Thorne approximation of the Kerr BH. However, as we approach the Schwarzschild limit, the regularity of the exterior perturbation functions $h_2$ and $k_2$ requires that $Q=J^2/M$, and the regularity of the exterior function $h_0$ requires that $\delta M=J^2/R^3$. Thus, in some sense, \emph{any} solution for which those perturbations remain small for highly compact configurations, and is matched to the vacuum Hartle-Thorne exterior, would need to have similar features.

For highly anisotropic, and compact, Bowers-Liang spheres, we also found that, in the limit when the compactness approaches its critical value, the auxiliary frame dragging function $\varpi\to 0$, and the moment of inertia approaches $MR^2$, which corresponds to the same values as the Hartle-Thorne approximation of the Kerr BH, analytic rotating gravastar, and ultracompact Schwarzschild stars. This ultimately stems from the fact that for all of these systems $R=2M$ and $\omega(R)=2J/R^3=\Omega$, at least to a close approximation. Interestingly, the constant $h_c$ turned out to go to the same limiting value as the sub-Buchdahl Schwarzschild star, but the analytic rotating gravastar and Hartle-Thorne approximation of the Kerr BH do not address the value of $h_c$, and we do not see an argument that requires $h_c$ to behave in a certain way in the Schwarzschild limit. Thus, tests on other systems may be required to elucidate whether this is coincidental or not.

One caveat to these results, regarding highly anisotropic and compact configurations, is that $\TT_0$ and $h_2,\PP_2,\TT_2,$ and $\Upsilon$ become large for configurations sufficiently close to their critical radii; thus, perturbation theory may be unreliable in this limit. However, there are certain configurations, for instance, $\xi=0.99, R/\Rs=1.5$, where the interior perturbation functions are small (so perturbation theory works well) and monotonic (so numeric problems with oscillatory or rapidly varying functions should not be an issue). Even for some cases, very close to divergences, there is little sensitivity to the $\epsilon$ integration parameters over a wide range, and codes written with a variety of schemes (four first-order ODEs, two second-order ODEs, using units of $r/Rs$, and $r/R$, etc.) all give very similar results. In principle, while highly anisotropic compact nonrotating BL spheres will, very closely, mimic a Schwarzschild BH, and the exterior spacetime of a slowly rotating BL sphere, according to this approximation, will approach a Kerr BH, the framework itself may be unreliable for the extremely compact configurations. 

Another situation where perturbation theory may be unreliable is near to the particular pairs of radius $R$ and anisotropy parameter $\xi$ associated with the boundaries of the negative quadrupole moment configurations where there are divergences in $Q$. These, in turn, imply regions where the integration constant $K$, and hence the surface values for $h_2$ or $k_2$, will be large. However, while the perturbative methods may be unreliable near to these divergences, high anisotropy configurations with prolate rather than oblate shapes exist for pairs of $(R,\xi)$ where all the perturbation functions remain small. 

One caveat to these results, regarding highly anisotropic and compact configurations, is that $\TT_0, h_2, \PP_2, \TT_2$ and $\Upsilon$ become large for configurations sufficiently close to their critical radii; thus, perturbation theory may be unreliable in this limit. In principle, a highly anisotropic compact nonrotating BL sphere will, very closely, mimic a Schwarzschild BH. While according to this approximation, the exterior spacetime of a slowly rotating BL sphere will approach a Kerr BH, the perturbative framework itself may be unreliable for the extremely compact configurations.

Another situation where perturbation theory may be unreliable is when we have particular pairs of radius $R$ and anisotropy parameter $\xi$ near to the $D_A, D_B$ boundaries of the negative quadrupole moment configurations where there are divergences in $Q$. These, in turn, imply regions where the integration constant $K$, and, hence, the surface values for $h_2$ or $k_2$, will be large. However, there are many values of $(R, \xi)$ for which all perturbation quantities remain small and the numerics appear to be well behaved, but the configuration is still predicted to be prolate.

We used the assumption that the four-velocity corresponded with a uniform angular velocity to be similar to the formalism in the isotropic case. Ultimately, this assumption of a uniform angular velocity comes from Ref.~\cite{Sharp:1967}, which showed that uniform rotation was the equilibrium configuration for rotating isotropic fluid made of baryons. However, it may be possible that other types of rotation are relevant for anisotropic systems; thus, it may be worthwhile to examine this point in future work.

\section{Acknowledgments}
We thank Paolo Gondolo and John C. Miller for valuable discussions. P.B. particularly thanks Paolo Gondolo for an old discussion about Ref.~\cite{Reina:2014fga} and possible interpretations of the $\delta M$ correction term. C.~P. acknowledges the support of the Research Centre for Theoretical Physics and Astrophysics, Institute of Physics, Silesian University in Opava.
\appendix

\section{Generalized change of mass}
\label{app:dM}
Here we generalize the result \eqref{dmMOD} to systems where anisotropy is present. Using Eqs.~(83) and (85) from Ref.~\cite{Reina:2014fga}, and moving the results into our conventions, we can deduce that
\beq
[h_0']=\frac{[2\nu_0'']}{[2\lambda']}\left[\frac{m_0}{r-2m}\right],
\label{h0p1}
\eeq
where, $[f]=f(R^+)-f(R^-)$ denotes a discontinuity across the surface $r=R$. From the zeroth-order Einstein equations \eqref{EE_stat}, we can obtain
\beq
[2\lambda']=8\pi [\rho]\frac{R^2}{R-2M},
\eeq
\beq
[2\nu_0'']=\frac{16\pi R}{R-2M}[p_\perp] + \frac{R-M}{R(R-2M)}[2\lambda'],
\eeq 
where we used the fact that $p_r$, $\lambda$, $m$, and $\nu'$ are continuous at the boundary $r=R$ and take the values $0$, $-(1/2)\ln(1-2M/R)$, $M$, and $M/R(R-2M)$, respectively. This implies that
\beq
\frac{[2\nu_0'']}{[2\lambda']}=\frac{2}{R}\frac{[p_\perp]}{[\rho]}+\frac{R-M}{R(R-2M)},
\label{vpplpX}
\eeq
and using Eqs.~\eqref{newh0} and \eqref{h0out}, as well as Eq.~\eqref{omegaout} and the fact that $\varpi'$ is continuous across $r=R$, we deduce
\beq
[h_0']=\frac{[m_0]}{(R-2M)^2}-\frac{4\pi R^2 \PP_0}{R-2M}.
\label{h0p2}
\eeq
Using Eqs.~\eqref{h0p1} and \eqref{h0p2} we get
\beq
\left(\frac{1}{R-2M}-\frac{[\nu_0'']}{[\lambda']}\right)[m_0]=4\pi R^2 \PP_0.
\eeq
Employing Eq.~\eqref{vpplpX}, we can solve for $[m_0]$ as
\beq
[m_0]=\frac{4\pi R^2 \PP_0}{\frac{M}{R(R-2M)}-\frac{2[p_\perp]}{R[\rho]}}.
\eeq
This equation is certainly useful, but it is possible to make a substitution which allows for an intuitive picture of the mass correction term. Including possible discontinuities in the generalization of the TOV equation~\eqref{tov}, we can obtain
\beq
[p_\perp]=\frac{1}{2}R[p_r']+\frac{M}{2(R-2M)}[\rho],
\eeq
such that
\begin{align}
[m_0] &=-\frac{4\pi r^2 \PP_0[\rho]}{[p_r']} \nonumber\\
&= -4\pi R^2 \rho(R^-) \frac{\PP_0}{p_r'(R^-)}=4\pi R^2 \rho(R^-) \xi_0,
\label{m0discfinal}
\end{align}
where the second equality holds, because $p_r'=0$ and $\rho=0$ everywhere in the exterior on the grounds that it is empty space, and the third equality uses the appropriate definition for the spherical deformation parameter $\xi_0$. In the isotropic case, authors sometimes substitute $dp/dr$ with $(\rho+p)\nu_{0}'$, such that $\xi_0=-\PP_0/(\rho+p)\nu_{0}'=\delta p_0/\nu_{0}'$ because in isotropic systems the TOV reduces to $p'=-(\rho+p)\nu_{0}'$. We cannot make a corresponding replacement here because the TOV equation for anisotropic systems [Eq.~\eqref{tov}] has an additional term. 

The form of Eq.~\eqref{m0discfinal} can be understood as including the mass of a thin spherical shell with thickness $\xi_0$, radius $R$, and density $\rho(R^-)$. Notice that this formula is correct regardless of whether $[p_\perp]=0$, which must be true for isotropic systems.

\section{Bowers-Liang sphere in the gravastar limit}
\label{app:grav}

The behavior of Bowers-Liang spheres, below their critical radius, has some features in common with the sub-Buchdahl Schwarzschild star~\cite{Mazur:2015kia, Beltracchi:2023qla}. The motivation to consider such configurations stems from the analysis already existing for the isotropic constant density configuration~\cite{Mazur:2015kia} or Schwarzschild star. As shown by Mazur and Mottola (MM-15)~\cite{Mazur:2015kia}, the Schwarzschild star allows compactness beyond the Buchdahl limit once anisotropic stresses are introduced. Moreover, in the limit as $R\to \Rs$, the Schwarzschild star becomes essentially the gravastar of Ref.~\cite{Mazur:2001fv,Mazur:2004fk}. The gravastar of~\cite{Mazur:2015kia} could be considered as the ``universal" limit of the gravastar proposed in~\cite{Mazur:2001fv, Mazur:2004fk} which is endowed with a thin-shell of ultrarelativistic fluid $p=\rho$ near the would-be horizon, surrounding the interior de Sitter region $p=-\rho$; thus, one is able to match the solution with the exterior Schwarzschild spacetime. The ``universality" of the solution can be understood in the sense that, when the thickness of the surrounding shell goes to zero, the resulting solution is independent of any EOS that one may introduce in the shell. Thus, in the MM-15 model, the matching of the interior de Sitter with the exterior Schwarzschild spacetime occurs exactly at the surface $R=\Rs=2M$ (up to possible Planckian corrections). This surface is endowed with anisotropic stresses, which produce a surface tension. 

A crucial element in the MM-15 model construction is that the pole in the radial pressure, for the Schwarzschild star, occurs exactly at the same point $R_0$ where the $g_{tt}$ metric component vanishes. Moreover, as the compactness of the Schwarzschild star increases, this pole moves outward from the origin. In the limit as $R\to \Rs$ (from above), the pole moves up to the surface (from below) where one finally obtains $R=R_0=\Rs$, producing a gravastar in the interior. Motivated by these results, here we examine the behavior of the nonrotating Bowers-Liang spheres when their compactness goes beyond the critical value.

\begin{figure}[ht!]
    \centering
    \includegraphics[width=\linewidth]{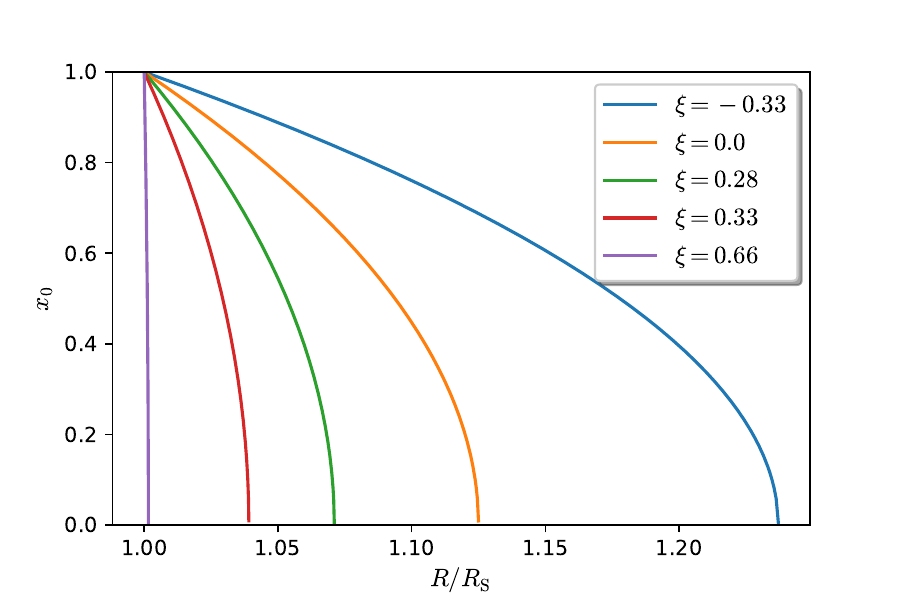}
    \caption{Zeros of the function $D$~\eqref{root}, as a function of $R/\Rs$, for various values of the anisotropy parameter $\xi$. Note that the pole appears first at the origin, and then moves toward 1. As the anisotropy increases, the pole moves to the left, approaching the Schwarzschild radius.}
    \label{fig:poles}
\end{figure}

A first glance at Eqs.~\eqref{prx} and \eqref{gttBLx} reveals some interesting features. We observe that the pressures are regular everywhere in the interior, except at some radius $x_0$ where the term
\beq
D\equiv 3y_1^{q}-y^q
\eeq
vanishes in the range $x\in[0,1]$. Note that this term appears in the denominator of $p_r$ as well as in the numerator of $g_{tt}$. Thus, in strict analogy with the MM-15 model, the pressure diverges at the same value where $g_{tt}=0$. The location of the root $x_0$ of the equation $D=0$, in $x\in[0,1]$, is found to be
\beq
x_0=\sqrt{\beta-3^{1/q}(\beta-1)}.
\label{root}
\eeq 
In the isotropic case $\xi=0$, Eq.~\eqref{root} reduces to $x_0=3\sqrt{1-(8/9)\beta}$, which corresponds to the pole in the isotropic case (see Eq.~(2.22) in~\cite{Mazur:2015kia}). In Fig.~\ref{fig:poles}, we show the various locations of the roots $x_0$, as a function of $R/\Rs$, for different values of the anisotropy $\xi$. It can be seen how the various poles appear first at the origin, and then move toward 1. For the isotropic case $\xi=0$, the pole appears first at the Buchdahl radius $R/\Rs=9/8$. As the anisotropy increases, the pole moves to the left, approaching the Schwarzschild radius. Thus, highly anisotropic configurations can support greater compactness, with respect to the isotropic case. It is worthwhile to mention that finding the pole for the configuration $\xi=0.99$ is rather awkward, because in such case we are in a regime beyond computational tolerances. Thus, the value shown in Table~\ref{tab:crit_radius} comes from a series expansion of Eq.~\eqref{root} in the quantity $\beta-1$.

\begin{figure*}[ht!]
\includegraphics[width=.44\linewidth]{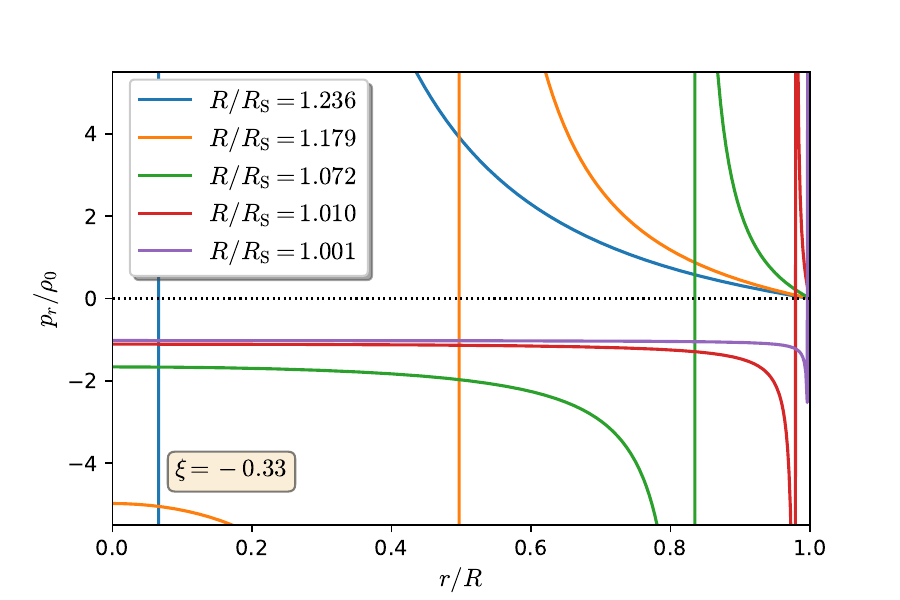}
\includegraphics[width=.44\linewidth]{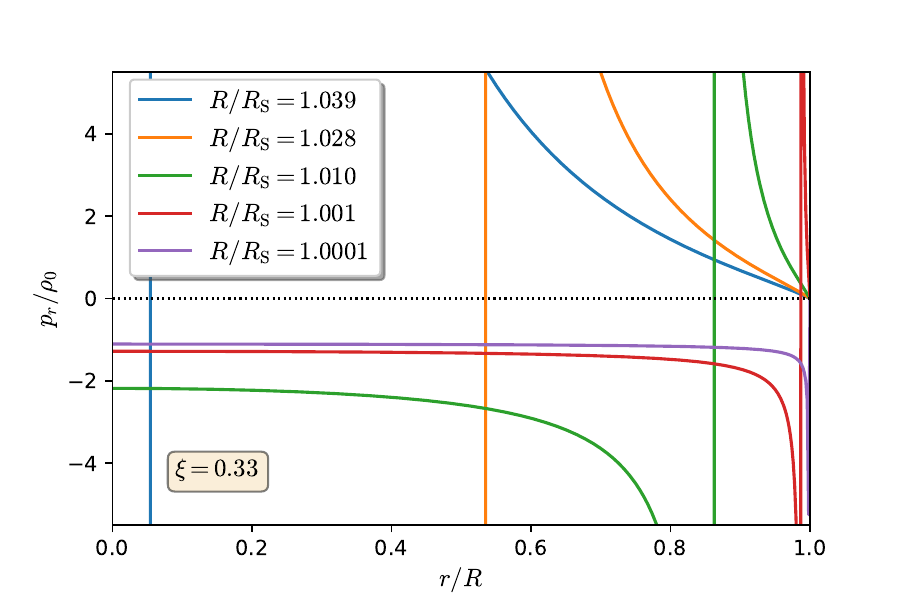}
\includegraphics[width=.44\linewidth]{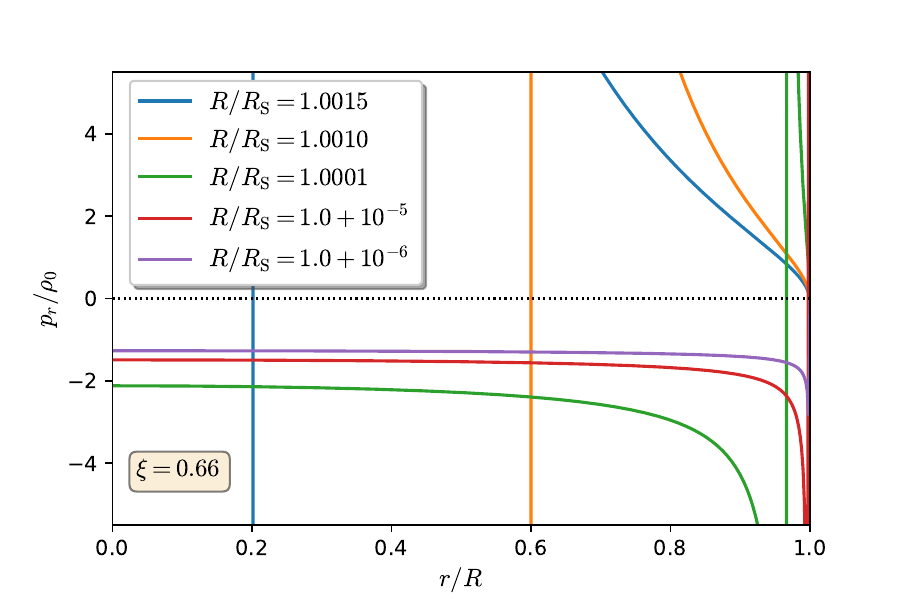}
\includegraphics[width=.44\linewidth]{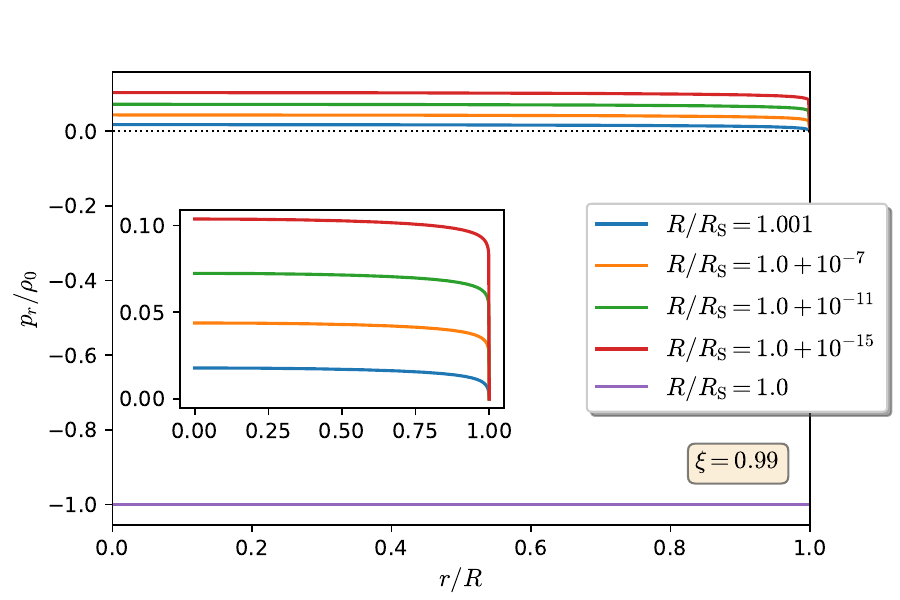}
\caption{Profiles of the radial pressure for the BL model, for various configurations, for different values of the parameter $\xi$. Note the emergence of a negative pressure region, when the compactness goes beyond the corresponding finite pressure limit. In the limit as the compactness approaches the Schwarzschild limit $\beta=1/2$, we observe that $p\to-\rho$, as in the gravastar of Ref.~\cite{Mazur:2015kia}.}
\label{BL_grav}
\end{figure*}

Figure~\ref{BL_grav} shows the radial pressure profiles of nonrotating BL spheres, for various anisotropies, for different values of the compactness below their critical value. We observe that, in strict analogy with the MM-15 model, there emerges a region with \emph{negative} pressure in the interval $0<x<x_0$, where $x_0$ is the pole in the pressure. Meanwhile, the region $x_0<x<1$ remains with positive pressure. As the compactness increases, the pole $x_0$ moves to the right, approaching the surface $x=1$. Interestingly, when the compactness approaches the Schwarzschild limit, the interior of the subcritical Bowers-Liang sphere approaches $p_r=p_\perp=-\rho$ throughout the interior and the pressure divergence moves to the star surface. Thus, in this limit, anisotropic configurations become a constant negative pressure fluid, similar to the gravastar proposed in~\cite{Mazur:2015kia}.

One result of particular interest to Ref.~\cite{Mazur:2015kia} was the $1/4$ time scaling factor for the sub-Buchdahl Schwarzschild star in the gravastar limit. This factor results in equal and opposite surface gravity parameters on either side of the gravastar surface, and it allows the entire manifold to be covered with a Rindler-like coordinate system. However, there is a difficulty in applying the same analysis to Bowers-Liang spheres: $g_{tt}$, at least when written in the form \eqref{gtt_BL}, will be complex inside the negative pressure region for certain values of the anisotropy parameter. Above the critical radius, there is no negative pressure region and $e^{2\nu}$ in the form of Eq.~\eqref{gtt_BL} is real and positive everywhere; therefore, it is perfectly sufficient for the calculations in the main body of the paper.

We can specifically look at $e^{2\nu}$ for $r=0,R=2M$ as an example, which using the form of Eq.~\eqref{gtt_BL} becomes
\beq
e^{2\nu_0}(r=0, R=2M)=\left(-\frac{1}{2}\right)^{1/q}.
\eeq
This is fully real only if $q=1/N$ and is real and positive (which allows for the conventional statement of $t$ being a timelike coordinate) only if $q=1/(2N)$, with $N\in Z$. Notice that the isotropic case corresponds to $q=1/2$, which is one of the values for which $e^{2\nu}$ is positive and real (specifically being $1/4$). However, it is not the only such value of $q$. For systems with negative pressure regions, it may be possible to redefine $g_{tt}$ in such a way as to remove an inappropriate complex phase factor inside the negative pressure region. This question is left open for future investigation.

\begin{figure*}
    \centering
    \includegraphics[width=0.3295\linewidth]{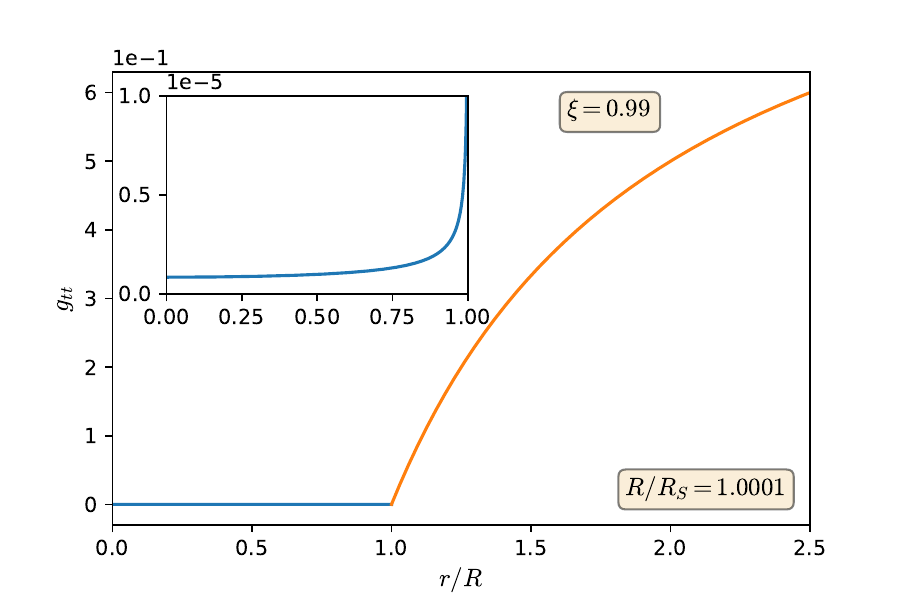}
    \includegraphics[width=0.3295\linewidth]{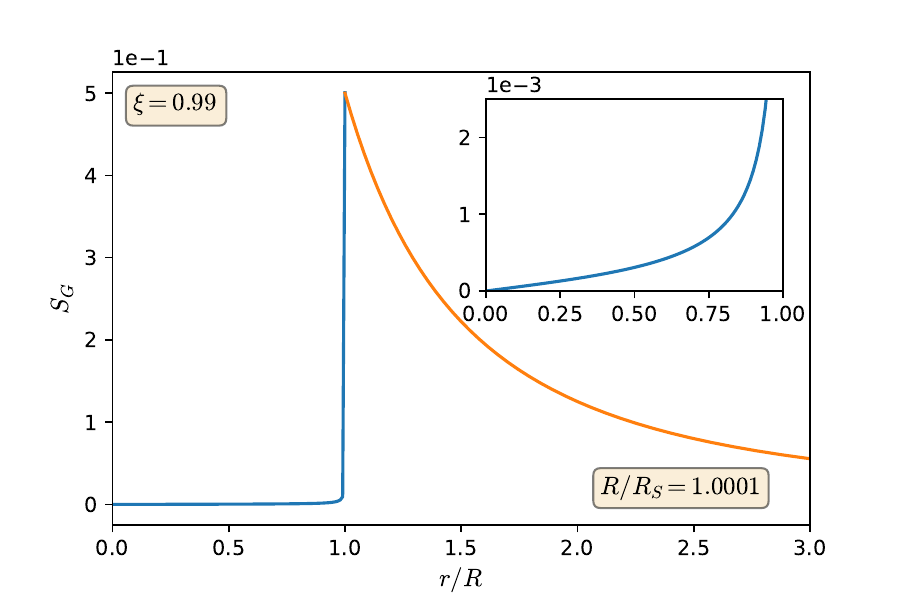}
    \includegraphics[width=0.3295\linewidth]{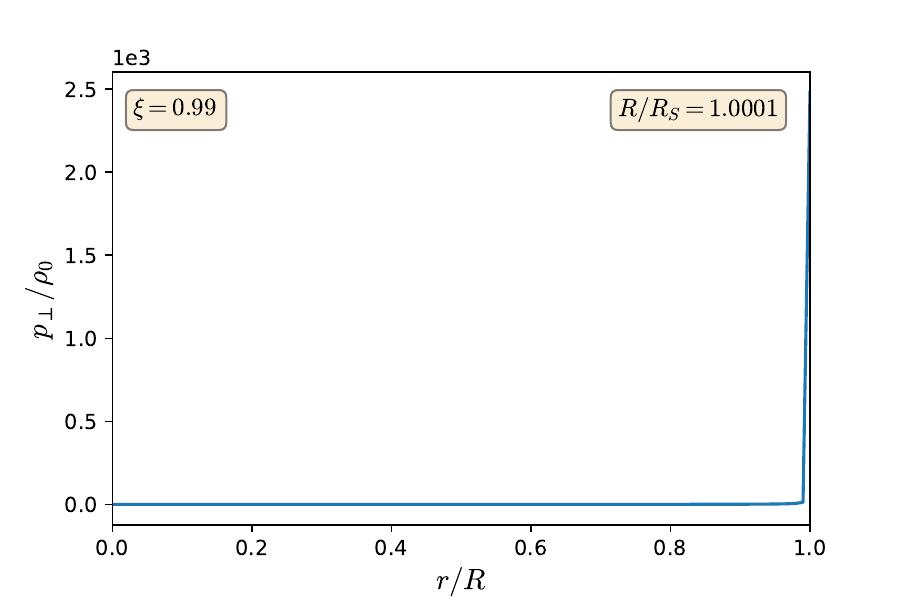}
    \caption{{\bf Left panel}: radial profile of the interior metric function $-g_{tt}$ (blue curve), and the exterior Schwarzschild metric (orange curve), for a highly anisotropic Bowers-Liang sphere, with $\xi=0.99$ and $R/\Rs=1.0001$. Observe how $g_{tt}$ is nearly zero throughout the interior of the configuration, similar to the quasi-black hole depicted in~\cite{Lemos:2007yh}. {\bf Middle panel:} interior surface gravity parameter~\eqref{surf_grav} (blue curve), and exterior surface gravity (orange curve), for a Bowers-Liang sphere, with the same $(\xi,R/\Rs)$ parameters as the left panel. {\bf Right panel:} radial profile of the transverse pressure, in units of the central energy density, for the same values of the parameters ($\xi,R/\Rs$) as in the left and middle panels. Note the approximate delta-function behavior of $p_{\perp}$ near the surface for BL spheres with these parameters.}
    \label{qbh}
\end{figure*}

It is noteworthy that, for highly anisotropic BL spheres, as they approach their critical radius, but stay formally above it, the configuration shows some of the features of a “quasi-black hole" as defined in \cite{Lemos:2007yh}. In particular, the metric function $g_{tt}$ is very close to $0$ throughout the interior of a high anisotropy BL sphere approaching the critical radius from above (see the left panel in Fig.~\ref{qbh}). However, the surface gravity
\beq
\label{surf_grav}
\frac{1}{2}e^{-\nu-\lambda}\frac{\partial e^{2 \nu}}{\partial r}
\eeq
(given in terms of the metric functions in \eqref{metric_stat}), as we approach the outside, is nonzero (see the middle panel in Fig.~\ref{qbh}). Thus, the quasihorizon is nonextremal, which is because the transverse pressure approximates a delta-function surface layer [see right panel in Fig.~\ref{qbh}, and case (2) from Section IV in \cite{Lemos:2007yh}].

\newpage
\section{Integral and surface properties of slowly rotating anisotropic homogeneous configurations in general relativity}
\label{app:table}
\begin{table*}[ht]
\begin{centering}
\begin{tabular}{c | ccccc | c | ccccc | c | ccccc}
\hline
\hline\noalign{\smallskip}
$\xi$ & \multicolumn{5}{ c| }{-0.66} & \multicolumn{6}{ c| }{-0.33} & \multicolumn{6}{ c| }{0.0}\\
\hline
\hline\noalign{\smallskip}
$R/\Rs$ & $\widetilde{\varpi}$ & $\bar{I}$ & $\widetilde{\delta M}$ & $\widetilde{Q}$ & $\epsilon$ & 
$R/\Rs$ & $\widetilde{\varpi}$ & $\bar{I}$ & $\widetilde{\delta M}$ & $\widetilde{Q}$ & $\epsilon$ & 
$R/\Rs$ & $\widetilde{\varpi}$ & $\bar{I}$ & $\widetilde{\delta M}$ & $\widetilde{Q}$ & $\epsilon$\\
\hline\noalign{\smallskip}
1.363 & 1.0770 & 0.5766 & 2.7800 & 1.3265 & 6.4577 & 
1.238 & 0.8792 & 0.6749 & 2.4552 & 1.1096 & 5.9590 & 
1.125 & 0.5727 & 0.7991 & 2.1783 & 1.0212 & 5.8035 \\
1.459 & 1.0422 & 0.5571 & 3.6968 & 1.4635 & 6.4767 &
1.327 & 0.9457 & 0.6303 & 3.2664 & 1.2207 & 6.5334 & 
1.260 & 0.8257 & 0.6900 & 3.1950 & 1.1596 & 7.4909 \\
1.527 & 1.0092 & 0.5460 & 4.1605 & 1.5864 & 6.5943 &
1.459 & 0.9507 & 0.5891 & 4.2297 & 1.4450 & 7.2950 & 
1.369 & 0.8864 & 0.6405 & 4.0243 & 0.8864 & 1.3438 \\
1.630 & 0.9536 & 0.5318 & 4.7475 & 1.7995 & 6.7716 &
1.545 & 0.9273 & 0.5700 & 4.7614 & 1.6240 & 7.6950 & 
1.428 & 0.8933 & 0.6206 & 4.4393 & 1.4665 & 9.2939 \\
1.749 & 0.8870 & 0.5185 & 5.3005 & 2.0792 & 6.9367 &
1.666 & 0.8792 & 0.5493 & 5.3951 & 1.9132 & 8.1344 & 
1.549 & 0.8766 & 0.5891 & 5.1948 & 1.7623 & 10.245 \\
1.818 & 0.8489 & 0.5120 & 5.5708 & 2.2555 & 7.0092 &
1.714 & 0.8577 & 0.5424 & 5.6125 & 2.0388 & 8.2730 & 
1.639 & 0.8490 & 0.5712 & 5.6711 & 2.0163 & 10.788 \\
1.900 & 0.8051 & 0.5051 & 5.8506 & 2.4769 & 7.0742 &
1.800 & 0.8178 & 0.5317 & 5.9579 & 2.2776 & 8.4771 & 
1.700 & 0.8265 & 0.5609 & 5.9543 & 2.2030 & 11.088 \\
\hline
\end{tabular}
\begin{tabular}{c | ccccc | c | ccccc | c | ccccc}
\noalign{\smallskip}
\hline\noalign{\smallskip}
$\xi$ & \multicolumn{5}{ c| }{0.33} & \multicolumn{6}{ c| }{0.66} & \multicolumn{6}{ c| }{0.99}\\
\hline
\hline\noalign{\smallskip}
$R/\Rs$ & $\widetilde{\varpi}$ & $\bar{I}$ & $\widetilde{\delta M}$ & $\widetilde{Q}$ & $\epsilon$ & 
$R/\Rs$ & $\widetilde{\varpi}$ & $\bar{I}$ & $\widetilde{\delta M}$ & $\widetilde{Q}$ & $\epsilon$ & 
$R/\Rs$ & $\widetilde{\varpi}$ & $\bar{I}$ & $\widetilde{\delta M}$ & $\widetilde{Q}$ & $\epsilon$\\
\noalign{\smallskip}\hline\noalign{\smallskip}
1.040 & 0.2185 & 0.9261 & 2.0335 & 1.0024 & 6.2866 & 
1.002 & 0.0119 & 0.9960 & 2.0002 & 1.0144 & 9.6662 & 
1.0001 & 0.0006 & 0.9998 & 2.0000 & 0.9784 & -2.2753 \\
1.125 & 0.5402 & 0.8124 & 2.4708 & 1.0665 & 8.5188 & 
1.019 & 0.1079 & 0.9639 & 2.0270 & 1.0596 & 14.771 & 
1.001 & 0.0059 & 0.9980 & 2.0006 & 0.9657 & -3.6196 \\
1.234 & 0.7473 & 0.7249 & 3.2960 & 1.2400 & 11.406 & 
1.125 & 0.5144 & 0.8234 & 2.6279 & 1.3908 & 27.528 & 
1.010 & 0.0576 & 0.9808 & 2.0193 & 0.9221 & -8.1027 \\
1.325 & 0.8224 & 0.6771 & 4.0056 & 1.4506 & 13.423 & 
1.280 & 0.7576 & 0.7133 & 3.8345 & 2.1904 & 39.336 & 
1.125 & 0.4951 & 0.8318 & 2.7137 & 0.3360 & -41.485 \\
1.416 & 0.8488 & 0.6421 & 4.6663 & 1.7128 & 15.063 & 
1.340 & 0.7944 & 0.6851 & 4.2894 & 2.5829 & 42.550 & 
1.234 & 0.6863 & 0.7502 & 3.5440 & -0.5552 & -65.498 \\
1.550 & 0.8396 & 0.6046 & 5.5070 & 2.1794 & 16.885 & 
1.459 & 0.8187 & 0.6423 & 5.1085 & 3.4746 & 47.336 & 
1.369 & 0.7786 & 0.6848 & 4.5138 & -2.000 & -85.630 \\
1.600 & 0.8279 & 0.5935 & 5.7779 & 2.3751 & 17.413 & 
1.500 & 0.8172 & 0.6304 & 5.3611 & 3.8127 & 48.590 & 
1.400 & 0.7868 & 0.6732 & 4.7170 & -2.3680 & -88.852 \\
\hline
\hline
\end{tabular}
\end{centering}
\caption{Surface physical properties of slowly rotating anisotropic Bowers-Liang spheres, for some selective values of the anisotropy parameter $\xi$, for various values of the ratio $R/\Rs$. The full results are displayed in the accompanying figures. We employ the same units as introduced by Refs.~\cite{Chandra:1974, Beltracchi:2023qla}, and we include here the corresponding factors of $c$ and $G$ in case one wishes to recover the physical parameters: the angular velocity $\widetilde{\varpi}\equiv \varpi/(GJ/c^2 \Rs^3)$; the normalized moment of inertia $\bar{I}\equiv I/MR^2$ is dimensionless; the change in mass $\widetilde{\delta M}\equiv \delta M/M$ is measured in the unit $(GJ/c^3 \Rs^2)^2$; the Kerr factor $\widetilde{Q}\equiv QM/J^2$ is dimensionless; the ellipticity $\epsilon$ is measured in the unit $(GJ/c^3 \Rs^2)^2$.}
\label{tab:surf}  
\end{table*}
\bibliography{ref}

\begin{thebibliography}{44}%
\makeatletter
\providecommand \@ifxundefined [1]{%
 \@ifx{#1\undefined}
}%
\providecommand \@ifnum [1]{%
 \ifnum #1\expandafter \@firstoftwo
 \else \expandafter \@secondoftwo
 \fi
}%
\providecommand \@ifx [1]{%
 \ifx #1\expandafter \@firstoftwo
 \else \expandafter \@secondoftwo
 \fi
}%
\providecommand \natexlab [1]{#1}%
\providecommand \enquote  [1]{``#1''}%
\providecommand \bibnamefont  [1]{#1}%
\providecommand \bibfnamefont [1]{#1}%
\providecommand \citenamefont [1]{#1}%
\providecommand \href@noop [0]{\@secondoftwo}%
\providecommand \href [0]{\begingroup \@sanitize@url \@href}%
\providecommand \@href[1]{\@@startlink{#1}\@@href}%
\providecommand \@@href[1]{\endgroup#1\@@endlink}%
\providecommand \@sanitize@url [0]{\catcode `\\12\catcode `\$12\catcode
  `\&12\catcode `\#12\catcode `\^12\catcode `\_12\catcode `\%12\relax}%
\providecommand \@@startlink[1]{}%
\providecommand \@@endlink[0]{}%
\providecommand \url  [0]{\begingroup\@sanitize@url \@url }%
\providecommand \@url [1]{\endgroup\@href {#1}{\urlprefix }}%
\providecommand \urlprefix  [0]{URL }%
\providecommand \Eprint [0]{\href }%
\providecommand \doibase [0]{http://dx.doi.org/}%
\providecommand \selectlanguage [0]{\@gobble}%
\providecommand \bibinfo  [0]{\@secondoftwo}%
\providecommand \bibfield  [0]{\@secondoftwo}%
\providecommand \translation [1]{[#1]}%
\providecommand \BibitemOpen [0]{}%
\providecommand \bibitemStop [0]{}%
\providecommand \bibitemNoStop [0]{.\EOS\space}%
\providecommand \EOS [0]{\spacefactor3000\relax}%
\providecommand \BibitemShut  [1]{\csname bibitem#1\endcsname}%
\let\auto@bib@innerbib\@empty
\bibitem [{\citenamefont {Buchdahl}(1959)}]{Buchdahl:1959zz}%
  \BibitemOpen
  \bibfield  {author} {\bibinfo {author} {\bibfnamefont {H.~A.}\ \bibnamefont
  {Buchdahl}},\ }\href {\doibase 10.1103/PhysRev.116.1027} {\bibfield
  {journal} {\bibinfo  {journal} {Phys. Rev.}\ }\textbf {\bibinfo {volume}
  {116}},\ \bibinfo {pages} {1027} (\bibinfo {year} {1959})}\BibitemShut
  {NoStop}%
\bibitem [{\citenamefont {Ruderman}(1972)}]{Ruderman:1972}%
  \BibitemOpen
  \bibfield  {author} {\bibinfo {author} {\bibfnamefont {M.}~\bibnamefont
  {Ruderman}},\ }\href {\doibase 10.1146/annurev.aa.10.090172.002235}
  {\bibfield  {journal} {\bibinfo  {journal} {Annu. Rev. Astron. Astrophys.}\
  }\textbf {\bibinfo {volume} {10}},\ \bibinfo {pages} {427} (\bibinfo {year}
  {1972})}\BibitemShut {NoStop}%
\bibitem [{\citenamefont {Kippenhahn}\ \emph {et~al.}(2012)\citenamefont
  {Kippenhahn}, \citenamefont {Weigert},\ and\ \citenamefont
  {Weiss}}]{Kippenhahn:2012qhp}%
  \BibitemOpen
  \bibfield  {author} {\bibinfo {author} {\bibfnamefont {R.}~\bibnamefont
  {Kippenhahn}}, \bibinfo {author} {\bibfnamefont {A.}~\bibnamefont {Weigert}},
  \ and\ \bibinfo {author} {\bibfnamefont {A.}~\bibnamefont {Weiss}},\ }\href
  {\doibase 10.1007/978-3-642-30304-3} {\emph {\bibinfo {title} {{Stellar
  structure and evolution}}}},\ Astronomy and Astrophysics Library\ (\bibinfo
  {publisher} {Springer},\ \bibinfo {address} {Berlin},\ \bibinfo {year}
  {2012})\BibitemShut {NoStop}%
\bibitem [{\citenamefont {Sawyer}(1972)}]{Sawyer:1972cq}%
  \BibitemOpen
  \bibfield  {author} {\bibinfo {author} {\bibfnamefont {R.~F.}\ \bibnamefont
  {Sawyer}},\ }\href {\doibase 10.1103/PhysRevLett.29.382} {\bibfield
  {journal} {\bibinfo  {journal} {Phys. Rev. Lett.}\ }\textbf {\bibinfo
  {volume} {29}},\ \bibinfo {pages} {382} (\bibinfo {year} {1972})}\BibitemShut
  {NoStop}%
\bibitem [{\citenamefont {Liebling}\ and\ \citenamefont
  {Palenzuela}(2023)}]{Liebling:2012fv}%
  \BibitemOpen
  \bibfield  {author} {\bibinfo {author} {\bibfnamefont {S.~L.}\ \bibnamefont
  {Liebling}}\ and\ \bibinfo {author} {\bibfnamefont {C.}~\bibnamefont
  {Palenzuela}},\ }\href {\doibase 10.1007/s41114-023-00043-4} {\bibfield
  {journal} {\bibinfo  {journal} {Living Rev. Rel.}\ }\textbf {\bibinfo
  {volume} {26}},\ \bibinfo {pages} {1} (\bibinfo {year} {2023})},\ \Eprint
  {http://arxiv.org/abs/1202.5809} {arXiv:1202.5809 [gr-qc]} \BibitemShut
  {NoStop}%
\bibitem [{\citenamefont {Yazadjiev}(2012)}]{Yazadjiev:2011ks}%
  \BibitemOpen
  \bibfield  {author} {\bibinfo {author} {\bibfnamefont {S.}~\bibnamefont
  {Yazadjiev}},\ }\href {\doibase 10.1103/PhysRevD.85.044030} {\bibfield
  {journal} {\bibinfo  {journal} {Phys. Rev. D}\ }\textbf {\bibinfo {volume}
  {85}},\ \bibinfo {pages} {044030} (\bibinfo {year} {2012})},\ \Eprint
  {http://arxiv.org/abs/1111.3536} {arXiv:1111.3536 [gr-qc]} \BibitemShut
  {NoStop}%
\bibitem [{\citenamefont {Folomeev}\ and\ \citenamefont
  {Dzhunushaliev}(2015)}]{Folomeev:2015aua}%
  \BibitemOpen
  \bibfield  {author} {\bibinfo {author} {\bibfnamefont {V.}~\bibnamefont
  {Folomeev}}\ and\ \bibinfo {author} {\bibfnamefont {V.}~\bibnamefont
  {Dzhunushaliev}},\ }\href {\doibase 10.1103/PhysRevD.91.044040} {\bibfield
  {journal} {\bibinfo  {journal} {Phys. Rev. D}\ }\textbf {\bibinfo {volume}
  {91}},\ \bibinfo {pages} {044040} (\bibinfo {year} {2015})},\ \Eprint
  {http://arxiv.org/abs/1501.06275} {arXiv:1501.06275 [gr-qc]} \BibitemShut
  {NoStop}%
\bibitem [{\citenamefont {Herrera}\ and\ \citenamefont
  {Santos}(1997)}]{Herrera:1997plx}%
  \BibitemOpen
  \bibfield  {author} {\bibinfo {author} {\bibfnamefont {L.}~\bibnamefont
  {Herrera}}\ and\ \bibinfo {author} {\bibfnamefont {N.~O.}\ \bibnamefont
  {Santos}},\ }\href {\doibase 10.1016/S0370-1573(96)00042-7} {\bibfield
  {journal} {\bibinfo  {journal} {Phys. Rept.}\ }\textbf {\bibinfo {volume}
  {286}},\ \bibinfo {pages} {53} (\bibinfo {year} {1997})}\BibitemShut
  {NoStop}%
\bibitem [{\citenamefont {Lemaitre}(1933)}]{Lemaitre:1933gd}%
  \BibitemOpen
  \bibfield  {author} {\bibinfo {author} {\bibfnamefont {G.}~\bibnamefont
  {Lemaitre}},\ }\href {\doibase 10.1023/A:1018855621348} {\bibfield  {journal}
  {\bibinfo  {journal} {Annales Soc. Sci. Bruxelles A}\ }\textbf {\bibinfo
  {volume} {53}},\ \bibinfo {pages} {51} (\bibinfo {year} {1933})}\BibitemShut
  {NoStop}%
\bibitem [{\citenamefont {Bowers}\ and\ \citenamefont
  {Liang}(1974)}]{Bowers:1974tgi}%
  \BibitemOpen
  \bibfield  {author} {\bibinfo {author} {\bibfnamefont {R.~L.}\ \bibnamefont
  {Bowers}}\ and\ \bibinfo {author} {\bibfnamefont {E.~P.~T.}\ \bibnamefont
  {Liang}},\ }\href {\doibase 10.1086/152760} {\bibfield  {journal} {\bibinfo
  {journal} {Astrophys. J.}\ }\textbf {\bibinfo {volume} {188}},\ \bibinfo
  {pages} {657} (\bibinfo {year} {1974})}\BibitemShut {NoStop}%
\bibitem [{\citenamefont {Bayin}(1982)}]{Bayin:1982vw}%
  \BibitemOpen
  \bibfield  {author} {\bibinfo {author} {\bibfnamefont {S.~S.}\ \bibnamefont
  {Bayin}},\ }\href {\doibase 10.1103/PhysRevD.26.1262} {\bibfield  {journal}
  {\bibinfo  {journal} {Phys. Rev. D}\ }\textbf {\bibinfo {volume} {26}},\
  \bibinfo {pages} {1262} (\bibinfo {year} {1982})}\BibitemShut {NoStop}%
\bibitem [{\citenamefont {Mak}\ and\ \citenamefont {Harko}(2003)}]{Mak:2001eb}%
  \BibitemOpen
  \bibfield  {author} {\bibinfo {author} {\bibfnamefont {M.~K.}\ \bibnamefont
  {Mak}}\ and\ \bibinfo {author} {\bibfnamefont {T.}~\bibnamefont {Harko}},\
  }\href {\doibase 10.1098/rspa.2002.1014} {\bibfield  {journal} {\bibinfo
  {journal} {Proc. Roy. Soc. Lond. A}\ }\textbf {\bibinfo {volume} {459}},\
  \bibinfo {pages} {393} (\bibinfo {year} {2003})},\ \Eprint
  {http://arxiv.org/abs/gr-qc/0110103} {arXiv:gr-qc/0110103} \BibitemShut
  {NoStop}%
\bibitem [{\citenamefont {Dev}\ and\ \citenamefont
  {Gleiser}(2002)}]{Dev:2000gt}%
  \BibitemOpen
  \bibfield  {author} {\bibinfo {author} {\bibfnamefont {K.}~\bibnamefont
  {Dev}}\ and\ \bibinfo {author} {\bibfnamefont {M.}~\bibnamefont {Gleiser}},\
  }\href {\doibase 10.1023/A:1020707906543} {\bibfield  {journal} {\bibinfo
  {journal} {Gen. Rel. Grav.}\ }\textbf {\bibinfo {volume} {34}},\ \bibinfo
  {pages} {1793} (\bibinfo {year} {2002})},\ \Eprint
  {http://arxiv.org/abs/astro-ph/0012265} {arXiv:astro-ph/0012265} \BibitemShut
  {NoStop}%
\bibitem [{\citenamefont {Herrera}\ \emph {et~al.}(2004)\citenamefont
  {Herrera}, \citenamefont {Di~Prisco}, \citenamefont {Martin}, \citenamefont
  {Ospino}, \citenamefont {Santos},\ and\ \citenamefont
  {Troconis}}]{Herrera:2004}%
  \BibitemOpen
  \bibfield  {author} {\bibinfo {author} {\bibfnamefont {L.}~\bibnamefont
  {Herrera}}, \bibinfo {author} {\bibfnamefont {A.}~\bibnamefont {Di~Prisco}},
  \bibinfo {author} {\bibfnamefont {J.}~\bibnamefont {Martin}}, \bibinfo
  {author} {\bibfnamefont {J.}~\bibnamefont {Ospino}}, \bibinfo {author}
  {\bibfnamefont {N.~O.}\ \bibnamefont {Santos}}, \ and\ \bibinfo {author}
  {\bibfnamefont {O.}~\bibnamefont {Troconis}},\ }\href {\doibase
  10.1103/PhysRevD.69.084026} {\bibfield  {journal} {\bibinfo  {journal} {Phys.
  Rev. D}\ }\textbf {\bibinfo {volume} {69}},\ \bibinfo {pages} {084026}
  (\bibinfo {year} {2004})}\BibitemShut {NoStop}%
\bibitem [{\citenamefont {{Hillebrandt}}\ and\ \citenamefont
  {{Steinmetz}}(1976)}]{Hillebrandt:1976}%
  \BibitemOpen
  \bibfield  {author} {\bibinfo {author} {\bibfnamefont {W.}~\bibnamefont
  {{Hillebrandt}}}\ and\ \bibinfo {author} {\bibfnamefont {K.~O.}\ \bibnamefont
  {{Steinmetz}}},\ }\href@noop {} {\bibfield  {journal} {\bibinfo  {journal}
  {Astron. Astrophys.}\ }\textbf {\bibinfo {volume} {53}},\ \bibinfo {pages}
  {283} (\bibinfo {year} {1976})}\BibitemShut {NoStop}%
\bibitem [{\citenamefont {Dev}\ and\ \citenamefont
  {Gleiser}(2003)}]{Dev:2003qd}%
  \BibitemOpen
  \bibfield  {author} {\bibinfo {author} {\bibfnamefont {K.}~\bibnamefont
  {Dev}}\ and\ \bibinfo {author} {\bibfnamefont {M.}~\bibnamefont {Gleiser}},\
  }\href {\doibase 10.1023/A:1024534702166} {\bibfield  {journal} {\bibinfo
  {journal} {Gen. Rel. Grav.}\ }\textbf {\bibinfo {volume} {35}},\ \bibinfo
  {pages} {1435} (\bibinfo {year} {2003})},\ \Eprint
  {http://arxiv.org/abs/gr-qc/0303077} {arXiv:gr-qc/0303077} \BibitemShut
  {NoStop}%
\bibitem [{\citenamefont {Doneva}\ and\ \citenamefont
  {Yazadjiev}(2012)}]{Doneva:2012rd}%
  \BibitemOpen
  \bibfield  {author} {\bibinfo {author} {\bibfnamefont {D.~D.}\ \bibnamefont
  {Doneva}}\ and\ \bibinfo {author} {\bibfnamefont {S.~S.}\ \bibnamefont
  {Yazadjiev}},\ }\href {\doibase 10.1103/PhysRevD.85.124023} {\bibfield
  {journal} {\bibinfo  {journal} {Phys. Rev. D}\ }\textbf {\bibinfo {volume}
  {85}},\ \bibinfo {pages} {124023} (\bibinfo {year} {2012})},\ \Eprint
  {http://arxiv.org/abs/1203.3963} {arXiv:1203.3963 [gr-qc]} \BibitemShut
  {NoStop}%
\bibitem [{\citenamefont {Yagi}\ and\ \citenamefont
  {Yunes}(2015)}]{Yagi:2015hda}%
  \BibitemOpen
  \bibfield  {author} {\bibinfo {author} {\bibfnamefont {K.}~\bibnamefont
  {Yagi}}\ and\ \bibinfo {author} {\bibfnamefont {N.}~\bibnamefont {Yunes}},\
  }\href {\doibase 10.1103/PhysRevD.91.123008} {\bibfield  {journal} {\bibinfo
  {journal} {Phys. Rev. D}\ }\textbf {\bibinfo {volume} {91}},\ \bibinfo
  {pages} {123008} (\bibinfo {year} {2015})},\ \Eprint
  {http://arxiv.org/abs/1503.02726} {arXiv:1503.02726 [gr-qc]} \BibitemShut
  {NoStop}%
\bibitem [{\citenamefont {Carloni}\ and\ \citenamefont
  {Vernieri}(2018)}]{Carloni:2017bck}%
  \BibitemOpen
  \bibfield  {author} {\bibinfo {author} {\bibfnamefont {S.}~\bibnamefont
  {Carloni}}\ and\ \bibinfo {author} {\bibfnamefont {D.}~\bibnamefont
  {Vernieri}},\ }\href {\doibase 10.1103/PhysRevD.97.124057} {\bibfield
  {journal} {\bibinfo  {journal} {Phys. Rev. D}\ }\textbf {\bibinfo {volume}
  {97}},\ \bibinfo {pages} {124057} (\bibinfo {year} {2018})},\ \Eprint
  {http://arxiv.org/abs/1709.03996} {arXiv:1709.03996 [gr-qc]} \BibitemShut
  {NoStop}%
\bibitem [{\citenamefont {Raposo}\ \emph {et~al.}(2019)\citenamefont {Raposo},
  \citenamefont {Pani}, \citenamefont {Bezares}, \citenamefont {Palenzuela},\
  and\ \citenamefont {Cardoso}}]{Raposo:2018rjn}%
  \BibitemOpen
  \bibfield  {author} {\bibinfo {author} {\bibfnamefont {G.}~\bibnamefont
  {Raposo}}, \bibinfo {author} {\bibfnamefont {P.}~\bibnamefont {Pani}},
  \bibinfo {author} {\bibfnamefont {M.}~\bibnamefont {Bezares}}, \bibinfo
  {author} {\bibfnamefont {C.}~\bibnamefont {Palenzuela}}, \ and\ \bibinfo
  {author} {\bibfnamefont {V.}~\bibnamefont {Cardoso}},\ }\href {\doibase
  10.1103/PhysRevD.99.104072} {\bibfield  {journal} {\bibinfo  {journal} {Phys.
  Rev. D}\ }\textbf {\bibinfo {volume} {99}},\ \bibinfo {pages} {104072}
  (\bibinfo {year} {2019})},\ \Eprint {http://arxiv.org/abs/1811.07917}
  {arXiv:1811.07917 [gr-qc]} \BibitemShut {NoStop}%
\bibitem [{\citenamefont {Ovalle}\ \emph {et~al.}(2019)\citenamefont {Ovalle},
  \citenamefont {Posada},\ and\ \citenamefont
  {Stuchl\'\i{}k}}]{Ovalle:2019lbs}%
  \BibitemOpen
  \bibfield  {author} {\bibinfo {author} {\bibfnamefont {J.}~\bibnamefont
  {Ovalle}}, \bibinfo {author} {\bibfnamefont {C.}~\bibnamefont {Posada}}, \
  and\ \bibinfo {author} {\bibfnamefont {Z.}~\bibnamefont {Stuchl\'\i{}k}},\
  }\href {\doibase 10.1088/1361-6382/ab4461} {\bibfield  {journal} {\bibinfo
  {journal} {Class. Quantum Grav.}\ }\textbf {\bibinfo {volume} {36}},\
  \bibinfo {pages} {205010} (\bibinfo {year} {2019})},\ \Eprint
  {http://arxiv.org/abs/1905.12452} {arXiv:1905.12452 [gr-qc]} \BibitemShut
  {NoStop}%
\bibitem [{\citenamefont {Pretel}(2020)}]{Pretel:2020xuo}%
  \BibitemOpen
  \bibfield  {author} {\bibinfo {author} {\bibfnamefont {J.~M.~Z.}\
  \bibnamefont {Pretel}},\ }\href {\doibase 10.1140/epjc/s10052-020-8301-3}
  {\bibfield  {journal} {\bibinfo  {journal} {Eur. Phys. J. C}\ }\textbf
  {\bibinfo {volume} {80}},\ \bibinfo {pages} {726} (\bibinfo {year} {2020})},\
  \Eprint {http://arxiv.org/abs/2008.05331} {arXiv:2008.05331 [gr-qc]}
  \BibitemShut {NoStop}%
\bibitem [{\citenamefont {Pretel}\ and\ \citenamefont
  {Duarte}(2022)}]{Pretel:2022plg}%
  \BibitemOpen
  \bibfield  {author} {\bibinfo {author} {\bibfnamefont {J.~M.~Z.}\
  \bibnamefont {Pretel}}\ and\ \bibinfo {author} {\bibfnamefont {S.~B.}\
  \bibnamefont {Duarte}},\ }\href {\doibase 10.1088/1361-6382/ac7a88}
  {\bibfield  {journal} {\bibinfo  {journal} {Class. Quant. Grav.}\ }\textbf
  {\bibinfo {volume} {39}},\ \bibinfo {pages} {155003} (\bibinfo {year}
  {2022})},\ \Eprint {http://arxiv.org/abs/2202.04467} {arXiv:2202.04467
  [gr-qc]} \BibitemShut {NoStop}%
\bibitem [{\citenamefont {Silva}\ \emph {et~al.}(2015)\citenamefont {Silva},
  \citenamefont {Macedo}, \citenamefont {Berti},\ and\ \citenamefont
  {Crispino}}]{Silva:2014fca}%
  \BibitemOpen
  \bibfield  {author} {\bibinfo {author} {\bibfnamefont {H.~O.}\ \bibnamefont
  {Silva}}, \bibinfo {author} {\bibfnamefont {C.~F.~B.}\ \bibnamefont
  {Macedo}}, \bibinfo {author} {\bibfnamefont {E.}~\bibnamefont {Berti}}, \
  and\ \bibinfo {author} {\bibfnamefont {L.~C.~B.}\ \bibnamefont {Crispino}},\
  }\href {\doibase 10.1088/0264-9381/32/14/145008} {\bibfield  {journal}
  {\bibinfo  {journal} {Class. Quantum Grav.}\ }\textbf {\bibinfo {volume}
  {32}},\ \bibinfo {pages} {145008} (\bibinfo {year} {2015})},\ \Eprint
  {http://arxiv.org/abs/1411.6286} {arXiv:1411.6286 [gr-qc]} \BibitemShut
  {NoStop}%
\bibitem [{\citenamefont {Hartle}(1967)}]{Hartle:1967he}%
  \BibitemOpen
  \bibfield  {author} {\bibinfo {author} {\bibfnamefont {J.~B.}\ \bibnamefont
  {Hartle}},\ }\href {\doibase 10.1086/149400} {\bibfield  {journal} {\bibinfo
  {journal} {Astrophys. J.}\ }\textbf {\bibinfo {volume} {150}},\ \bibinfo
  {pages} {1005} (\bibinfo {year} {1967})}\BibitemShut {NoStop}%
\bibitem [{\citenamefont {Hartle}\ and\ \citenamefont
  {Thorne}(1968)}]{Hartle:1968si}%
  \BibitemOpen
  \bibfield  {author} {\bibinfo {author} {\bibfnamefont {J.~B.}\ \bibnamefont
  {Hartle}}\ and\ \bibinfo {author} {\bibfnamefont {K.~S.}\ \bibnamefont
  {Thorne}},\ }\href {\doibase 10.1086/149707} {\bibfield  {journal} {\bibinfo
  {journal} {Astrophys. J.}\ }\textbf {\bibinfo {volume} {153}},\ \bibinfo
  {pages} {807} (\bibinfo {year} {1968})}\BibitemShut {NoStop}%
\bibitem [{\citenamefont {Pattersons}\ and\ \citenamefont
  {Sulaksono}(2021)}]{Pattersons:2021lci}%
  \BibitemOpen
  \bibfield  {author} {\bibinfo {author} {\bibfnamefont {M.~L.}\ \bibnamefont
  {Pattersons}}\ and\ \bibinfo {author} {\bibfnamefont {A.}~\bibnamefont
  {Sulaksono}},\ }\href {\doibase 10.1140/epjc/s10052-021-09481-2} {\bibfield
  {journal} {\bibinfo  {journal} {Eur. Phys. J. C}\ }\textbf {\bibinfo {volume}
  {81}},\ \bibinfo {pages} {698} (\bibinfo {year} {2021})}\BibitemShut
  {NoStop}%
\bibitem [{\citenamefont {Beltracchi}(2022)}]{Beltracchi:2022vvn}%
  \BibitemOpen
  \bibfield  {author} {\bibinfo {author} {\bibfnamefont {P.}~\bibnamefont
  {Beltracchi}},\ }\href {\doibase 10.1103/PhysRevD.106.104038} {\bibfield
  {journal} {\bibinfo  {journal} {Phys. Rev. D}\ }\textbf {\bibinfo {volume}
  {106}},\ \bibinfo {pages} {104038} (\bibinfo {year} {2022})},\ \Eprint
  {http://arxiv.org/abs/2206.04139} {arXiv:2206.04139 [gr-qc]} \BibitemShut
  {NoStop}%
\bibitem [{\citenamefont {{Chandrasekhar}}\ and\ \citenamefont
  {{Miller}}(1974)}]{Chandra:1974}%
  \BibitemOpen
  \bibfield  {author} {\bibinfo {author} {\bibfnamefont {S.}~\bibnamefont
  {{Chandrasekhar}}}\ and\ \bibinfo {author} {\bibfnamefont {J.~C.}\
  \bibnamefont {{Miller}}},\ }\href {\doibase 10.1093/mnras/167.1.63}
  {\bibfield  {journal} {\bibinfo  {journal} {Mon. Not. R. Astron. Soc.}\
  }\textbf {\bibinfo {volume} {167}},\ \bibinfo {pages} {63} (\bibinfo {year}
  {1974})}\BibitemShut {NoStop}%
\bibitem [{\citenamefont {Beltracchi}\ and\ \citenamefont
  {Posada}(2024)}]{Beltracchi:2023qla}%
  \BibitemOpen
  \bibfield  {author} {\bibinfo {author} {\bibfnamefont {P.}~\bibnamefont
  {Beltracchi}}\ and\ \bibinfo {author} {\bibfnamefont {C.}~\bibnamefont
  {Posada}},\ }\href {\doibase 10.1088/1361-6382/ad1a52} {\bibfield  {journal}
  {\bibinfo  {journal} {Class. Quantum Grav.}\ }\textbf {\bibinfo {volume}
  {41}},\ \bibinfo {pages} {045001} (\bibinfo {year} {2024})},\ \Eprint
  {http://arxiv.org/abs/2305.09544} {arXiv:2305.09544 [gr-qc]} \BibitemShut
  {NoStop}%
\bibitem [{\citenamefont {Uchikata}\ and\ \citenamefont
  {Yoshida}(2014)}]{Uchikata:2014kwm}%
  \BibitemOpen
  \bibfield  {author} {\bibinfo {author} {\bibfnamefont {N.}~\bibnamefont
  {Uchikata}}\ and\ \bibinfo {author} {\bibfnamefont {S.}~\bibnamefont
  {Yoshida}},\ }\href {\doibase 10.1103/PhysRevD.90.064042} {\bibfield
  {journal} {\bibinfo  {journal} {Phys. Rev. D}\ }\textbf {\bibinfo {volume}
  {90}},\ \bibinfo {pages} {064042} (\bibinfo {year} {2014})},\ \Eprint
  {http://arxiv.org/abs/1506.06478} {arXiv:1506.06478 [gr-qc]} \BibitemShut
  {NoStop}%
\bibitem [{\citenamefont {Uchikata}\ and\ \citenamefont
  {Yoshida}(2016)}]{Uchikata:2015yma}%
  \BibitemOpen
  \bibfield  {author} {\bibinfo {author} {\bibfnamefont {N.}~\bibnamefont
  {Uchikata}}\ and\ \bibinfo {author} {\bibfnamefont {S.}~\bibnamefont
  {Yoshida}},\ }\href {\doibase 10.1088/0264-9381/33/2/025005} {\bibfield
  {journal} {\bibinfo  {journal} {Class. Quantum Grav.}\ }\textbf {\bibinfo
  {volume} {33}},\ \bibinfo {pages} {025005} (\bibinfo {year} {2016})},\
  \Eprint {http://arxiv.org/abs/1506.06485} {arXiv:1506.06485 [gr-qc]}
  \BibitemShut {NoStop}%
\bibitem [{\citenamefont {Pani}(2015)}]{Pani:2015tga}%
  \BibitemOpen
  \bibfield  {author} {\bibinfo {author} {\bibfnamefont {P.}~\bibnamefont
  {Pani}},\ }\href {\doibase 10.1103/PhysRevD.95.049902} {\bibfield  {journal}
  {\bibinfo  {journal} {Phys. Rev. D}\ }\textbf {\bibinfo {volume} {92}},\
  \bibinfo {pages} {124030} (\bibinfo {year} {2015})},\ \bibinfo {note}
  {[Erratum: Phys.Rev.D 95, 049902 (2017)]},\ \Eprint
  {http://arxiv.org/abs/1506.06050} {arXiv:1506.06050 [gr-qc]} \BibitemShut
  {NoStop}%
\bibitem [{\citenamefont {Uchikata}\ \emph {et~al.}(2016)\citenamefont
  {Uchikata}, \citenamefont {Yoshida},\ and\ \citenamefont
  {Pani}}]{Uchikata:2016qku}%
  \BibitemOpen
  \bibfield  {author} {\bibinfo {author} {\bibfnamefont {N.}~\bibnamefont
  {Uchikata}}, \bibinfo {author} {\bibfnamefont {S.}~\bibnamefont {Yoshida}}, \
  and\ \bibinfo {author} {\bibfnamefont {P.}~\bibnamefont {Pani}},\ }\href
  {\doibase 10.1103/PhysRevD.94.064015} {\bibfield  {journal} {\bibinfo
  {journal} {Phys. Rev. D}\ }\textbf {\bibinfo {volume} {94}},\ \bibinfo
  {pages} {064015} (\bibinfo {year} {2016})},\ \Eprint
  {http://arxiv.org/abs/1607.03593} {arXiv:1607.03593 [gr-qc]} \BibitemShut
  {NoStop}%
\bibitem [{\citenamefont {Reina}\ and\ \citenamefont
  {Vera}(2015)}]{Reina:2014fga}%
  \BibitemOpen
  \bibfield  {author} {\bibinfo {author} {\bibfnamefont {B.}~\bibnamefont
  {Reina}}\ and\ \bibinfo {author} {\bibfnamefont {R.}~\bibnamefont {Vera}},\
  }\href {\doibase 10.1088/0264-9381/32/15/155008} {\bibfield  {journal}
  {\bibinfo  {journal} {Class. Quantum Grav.}\ }\textbf {\bibinfo {volume}
  {32}},\ \bibinfo {pages} {155008} (\bibinfo {year} {2015})},\ \Eprint
  {http://arxiv.org/abs/1412.7083} {arXiv:1412.7083 [gr-qc]} \BibitemShut
  {NoStop}%
\bibitem [{\citenamefont {{Miller}}(1977)}]{Miller:1977}%
  \BibitemOpen
  \bibfield  {author} {\bibinfo {author} {\bibfnamefont {J.~C.}\ \bibnamefont
  {{Miller}}},\ }\href {\doibase 10.1093/mnras/179.3.483} {\bibfield  {journal}
  {\bibinfo  {journal} {Mon. Not. R. Astron. Soc.}\ }\textbf {\bibinfo {volume}
  {179}},\ \bibinfo {pages} {483} (\bibinfo {year} {1977})}\BibitemShut
  {NoStop}%
\bibitem [{\citenamefont {Schwarzschild}(1916)}]{Schwarzschild:1916inc}%
  \BibitemOpen
  \bibfield  {author} {\bibinfo {author} {\bibfnamefont {K.}~\bibnamefont
  {Schwarzschild}},\ }\href@noop {} {\bibfield  {journal} {\bibinfo  {journal}
  {Sitzungsber. Preuss. Akad. Wiss. Berlin (Math. Phys.)}\ }\textbf {\bibinfo
  {volume} {1916}},\ \bibinfo {pages} {424} (\bibinfo {year} {1916})},\ \Eprint
  {http://arxiv.org/abs/physics/9912033} {arXiv:physics/9912033
  [physics.hist-ph]} \BibitemShut {NoStop}%
\bibitem [{\citenamefont {Mazur}\ and\ \citenamefont
  {Mottola}(2015)}]{Mazur:2015kia}%
  \BibitemOpen
  \bibfield  {author} {\bibinfo {author} {\bibfnamefont {P.~O.}\ \bibnamefont
  {Mazur}}\ and\ \bibinfo {author} {\bibfnamefont {E.}~\bibnamefont
  {Mottola}},\ }\href {\doibase 10.1088/0264-9381/32/21/215024} {\bibfield
  {journal} {\bibinfo  {journal} {Class. Quantum Grav.}\ }\textbf {\bibinfo
  {volume} {32}},\ \bibinfo {pages} {215024} (\bibinfo {year} {2015})},\
  \Eprint {http://arxiv.org/abs/1501.03806} {arXiv:1501.03806 [gr-qc]}
  \BibitemShut {NoStop}%
\bibitem [{\citenamefont {Beltracchi}\ \emph {et~al.}(2022)\citenamefont
  {Beltracchi}, \citenamefont {Gondolo},\ and\ \citenamefont
  {Mottola}}]{Beltracchi:2021lez}%
  \BibitemOpen
  \bibfield  {author} {\bibinfo {author} {\bibfnamefont {P.}~\bibnamefont
  {Beltracchi}}, \bibinfo {author} {\bibfnamefont {P.}~\bibnamefont {Gondolo}},
  \ and\ \bibinfo {author} {\bibfnamefont {E.}~\bibnamefont {Mottola}},\ }\href
  {\doibase 10.1103/PhysRevD.105.024002} {\bibfield  {journal} {\bibinfo
  {journal} {Phys. Rev. D}\ }\textbf {\bibinfo {volume} {105}},\ \bibinfo
  {pages} {024002} (\bibinfo {year} {2022})},\ \Eprint
  {http://arxiv.org/abs/2107.00762} {arXiv:2107.00762 [gr-qc]} \BibitemShut
  {NoStop}%
\bibitem [{\citenamefont {Reina}(2016)}]{Reina:2015jia}%
  \BibitemOpen
  \bibfield  {author} {\bibinfo {author} {\bibfnamefont {B.}~\bibnamefont
  {Reina}},\ }\href {\doibase 10.1093/mnras/stv2599} {\bibfield  {journal}
  {\bibinfo  {journal} {Mon. Not. R. Astron. Soc.}\ }\textbf {\bibinfo {volume}
  {455}},\ \bibinfo {pages} {4512} (\bibinfo {year} {2016})},\ \Eprint
  {http://arxiv.org/abs/1503.07835} {arXiv:1503.07835 [gr-qc]} \BibitemShut
  {NoStop}%
\bibitem [{\citenamefont {{Hartle}}\ and\ \citenamefont
  {{Sharp}}(1967)}]{Sharp:1967}%
  \BibitemOpen
  \bibfield  {author} {\bibinfo {author} {\bibfnamefont {J.~B.}\ \bibnamefont
  {{Hartle}}}\ and\ \bibinfo {author} {\bibfnamefont {D.~H.}\ \bibnamefont
  {{Sharp}}},\ }\href {\doibase 10.1086/149002} {\bibfield  {journal} {\bibinfo
   {journal} {Atrophys. J.}\ }\textbf {\bibinfo {volume} {147}},\ \bibinfo
  {pages} {317} (\bibinfo {year} {1967})}\BibitemShut {NoStop}%
\bibitem [{\citenamefont {Mazur}\ and\ \citenamefont
  {Mottola}(2023)}]{Mazur:2001fv}%
  \BibitemOpen
  \bibfield  {author} {\bibinfo {author} {\bibfnamefont {P.~O.}\ \bibnamefont
  {Mazur}}\ and\ \bibinfo {author} {\bibfnamefont {E.}~\bibnamefont
  {Mottola}},\ }\href {\doibase 10.3390/universe9020088} {\bibfield  {journal}
  {\bibinfo  {journal} {Universe}\ }\textbf {\bibinfo {volume} {9}},\ \bibinfo
  {pages} {88} (\bibinfo {year} {2023})},\ \Eprint
  {http://arxiv.org/abs/gr-qc/0109035} {arXiv:gr-qc/0109035} \BibitemShut
  {NoStop}%
\bibitem [{\citenamefont {Mazur}\ and\ \citenamefont
  {Mottola}(2004)}]{Mazur:2004fk}%
  \BibitemOpen
  \bibfield  {author} {\bibinfo {author} {\bibfnamefont {P.~O.}\ \bibnamefont
  {Mazur}}\ and\ \bibinfo {author} {\bibfnamefont {E.}~\bibnamefont
  {Mottola}},\ }\href {\doibase 10.1073/pnas.0402717101} {\bibfield  {journal}
  {\bibinfo  {journal} {Proc. Nat. Acad. Sci.}\ }\textbf {\bibinfo {volume}
  {101}},\ \bibinfo {pages} {9545} (\bibinfo {year} {2004})},\ \Eprint
  {http://arxiv.org/abs/gr-qc/0407075} {arXiv:gr-qc/0407075} \BibitemShut
  {NoStop}%
\bibitem [{\citenamefont {Lemos}\ and\ \citenamefont
  {Zaslavskii}(2007)}]{Lemos:2007yh}%
  \BibitemOpen
  \bibfield  {author} {\bibinfo {author} {\bibfnamefont {J.~P.~S.}\
  \bibnamefont {Lemos}}\ and\ \bibinfo {author} {\bibfnamefont {O.~B.}\
  \bibnamefont {Zaslavskii}},\ }\href {\doibase 10.1103/PhysRevD.76.084030}
  {\bibfield  {journal} {\bibinfo  {journal} {Phys. Rev. D}\ }\textbf {\bibinfo
  {volume} {76}},\ \bibinfo {pages} {084030} (\bibinfo {year} {2007})},\
  \Eprint {http://arxiv.org/abs/0707.1094} {arXiv:0707.1094 [gr-qc]}
  \BibitemShut {NoStop}%
\end{thebibliography}%
\bibliographystyle{apsrev4-1}
\end{document}